\documentclass[%
reprint,
superscriptaddress,
 amsmath,amssymb,
 aps,
prb,
floatfix,
]{revtex4-1}

\usepackage{graphicx}
\usepackage{dcolumn}
\usepackage{bm}
\usepackage[bookmarks=true,colorlinks=true,linkcolor={blue},citecolor=blue,urlcolor=blue]{hyperref} 
\usepackage[mathlines]{lineno}

\usepackage{longtable}	
\usepackage{tabularx}	
\usepackage{array}		
\usepackage{multirow}	
\usepackage{array}		

\begin{document}


\title{A theoretical investigation of structural, electronic and optical properties of bulk copper nitrides}


\author{Mohammed S. H. Suleiman}
\email[Corresponding author: ]{suleiman@aims.ac.za}
\affiliation{School of Physics, University of the Witwatersrand, Johannesburg, South Africa.}
\affiliation{Department of Physics, Sudan University of Science and Technology, Khartoum, Sudan.}

\author{Mahlaga P. Molepo}
\altaffiliation[Current address: ]{Department of Physics, University of Pretoria, Pretoria, South Africa.}
\affiliation{School of Physics, University of the Witwatersrand, Johannesburg, South Africa.}
 
\author{Daniel P. Joubert}
\affiliation{School of Physics, University of the Witwatersrand, Johannesburg, South Africa.}

\date{\today}

\begin{abstract}
We present a detailed first-principles DFT study of the equation of state (EOS), energy-optimized geometries, phase stabilities and electronic properties of bulk crystalline Cu$_3$N, CuN and CuN$_2$ in a set of twenty different structural phases. We analyzed different structural preferences for these three stoichiometries and determined their equilibrium structural parameters. Band-structure and density of states of the relatively most stable phases were carefully investigated. Further, we carried out GW$_{0}$ calculations within the random-phase approximation (RPA) to the dielectric tensor to investigate the optical spectra of the experimentally synthesized phase Cu$_{3}$N(D0$_{9}$). Obtained results are compared with experiment and with previous calculations.
\end{abstract}


\maketitle

\tableofcontents	
\section{Introduction}\label{Introduction} 
In 1939, Juza and Hahn succeeded to produce Cu$_3$N \cite{the_first_Cu3N_1939_exp} for the first time \cite{Cu3N_1996_comp,Cu3N_2006_exp,nano_Cu3N_2005_exp}. Since then, copper nitride has been prepared in various techniques \cite{Cu3N_2006_exp,Cu3N_Cu4N_Cu8N_2007_comp,copper_nitrides_2001_exp,Cu3N_2009_exp_optical_properties}, its properties and applications have been researched, both theoretically and experimentally, and it may now be considered as the most accessible among the noble metal nitrides \cite{nano_Cu3N_2005_exp}.

Synthesis and reliable characterization of the properties of a stoichiometric copper nitride constitute a big challenge because it is thermally unstable material \cite{Cu3N_2006_exp}. However, this low thermal stability results in promising applications in optical memories and  laser writing \cite{copper_nitrides_2001_exp,nano_Cu3N_2005_exp}.

The viability of using the simple cubic stoichiometric Cu$_3$N films for write-once optical data storage has been widely explored and confirmed \cite{Cu3N_2000_applications,Cu3N_1996_exp,Cu3N_1990_exp}, superior to other toxic and unstable materials in air at room temperature which are used for the same purpose \cite{Cu3N_1996_exp}. Also, the feasibility of using Cu$_3$N as a coating to generate metal lines by maskless laser writing has been studied; where conducting lines of a few micron in width could be generated with resistivities within an order of magnitude of the bulk Cu metal \cite{Cu3N_Ni3N_1993_exp_conference}. This interesting material has been suggested for usage in a number of nano-electronic and nano-photonic devices \cite{Cu3N_2009_exp_optical_properties}.

Depending on the total sputtering pressure and on the content of nitrogen gas, Hayashi et al. \cite{metallic_Cu3N_1998_exp} prepared four categories of sputter-deposited Cu--N films: metallic Cu--rich Cu$_3$N, semiconducting Cu-rich Cu$_3$N, semiconducting stoichiometric Cu$_3$N and semiconducting N--rich Cu3N 
films. In general, it has been reported that it is possible to achieve sub-, over- and stoichiometric copper nitrides \cite{Cu3N_Cu4N_2006_exp_just_read_it,copper_nitrides_optical_properties_2001_exp}, and the effect of the nitrogen to copper ratio on the physical properties has been studied by many researchers \cite{Cu3N_Cu4N_Cu8N_2007_comp,Cu3N_Cu4N_Cu3N2_2008_comp,Cu3N_Cu4N_2006_exp_just_read_it,copper_nitrides_optical_properties_2001_exp}.

The structural properties of Cu$_3$N in the experimentally reported cubic anti-ReO$_3$ phase are interesting on their own. This structure has many vacant interstitial sites like WO$_3$. The latter could be made into a conductor by doping it with some metal ions \cite{Cu3N_1989_exp}. This is very suggestive, since one may be able to engineer the physical properties of such technologically important material \cite{Cu3N_Cu4N_Cu3N2_2008_comp,Cu3N_2004_comp}. In fact, the study of \textit{possible} intercalated copper nitride alloys has been an active subject of research on its own (cf. Ref. [\onlinecite{Cu3N_Cu4N_Cu8N_2007_comp}] and references therein).

Although copper nitride possesses interesting properties leading to different technological applications, there is still a large discrepancy in the formation mechanism and inconsistency in the experimentally reported and in the theoretically predicted physical properties of copper nitrides \cite{Cu3N_2006_exp,Cu3N_Cu4N_Cu8N_2007_comp,Cu3N_Cu4N_Cu3N2_2008_comp,Cu3N_2009_exp_optical_properties,CuxNy_2009_ooooh}. These differences and contradictions are stemming mainly from the unstable nature (i.e. the metastability and low decomposition temperature) of copper nitride \cite{Cu3N_2006_exp,Cu3N_2009_exp_optical_properties,Cu3N_thin_films_2005_exp}, the experimental conditions \cite{Cu3N_Cu4N_Cu3N2_2008_comp}, the experimental analysis methods \cite{Cu3N_2006_exp}, the non-stoichiometry of the prepared samples \cite{metallic_Cu3N_1998_exp} or the lack of knowledge of the real stoichiometry of the prepared samples \cite{Cu3N_2009_exp_optical_properties}; and from the different theoretical calculation methods and approximations \cite{Cu3N_Cu4N_Cu3N2_2008_comp}.

Thus, the emerging potential technological applications of copper nitride are faced by the inconsistency in its basic physical properties. This may explain the tremendously increasing interest in further studying this material, especially within first-principles quantum mechanical approaches. Moreover, concerning its optical properties, only a few experiments are available in the literature \cite{copper_nitride_films_2003_exp} and there are very few theoretical studies.

Motivated by all these, and searching for a wider range of possible applications, we present in the current work first-principles calculations on bulk crystalline Cu$_3$N, CuN and CuN$_2$ over a series of reported and theoretical structural phases. The studied structural properties include energy-volume equation of state (EOS), equilibrium lattice structural parameters, cohesive and formation energies, relative phase stabilities, bulk modulus and its pressure derivative. Electronic characterization of the energetically most stable phases was done via the analysis of their band structure and their total and partial density of states (DOS). In order to improve the calculated electronic structure, and to investigate the optical spectra, we carried out $GW0$ calculations within the the random-phase approximation (RPA) to the dielectric tensor. The frequency-dependent optical constants (absorption coefficient, reflectivity and refractive index spectra) of the experimentally reported phase Cu$_{3}$N(D0$_{9}$) were derived from the calculated frequency-dependent microscopic dielectric tensor.

We hope that the present work would serve as a reference source for meaningful comparisons which may be made among the largely different calculations.
%
\section{\label{Stoichiometries and Crystal Structures}Stoichiometries and Crystal Structures}
There had been no known binary nitrides of the noble metals until Gregoryanz et al. \cite{1stPtN2004} reported the discovery and characterization of crystalline PtN. From their results, they strongly suggested that it would be possible to synthesize other novel nitrides with late transition metals such as those in the Ni and Cu groups. They claimed that such nitrides would have potentially intriguing physical properties, and that their results should stimulate further theoretical studies \cite{1stPtN2004}.

To the best of our knowledge, the only experimentally reported stoichiometries of copper nitride are Cu$_3$N \cite{the_first_Cu3N_1939_exp,Cu3N_Cu4N_2006_exp_just_read_it,Cu3N_Cu4N_Cu6N_2004_exp} and Cu$_4$N \cite{Cu4N_1989_exp,Cu3N_Cu4N_2006_exp_just_read_it,Cu3N_Cu4N_Cu6N_2004_exp}, while CuN and CuN$_2$ have not been observed yet. However, many transition-metal nitrides (TMNs) are known to form more than one nitride \cite[p. 835]{StructuralInChem}. Our interest in the latter two nitride stoichiometries is based on the fact that for other late transition metals close to Cu in the periodic table these 1:1 and 1:2 nitrides have been reported; as will shortly be mentioned.

In general, the recent success in experimentally producing some late TMNs has motivated many researchers to further investigating the possibility of synthesizing other TMNs \cite{CuN_AgN_AuN_2007_comp} in those reported stoichiometries and structures. Thus, it is of interest to know whether copper can form CuN and CuN$_2$ with the reported structures of other TMNs.

In this work, we investigate Cu$_3$N in the following seven structures:
D0$_3$ (the fcc structure of AlFe$_3$, space group Fm$\bar{3}$m No. 225, $Z = 1$)\footnote{$Z$ here, and in what follows, refers to the number of chemical formula units per unit cell.},              
A15 (the sc structure of Cr$_3$Si, space group Pm$\bar{3}$n No. 223, $Z = 2$),                  
D0$_9$ (the sc structure of the anti-ReO$_3$ ($\alpha$), space group Pm$\bar{3}$m No. 221, $Z = 1$),                    
L1$_2$ (the sc structure of Cu$_3$Au, space group Pm$\bar{3}$m No. 221, $Z = 1$),               
D0$_2$ (the bbc structure of CoAs$_3$ skutterudite, space group Im$\bar{3}$ No. 204, $Z = 4$),
$\epsilon$-Fe$_3$N (the hexagonal structure of $\epsilon$-Fe$_3$N, space group P6$_{3}$22 No. 182, $Z = 2$)
and RhF$_3$  (the trigonal (rhombohedric) structure of RhF$_3$, space group R$\bar{3}$c No. 167, $Z = 2$).    

CuN was investigated in the following nine structures:
B1 (the fcc structure of NaCl, space group Fm$\bar{3}$m No. 225, $Z = 1$) which is the most popular structure for TMNs \cite{Pearson,AgN2_AuN2_PtN2_2005_comp} and many early TMNs have been synthesized in this structure \cite{CuN_1997_comp},
B2 (the sc structure of CsCl, space group Pm$\bar{3}$m No. 221, $Z = 1$),
B3 (the fcc structure of ZnS zincblende, space group    F$\bar{4}3$m No. 216, $Z = 1$) which is the structure of the first synthesized binary nitride of the noble metals group PtN \cite{1stPtN2004} ,
B8$_{1}$ (the hexagonal structure of NiAs, space group P$6_{3}$/mmc No. 194, $Z = 2$),
B$_{\text{k}}$ (the hexagonal structure of BN, space group P$6_{3}$/mmc No. 194, $Z = 2$),
B$_{\text{h}}$ (the hexagonal structure of WC, space group P$\bar{6}$m$2$ No. 187, $Z = 1$),
B4 (the hexagonal structure of ZnS wurtzite, space group P$6_{3}$mc No. 186, $Z = 2$),
B17 (the s tetragonal structure of PtS cooperite, space group P$4_{2}$/mmc No. 131, $Z = 2$) which was theoretically predicted to be the ground-state structure of PtN \cite{PtN_2006_comp_B17_structure_important},    
and B24 (the fc orthorhombic structure of TlF, space group Fmmm No. 69, $Z = 1$).  

For CuN$_2$, the following four structures were considered:
C1 (the fcc structure of CaF$_{2}$ fluorite, space group Fm$\bar{3}$m No. 225, $Z = 1$) in which PtN$_{2}$ was theoretically predicted to be stabilized \cite{PtN2_2005_comp},
C2 (the sc structure of FeS$_{2}$ pyrite, space group Pa$\bar{3}$ No. 205, $Z = 4$),    
C18 (the s orthorhombic structure of FeS$_{2}$ marcasite, space group Pnnm No. 58, $Z = 2$)
and CoSb$_{2}$ (the s monoclinc structure of CoSb$_{2}$, space group P2$_1$/c No. 14, $Z = 4$).

Apart from the synthesized Cu$_{3}$N(D0$_{9}$), we consider these stoichiometries and most of these structures because there have been some experimental reports or previous theoretical investigations on copper nitrides or on other TMNs, as  will be pointed out in the text. Nevertheless, our first aim is to examine some possible structures of the infinite parameter space in order to identify the most energetically favorable candidates.
%
\section{\label{Electronic Optimization Details}Electronic Optimization Details}
Our electronic structure calculations were based on spin density functional theory (SDFT) \cite{SDFT_1972,SDFT_Pant_1972} as implemented in the all-electron Vienna \textit{ab initio} Simulation Package (VASP) \cite{Vasp_ref_PhysRevB.47.558_1993,Vasp_ref_PhysRevB.49.14251_1994,Vasp_cite_Kressw_1996,Vasp_PWs_Kresse_1996,DFT_VASP_Hafner_2008,PAW_Kresse_n_Joubert}. SDFT is the most widely used \cite{A_birds_eye} form of density functional theory (DFT) \cite{HK_1964,KS_1965}. Its main advantage over the original Hohenberg-Kohn-Sham DFT formalism is that it enables us to build in more of the actual physics into the approximate functionals with greater flexibility \cite{XC_SDFT_1976}.

In solving the self-consistent Kohn-Sham (KS) Schr\"{o}dinger-like eigenvalue equations \cite{KS_1965}
\begin{eqnarray}	\label{KS equations}
\begin{split}
  \Bigg \{ - \frac{\hbar^{2}} {2m_{e}}  \nabla^{2} + \int d\mathbf{r}^{\prime} \frac{n(\mathbf{r}^{\prime})}{|\mathbf{r}-\mathbf{r}^{\prime}|} + V_{ext}(\mathbf{r}) \\  + V_{XC}^{\sigma, \mathbf{k}}[n(\mathbf{r})] \Bigg \} \psi_{i}^{\sigma, \mathbf{k}}(\mathbf{r})  =  
   \epsilon_{i}^{\sigma, \mathbf{k}} \psi_{i}^{\sigma, \mathbf{k}}(\mathbf{r}),
\end{split}
\end{eqnarray}
VASP expands the pseudo part of the Kohn-Sham one-electron spin orbitals $\psi_{i}^{\sigma , \mathbf{k}}(\mathbf{r})$ on a basis set of plane-waves (PWs). We included only those PWs with kinetic energy $\frac{\hbar^{2}} {2m_{e}} |\mathbf{k} + \mathbf{G}| < E_{cut}$, such that the change in total electronic energy and in the so-called Fermi energy $E_{F}$ corresponding to an increase in this energy cut-off $E_{cut}$ by $100 \; eV$ is less than $3 \; \text{m} eV/ \text{atom}$ and $2 \; \text{m}eV$, respectively. This is always met by $E_{cut} = 600 \; eV$ for the systems investigated.

The lattice translation symmetry of a crystalline solid manifests itself in the quantum number $\mathbf{k}$. In principle, there is one equation of type (\ref{KS equations}) for each $\mathbf{k}$-point in the first Brillouin zone (BZ), and the expectation value $\langle O \rangle$ of any operator $O$ is obtained by integrating its matrix elements $\left\langle    \psi_{i}^{\sigma , \mathbf{k}}(\mathbf{r})   |   O    |   \psi_{i}^{\sigma , \mathbf{k}}(\mathbf{r})  \right\rangle$ over all occupied bands in the $\mathbf{k}$-space. For example, the electronic spin density $n_{\sigma}(\mathbf{r})$, which also couples equations (\ref{KS equations}) above, is given by
\begin{eqnarray}	\label{electronic spin density}
 n_{\sigma}(\mathbf{r}) = \displaystyle\sum_{\mathbf{k} \in BZ} \omega_{\mathbf{k}}  \displaystyle\sum_{i=1}^{N_{\mathbf{k}, \sigma}} f_{i}^{\sigma, \mathbf{k}} |\psi_{i}^{\sigma , \mathbf{k}}(\mathbf{r})|^{2} \text{;}
\end{eqnarray}
where the index $\sigma$ indicates the spin component ($\uparrow$ or $\downarrow$), $N_{\mathbf{k}, \sigma}$ is the number of the occupied single-electron eigenstates $\psi_{i}^{\sigma , \mathbf{k}}(\mathbf{r})$ (with spin projection $\sigma$) at each $\mathbf{k}$-point of the sampled BZ, and $f_{i}^{\sigma, \mathbf{k}}$ are the corresponding occupation numbers. The weights $\omega_{\mathbf{k}}$ should satisfy
\begin{eqnarray}
 \displaystyle\sum_{\mathbf{k} \in BZ} \omega_{\mathbf{k}}  N_{\mathbf{k}, \sigma} = N_{\sigma};
\end{eqnarray}
where $N_{\sigma}$ is the total number of electrons with spin $\sigma$.

For performing BZ integrations, our BZs were sampled using $\mathbf{\Gamma}$-centered Monkhorst-Pack meshes \cite{MP_k_mesh_1976}. We found that a $17 \times 17 \times 17$ mesh corresponds to a number of $\mathbf{k}$-points in the irreducible wedge of the Brillouin zone (IBZ) which is always dense enough such that any increase in the density of the mesh produces a change in the total energy less than $2$ meV/atom, accompanied by a change in $E_{F}$ of less than $0.02$ eV.

For static calculations of the total electronic energy and the density of states (DOS), partial occupancies $f_{i}^{\sigma, \mathbf{k}}$ were set using the tetrahedron method with Bl\"{o}chl corrections \cite{tetrahedron_method_theory_1971,tetrahedron_method_theory_1972,ISMEAR5_1994}; while in the ionic relaxation, the smearing method of Methfessel-Paxton (MP) \cite{MP_smearing_1989} was used. In doing this, the Fermi surface has been carefully treated and the smearing width was chosen such that the fictitious entropy - introduced by the smearing occupation scheme - has been kept always below $1 \; \text{m}eV/\text{atom}$.

The generalized gradient approximation (GGA) \cite{XC_GGA_1988,XC_GGA_applications_1992,XC_GGA_applications_1992_ERRATUM} was used for the exchange-correlation potentials $V_{XC}^{\sigma, \mathbf{k}}[n(\mathbf{r})]$, where the Perdew-Burke-Ernzerhof (PBE) parametrization \cite{PBE_GGA_1996,PBE_GGA_Erratum_1997,XC_PBE_1999} is applied. The electron-ion interactions - the third term in Eq.(\ref{KS equations}) - were described by the projector augmented wave (PAW) method \cite{PAW_Blochl, PAW_Kresse_n_Joubert}. The PAW potential explicitly treats the $5$ electrons of $2s^{2}2p^{3}$ as valence electrons for nitrogen; while the $11$ electrons of $3d^{10}4p^{1}$ are treated as valence electrons in the Cu case, assuming completely filled $d$ shell and placing all other electrons into the atomic core. Concerning relativistic effects, VASP performs a fully relativistic calculation for the core electrons, while for valence electrons only scalar kinematic relativistic effects are incorporated in the PAW potential via mass-velocity and Darwin corrections \cite{DFT_VASP_Hafner_2008}. We have not considered spin-orbit interaction (SOI) of the valence electrons.

The implemented blocked Davidson iteration scheme \cite{Davidson_original_article_1975} was chosen for the relaxation of the electronic degrees of freedom. Convergence was considered to be achieved when the change in the total energy and in the eigenvalues between two successive self-consistent (SC) steps are both smaller than $1 \times 10^{-4} \; eV$.
%
%
\section{Geometry Relaxation and Relative Stabilities} \label{Geometry Relaxation and Relative Stabilities}
To study the energy-volume $E(V)$ equation of state (EOS), and to determine the equilbrium parameters of each structure, we make isotropic variation of the cell volume while ions with free internal parameters are allowed to search for local minima on the Born-Oppenheimer potential hyper-surface \cite{Born_Oppenheimer_1927}, following the implemented conjugate-gradient (CG) algorithm \cite{Richard_Martin}, untill all Hellmann-Feynman force components \cite{Hellmann–Feynman_theorem} on each ion are smaller than $1 \times 10^{-2} \; eV/\text{\AA}$.

Cohesive energy $E_{coh}$ of a solid is defined relative to a state with all atoms neutral and infinitely separated \cite{Grimvall}. Thus, in practice, it corresponds to the difference between the crystal energy per unit cell and the total energy of the isolated atoms \cite{Cu3N_Cu4N_Cu8N_2007_comp}
\begin{eqnarray} \label{general_E_coh equation}
E_{coh}  =   E(\text{crystal}) - E(\text{atoms}).
\end{eqnarray}
Thus, cohesive energy per atom can be expressed, in our case, as\footnote{Eq. \ref{E_coh equation} results in a negative $E_{coh}$. However, another convention with positive $E_{coh}$ is also common, where energy signs in Eq. \ref{E_coh equation} change.} 
\begin{eqnarray} \label{E_coh equation}
E_{coh}^{\text{Cu}_{m}\text{N}_{n}}  =   \frac{  E_{\text{solid}}^{\text{Cu}_{m}\text{N}_{n}} - Z \times \left( m E_{\text{atom}}^{\text{Cu}} + n E_{\text{atom}}^{\text{N}} \right) }{Z \times (m + n)},
\end{eqnarray}
where $Z$ is the number of Cu$_{m}$N$_{n}$ per unit cell, $E_{\text{atom}}^{\text{Cu}}$ and $E_{\text{atom}}^{\text{N}}$ are the atomic energies, and $m,n = 1,2 \text{ or } 3$ are the stoichiometric weights.

Both crystal and atomic energies must be calculated at the same level of accuracy \cite{Cu3N_Cu4N_Cu8N_2007_comp,interatomic_potentials}. VASP, however, calculates cohesive energies with respect to spherical non spin-polarised reference atoms \cite{VASPguide}. Moreover, Vasp, in principle, allows only for the use of periodic systems. Thus, after being placed in an orthorhombic cell with $13 \; \text{\AA} \times 14 \; \text{\AA} \times 15 \; \text{\AA}$ dimensions, the energy of each isolated spin polarised pseudo-atom (with the same foregoing electronic configuration) was calculated. The large dimensions of the cell ensures that there is no significant interaction between the atom and its images; while the physically incorrect spherical states are avoided by means of the orthorhombic symmetry (cf. Ref. [\onlinecite{B_prime_Aug_2012_comp}] and Ref. 28 therein.)\footnote{It is also well known that GGA may slightly lower the ground-state energy when a nonspherical ground-state density is allowed for (cf. Ref. [\onlinecite{PhysRevB.65.245212}] and Ref. 46 therein).}. $\Gamma$ point and Gaussian smearing method with a small width of $0.002 \; eV$ were used, and the obtained atomic energies were subtracted manually from the cohesive energies $E_{\text{solid}}^{\text{Cu}_{m}\text{N}_{n}}$ calculated by VASP.

$E_{coh}$ represents the energy needed to decompose the solid into its atomic constituents \cite{Ashcroft1976}. Hence, those phases with the lower $E_{coh}$ are the relatively most stable. So, in order to investigate the relative stabilities of the phases under consideration, the obtained $E_{coh}$ as a function of volume $V$ per atom were fitted \cite{eos_f90_code} to a Birch-Murnaghan 3rd-order equation of state (EOS) \cite{BM_3rd_eos}
\begin{eqnarray} \label{3rd_BM_eos}
\begin{split}
E(V) = E_{0} +  \frac{9V_{0}B_{0}}{16}  \left(    \left[ \left( \frac{V_{0}}{V} \right)^{\frac{2}{3} } - 1 \right]^{3} B_{0}^{\prime}      \right.  \\ \left.      +  \left[ \left( \frac{V_{0}}{V} \right)^{\frac{2}{3} } - 1 \right]^{2} \left[6 - 4 \left( \frac{V_{0}}{V} \right)^{\frac{2}{3} }\right]    \right) ,
\end{split}
\end{eqnarray}
where $V_{0}$, $E_{0}$, $B_{0}$ and $B_{0}^{\prime}$ are the equilibrium volume, the equilibrium cohesive energy, the equilibrium bulk modulus  and its pressure derivative, respectively. These four equilibrium fitting parameters were determined by a least-squares method.

\begin{figure}[!]
\includegraphics[width=0.4\textheight]{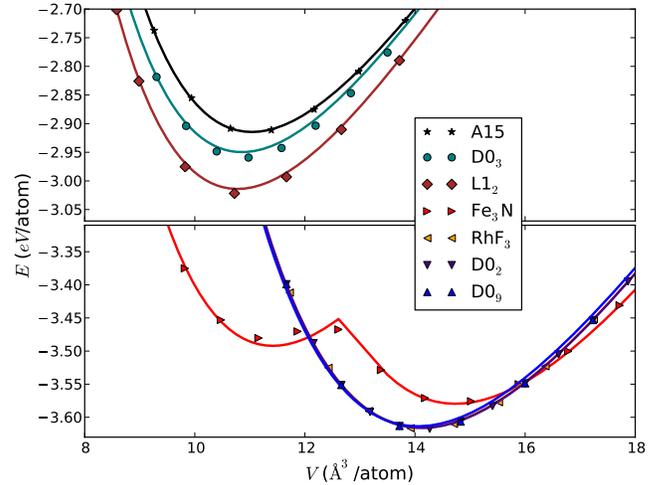}
\caption{\label{Cu3N1_ev_EOS}(Color online.) Cohesive energy $E_{\text{coh}} (eV/\text{atom})$ versus atomic volume $V$ (\AA$^{3}$/\text{atom}) for Cu$_3$N in seven different structural phases.}
\end{figure}

\begin{figure}[!]
\includegraphics[width=0.4\textheight]{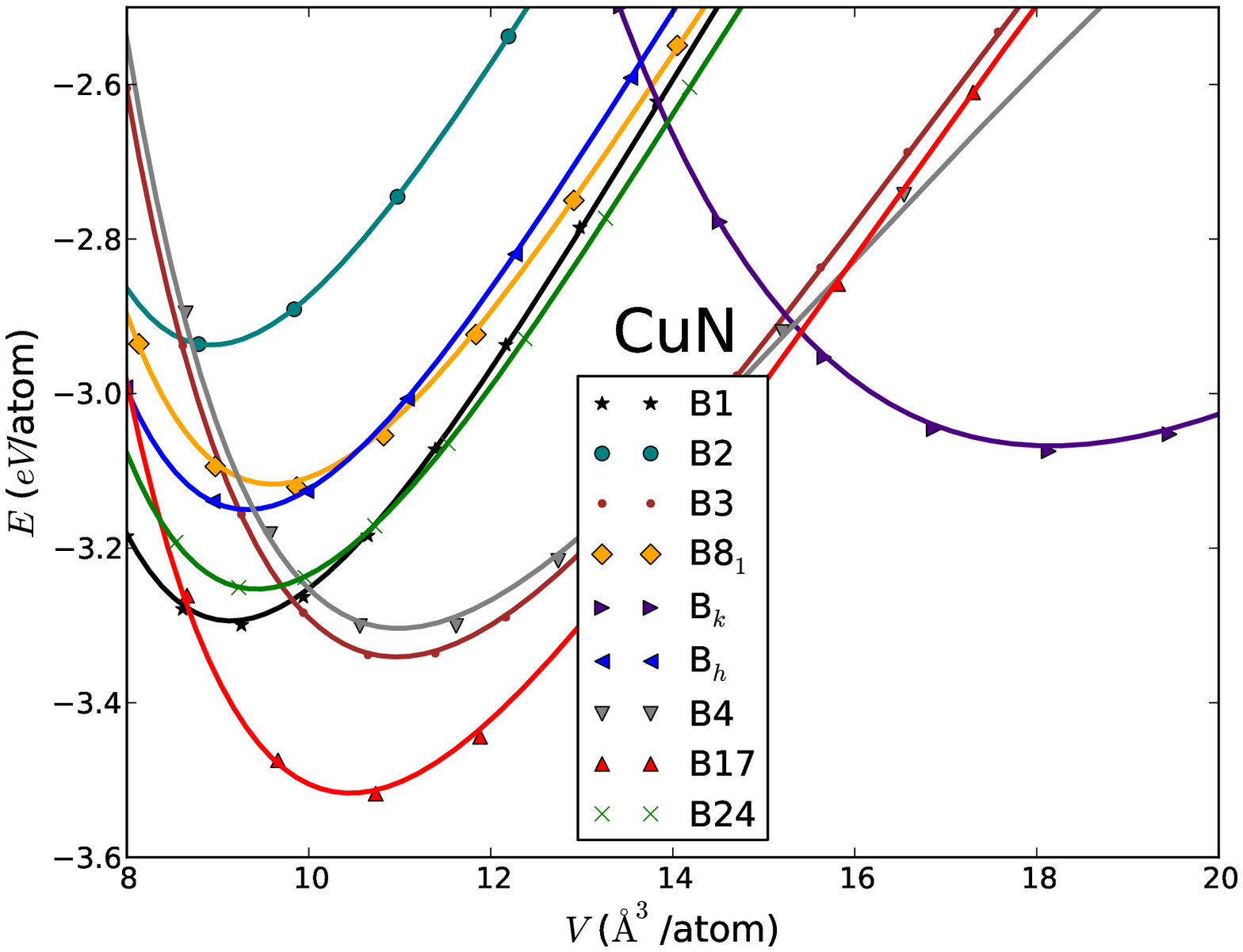}
\caption{\label{Cu1N1_ev_EOS}(Color online.) Cohesive energy $E_{\text{coh}} (eV/\text{atom})$ versus atomic volume $V$ (\AA$^{3}$/\text{atom}) for CuN in nine different structural phases.}
\end{figure}

\begin{figure}[!]
\includegraphics[width=0.4\textheight, height=0.4\textheight]{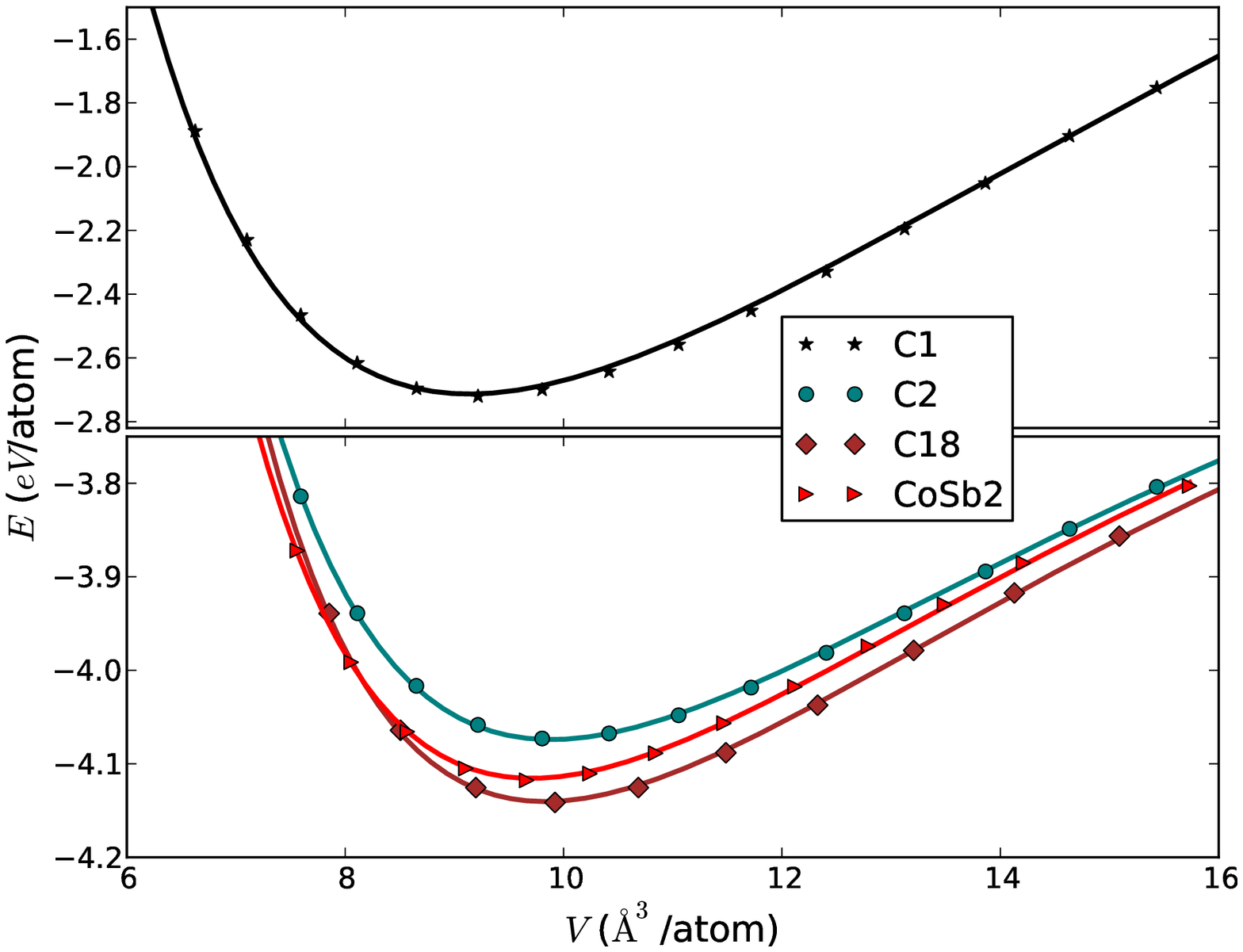}
\caption{\label{Cu1N2_ev_EOS}(Color online.) Cohesive energy $E_{coh} (eV/\text{atom})$ versus atomic volume $V (\AA^{3}/\text{atom})$ for CuN$_2$ in four different structural phases.}
\end{figure}

Cohesive energy versus atomic volume data for the different phases of Cu$_3$N, CuN$_2$ and CuN are visualized graphically in Fig. \ref{Cu3N1_ev_EOS}, Fig. \ref{Cu1N1_ev_EOS} and Fig. \ref{Cu1N2_ev_EOS}, respectively. The corresponding obtained  equilibrium structural parameters and energetic and elastic properties are presented in Table \ref{copper_nitrides_equilibrium_structural_properties}. In this table, phases are first grouped according to the nitrogen content, starting with the stoichiometry with the lowest nitrogen content Cu$_3$N, followed by the 1:1 phases and ending with the nitrogen-richest CuN$_2$ ones. Within each group, phases are ordered according to their structural symmetry, starting from the highest symmetry (i.e. space group) to the least symmetry.  Our results are compared with available experiment and with previous theoretical calculations; with the calculation methods and $XC$ functional pointed out in the Table footnotes whenever appropriate.
%
%
%
\begin{table*} 
\caption{\label{copper_nitrides_equilibrium_structural_properties}Calculated and experimental zero-pressure properties of the twenty studied phases of Cu$_{3}$N, CuN and CuN$_2$\textbf{:} Lattice constants ($a(\text{\AA})$, $b(\text{\AA})$, $c(\text{\AA})$, $\alpha(^{\circ})$ and $\beta(^{\circ})$), equilibrium atomic volume $V_{0}(\text{\AA}^{3}/$atom$)$, cohesive energy $E_{\text{coh}} (eV/$atom$)$, bulk modulus $B_{0} (GPa)$ and its pressure derivative $B_{0}^{\prime}$, and formation energy $E_{f}(eV/\text{atom})$. The presented data are of the current work (\textit{Pres.}), experimentally reported (\textit{Exp.}) and of previous calculations (\textit{\textit{Comp.}}).}
\resizebox{1.0\textwidth}{!}{
\setlength{\extrarowheight}{3pt}
\begin{tabular}{lllllllllll}
\hline
\textbf{Structure}	&		& $a(\AA)$		   & $b(\AA)$			& $c(\AA)$		  & $\alpha(^{\circ})$ or $\beta(^{\circ})$		& $V_{0} (\AA^{3}/$atom$)$   & $E_{\text{coh}}(eV/\text{atom})$	& $B_{0}(\text{GPa})$		& $B_{0}^{\prime}$ & $E_{f}(eV/\text{atom})$\\
\hline \hline 
                		\multicolumn{11}{c}{\textbf{Cu}} \\
\hline 
\multirow{3}{*}{\textbf{A1}\footnotemark[1]\textsuperscript{,}\footnotemark[2]}	& \textit{\textit{Pres.}}.	 &	 $3.636$  &--  &-- &--  &  $12.02$        &  $-3.474$   &   $136.351$          &  $5.032$ &  --\\
                               		& \textit{Exp.}	&	 ($3.6148 \pm 0.0003$)\footnotemark[3], ($3.6077 \pm 0.0002)$\footnotemark[6]	&--  &-- &--                            &  $11.811$\footnotemark[7], $11.810$\footnotemark[6]	&	 $-3.49$\footnotemark[9]  &  $137$\footnotemark[9], $137$\footnotemark[16] &  $5.48$\footnotemark[17] & --\\
                               		& \textit{Comp.}	&	 $3.52$\footnotemark[4]\textsuperscript{,}\footnotemark[5] &--  &-- &--  &  $11.009$\footnotemark[8] &  $-4.29$\footnotemark[10], $-3.12$\footnotemark[11], $-3.30$\footnotemark[12],  $-4.66$\footnotemark[13], $-3.69$\footnotemark[14]\textsuperscript{,}\footnotemark[15] &  $189$\footnotemark[4], $190$\footnotemark[5] &   $4.46$\footnotemark[18], $5.20$\footnotemark[19], $5.14$\footnotemark[20]	& --\\     
\hline 
                		\multicolumn{11}{c}{\textbf{Cu$_3$N}} \\
\hline 
\textbf{D0$_3$}        &\textit{\textit{Pres.}}.	&	 $5.585$  & --    & --  & --         &  $10.89$      &  $-2.960$    & $142.829$  &\textit{\textit{Pres.}}.: $4.845$	&  $0.944$ \\

\hline 
\textbf{A15}           &\textit{Pres.}	&	 $4.455$     & --  & --    & --       &  $11.05$                     &  $-2.915$    &  $138.164$  &  $4.845$ &  $0.989$ \\

\hline 
\multirow{4}{*}{\textbf{D0$_9$}}        &\textit{Pres.}	&	 $3.827$   & --  & -- & --         &  $14.05$   &  $-3.614$                     &  $112.5$     &  $4.899$ &  $0.287$ \\
                                        &\multirow{2}{*}{\textit{Exp.}}	&	 ($3.810 \sim 3.830$)\footnotemark[21], $3.815$\footnotemark[24], $3.83$\footnotemark[25], $3.82$\footnotemark[33]
  & \multirow{2}{*}{--} & \multirow{2}{*}{--} & \multirow{2}{*}{--}                     &                      &                      &           &  \\

&	&	 $(3.830 \pm 0.005)$\footnotemark[26], $<3.868$\footnotemark[30], $3.855$\footnotemark[27]	& &&&&&&&\\
	
                                        &\textit{Comp.}	&	 $3.846$\footnotemark[22], $3.82$\footnotemark[23]\textsuperscript{,}\footnotemark[31], $3.826$\footnotemark[28], $3.841$\footnotemark[29], $3.83$\footnotemark[32] & -- & -- & --                    &  $13.94$\footnotemark[23], $14.02$\footnotemark[31]  &  $-4.863$\footnotemark[23], $-4.865$\footnotemark[31]       &  $115.2$\footnotemark[23], $116$\footnotemark[28], $104$\footnotemark[31] &  $4.066$\footnotemark[23], $4.47$\footnotemark[28], $5.26$\footnotemark[31]\\

\hline 
\multirow{2}{*}{\textbf{L1$_2$}}        &\textit{Pres.}	&	 $3.507$  & -- & -- & --  &  $10.78$  &  $-3.022$                     &  $147.516$         &  $4.817$\\
                                        &\textit{Comp.}	&	 $3.50$\footnotemark[28]   &--  & -- & --                   &                      &                      &  $153$\footnotemark[28]    &  $4.74$\footnotemark[28]	&  $0.882$\\

\hline 
 \textbf{D0$_2$}        &\textit{Pres.}	&	 $7.674$  & -- & --  & --                     &  $14.12$     &  $-3.616$  &  $111.776$  &  $4.757$	&  $0.286$ \\

\hline 
 \textbf{$\epsilon$-Fe$_3$N}	&	\textit{Pres.}	&	 $5.263$  & --     &  $4.905$  & --                     &  $14.71$   &  $-3.579$  &  $109.798$  &  $4.819$ 	&  $0.325$ \\

\hline 
 \textbf{RhF$_3$}       &\textit{Pres.}	&	 $5.426$ & --  & --      &  $\alpha=60.003$                    &  $14.12$ &  $-3.615$ &  $111.192$   &  $4.758$ 	&  $0.286$ \\
%
%
\hline 
                   \multicolumn{11}{c}{\textbf{CuN}} \\
\hline 
\multirow{2}{*}{\textbf{B1}}   		&\textit{Pres.}	&	 $4.182$        &  --                   & --                  &  --                  &  $9.143$           &  $-3.300$          &  $200.770$  &  $4.687$	&  $1.035$\\
             		& \textit{Comp.}       		& 	$4.185$\footnotemark[34], $4.05$\footnotemark[35], $4.336$\footnotemark[36], $4.074$\footnotemark[37], $4.17$\footnotemark[38] & --		       & --			   & --		       &			   &   & 	$201.60$\footnotemark[34], $307$\footnotemark[35], $244.27$\footnotemark[36], $257.46$\footnotemark[37], $207$\footnotemark[38]	&  $3.811$\footnotemark[34], $4.491$\footnotemark[37] 	 \\

\hline 
\multirow{2}{*}{\textbf{B2}}   		&\textit{Pres.}	&	 $2.615$           & --                  & --             & --        &  $8.936$              &  $-2.937$             &  $195.896$      &  $4.775$ &  $1.398$\\
                               		&\textit{Comp.}	&	 $2.61$\footnotemark[34], $2.54$\footnotemark[37], $2.51$\footnotemark[38]  & --                  & --                 & --                &                      &    &  $200.01$\footnotemark[34], $265.40$\footnotemark[37], $196$\footnotemark[38] &  $4.352$\footnotemark[34], $4.373$\footnotemark[37] 		& \\     

\hline 
\multirow{2}{*}{\textbf{B3}}   		&\textit{Pres.}	&	 $4.445$            & --         & --           & --                   &  $10.98$              &  $-3.343$          &  $161.726$     &  $4.677$ &  $0.992$\\
                               		&\textit{Comp.}	&	$4.447$\footnotemark[34], $4.34$\footnotemark[35],  $4.078$\footnotemark[36], $4.341$\footnotemark[37], $4.44$\footnotemark[38]    & --          & --         & --                    &                      &                      &  $164.96$\footnotemark[34], $305$\footnotemark[35], $240.66$\footnotemark[36], $212.16$\footnotemark[37], $158$\footnotemark[38] & $4.534$\footnotemark[34], $4.311$\footnotemark[37] 	& \\     

\hline 
\multirow{2}{*}{\textbf{B8$_{1}$}}   		&	\textit{Pres.}	&	 $3.174$         & --        &  $4.415$            & --                 &  $9.603$            &  $-3.128$       &  $184.371$     &  $4.850$ &  $1.211$\\
                              		&	\textit{Comp.}	&	 $3.08$\footnotemark[38] & --    &   $5.020$\footnotemark[38]              & --        &                      &                      &  $227$\footnotemark[38]   &   \\     

\hline 
 \textbf{B$_{\text{k}}$}	&	\textit{Pres.}	&	 $3.160$      & --              &  $8.406$    & --            &  $18.17$        &  $-3.074$  &  $86.124$    &  $4.494$	&  $1.261$\\

\hline 
 \textbf{B$_{\text{h}}$}	&	\textit{Pres.}	&	 $2.805$      & --              &  $2.738$       & --          &  $9.327$         &  $-3.149$     &  $192.9$     &  $4.779$ 	&  $1.186$\\

\hline 
\multirow{2}{*}{\textbf{B4}}   		&	\textit{Pres.}	&	 $3.148$    & --        &  $5.155$     & --                   &  $11.06$        &  $-3.309$   &  $152.956$      &  $4.963$ &  $1.026$ \\
                               		&\textit{Comp.}	&	 $3.17$\footnotemark[34], $3.077$\footnotemark[37], $3.16$\footnotemark[38]     & --         &  $5.16$\footnotemark[34], $5.016$\footnotemark[37], $5.151$\footnotemark[38]  & --     &                      &                      &  $157.85$\footnotemark[34], $202.10$\footnotemark[37], $155$\footnotemark[38]  &  $4.41$\footnotemark[34], $4.35$\footnotemark[37] \\     

\hline 
 \textbf{B17}   	&\textit{Pres.}	&	 $2.870$  & --  &  $5.052$   & --       &  $10.40$   &  $-3.509$  &  $174.324$  &  $4.948$ 	&  $0.818$\\

\hline 
 \textbf{B24}   	&\textit{Pres.}	&	 $3.928$   &  $4.167$    &  $4.611$    & --   &  $9.435$  &  $-3.253$   &  $189.745$    &  $4.708$ &  $1.082$\\
\hline 
            		\multicolumn{11}{c}{\textbf{CuN$_2$}} \\
\hline 
\multirow{2}{*}{\textbf{C1}}            &\textit{Pres.}	&	 $4.8$  & --      & --  & --   &  $9.214$    &  $-2.712$  &  $198.265$     &  $4.652$ 	&  $1.910$\\
                                        &\textit{Comp.}	&	 $4.694$\footnotemark[36] & --   & --  & --            &                      &                      &  $258.94$\footnotemark[36] &   \\

\hline 
 \textbf{C2}            &\textit{Pres.}	&	 $4.919$  & --  & --        & --   &  $9.920$  &  $-4.065$  &  $80.907$ &  $6.170$		&  $0.557$\\

\hline 
 \textbf{C18}           &\textit{Pres.}	&	 $3.039$      &  $3.988$  &  $4.867$  & -- &  $9.831$  &   $-4.132$ &  $92.680$ &  $6.317$ 	&  $0.490$\\

\hline 
 \textbf{CoSb$_2$}      &\textit{Pres.}	&	 $5.303$  &  $5.015$ &  $9.106$  &  $\beta=151.225$                 &  $9.714$ &  $-4.110$ &  $92.028$ &  $6.167$ 	&  $0.512$\\
%
%
\hline \hline	
\end{tabular}
}	

\footnotetext[1]{Ref. [\onlinecite{David_Young_Phase_Diagrams_1991}]: Information is given at RTP.}
\footnotetext[2]{Ref. [\onlinecite{CRC_Handbook_82ed}].}

\footnotetext[3]{Ref. [\onlinecite{Jerry_1974}]: This is an average of 66 experimental values, at $20 ^{\circ} C$.}
\footnotetext[4]{Ref. [\onlinecite{elemental_metals_1996_comp}]: using LAPW-TB.}
\footnotetext[5]{Ref. [\onlinecite{elemental_metals_1996_comp}]: using LAPW-LDA.}
\footnotetext[6]{Ref. [\onlinecite{Wyckoff}].}

\footnotetext[7]{See Ref. 15 in [\onlinecite{PhysRevB.45.5777}].}
\footnotetext[8]{Ref. [\onlinecite{PhysRevB.45.5777}]: using APW-MT-LDA.}

\footnotetext[9]{Ref. [\onlinecite{Kittel1996}]: Cohesive energies are given at $0 \; K$ and $1 \text{ atm} = 0.00010 \text{ GPa}$; while bulk mudulii are given at room temperature.}
\footnotetext[10]{Ref. [\onlinecite{cohesive_energy_DFT}]: using LDA}
\footnotetext[11]{Ref. [\onlinecite{cohesive_energy_DFT}]: using BP-GGA.}
\footnotetext[12]{Ref. [\onlinecite{cohesive_energy_DFT}]: using PW-GGA.}
\footnotetext[13]{Ref. [\onlinecite{elemental_metals_2008_comp}]: using PAW-LDA.}
\footnotetext[14]{Ref. [\onlinecite{elemental_metals_2008_comp}]: using PAW-PW91.}
\footnotetext[15]{Ref. [\onlinecite{elemental_metals_2008_comp}]: using PAW-GGA(PBE).}
\footnotetext[16]{Ref. (25) in [\onlinecite{B_prime_1997_theory_comp_n_exp}]: at room temperature.}
\footnotetext[17]{See Refs. (8)--(11) in [\onlinecite{B_prime_1997_theory_comp_n_exp}].}
\footnotetext[18]{Ref. [\onlinecite{B_prime_1997_theory_comp_n_exp}]: using the so-called method of transition metal pseudopotential theory; a modified form of a method proposed by Wills and Harrison to represent the effective interatomic interaction.}
\footnotetext[19]{Ref. [\onlinecite{B_prime_1997_theory_comp_n_exp}]: using a semiempirical estimate based on the calculation of the slope of the shock velocity \textit{vs.} particle velocity curves obtained from the dynamic high-pressure experiments.  The given values are estimated at $\sim 298 \; K$.}
\footnotetext[20]{Ref. [\onlinecite{B_prime_1997_theory_comp_n_exp}]: using a semiempirical method in which the experimental static $P-V$ data are fitted to an EOS form where $B_{0}$ and $B_{0}^{\prime}$ are adjustable parameters.  The given values are estimated at $\sim 298 \; K$.}

\footnotetext[21]{Values obtained in the experimental work by Gallardo-Vega and Cruz \onlinecite{Cu3N_Cu4N_2006_exp_just_read_it} are between $3.810$ \AA and $3.830$ \AA.}
\footnotetext[22]{Ref. [\onlinecite{Cu3N_Cu4N_Cu3N2_2008_comp}]: using PAW-GGA(Perdew-Wang).}
\footnotetext[23]{Ref. [\onlinecite{Cu3MN_2007_comp}]: using FP-LAPW-GGA(PBE). Only the total energy ($-19.45 \; eV$) is given!}
\footnotetext[24]{Ref. [\onlinecite{Cu3N_1989_exp}].}
\footnotetext[25]{Ref. [\onlinecite{Cu3N_2006_exp}].}
\footnotetext[26]{Ref. [\onlinecite{Cu3N_2000_applications}].}
\footnotetext[27]{Ref. [\onlinecite{optical_properties_Cu3N_2011_exp}].}
\footnotetext[28]{Ref. [\onlinecite{Cu3N_2005_comp}]: using FP-LAPW-GGA(PBE).}
\footnotetext[29]{Ref. [\onlinecite{Cu3N_Cu4N_2011_comp}]: using UPP-GGA.}
\footnotetext[30]{Ref. [\onlinecite{Cu3N_1995_exp}].}
\footnotetext[31]{Ref. [\onlinecite{Cu3N_2004_comp}]: using FP-LAPW-GGA(PBE). Only the total energy ($-19.46 \; eV$) is given!} 
\footnotetext[32]{Ref. [\onlinecite{Cu3N_Cu4N_Cu8N_2007_comp}]: using FP-LAPW+lo-GGA(PBE)}
\footnotetext[33]{Ref. [\onlinecite{Cu3N_Ni3N_1993_exp}]}

\footnotetext[34]{Ref. [\onlinecite{CuN_AgN_AuN_2007_comp}]: using FP-LAPW+lo method within GGA(PBE).}
\footnotetext[35]{Ref. [\onlinecite{CuN_1997_comp}]: using FLAPW-LDA.}
\footnotetext[36]{Ref. [\onlinecite{CuN_CuN2_2011_comp}]: using UPP-GGA(PBE).}
\footnotetext[37]{Ref. [\onlinecite{CuN_AgN_AuN_2007_comp}]: using FP-LAPW+lo method within LDA.}
\footnotetext[38]{Ref. [\onlinecite{CuN_NiN_2004_comp}]: using full-potential linear muffin-tin orbital (FP-LMTO) method within GGA(PBE).}

\end{table*}
%

\begin{figure*}[H!]
\includegraphics[width=0.95\textwidth]{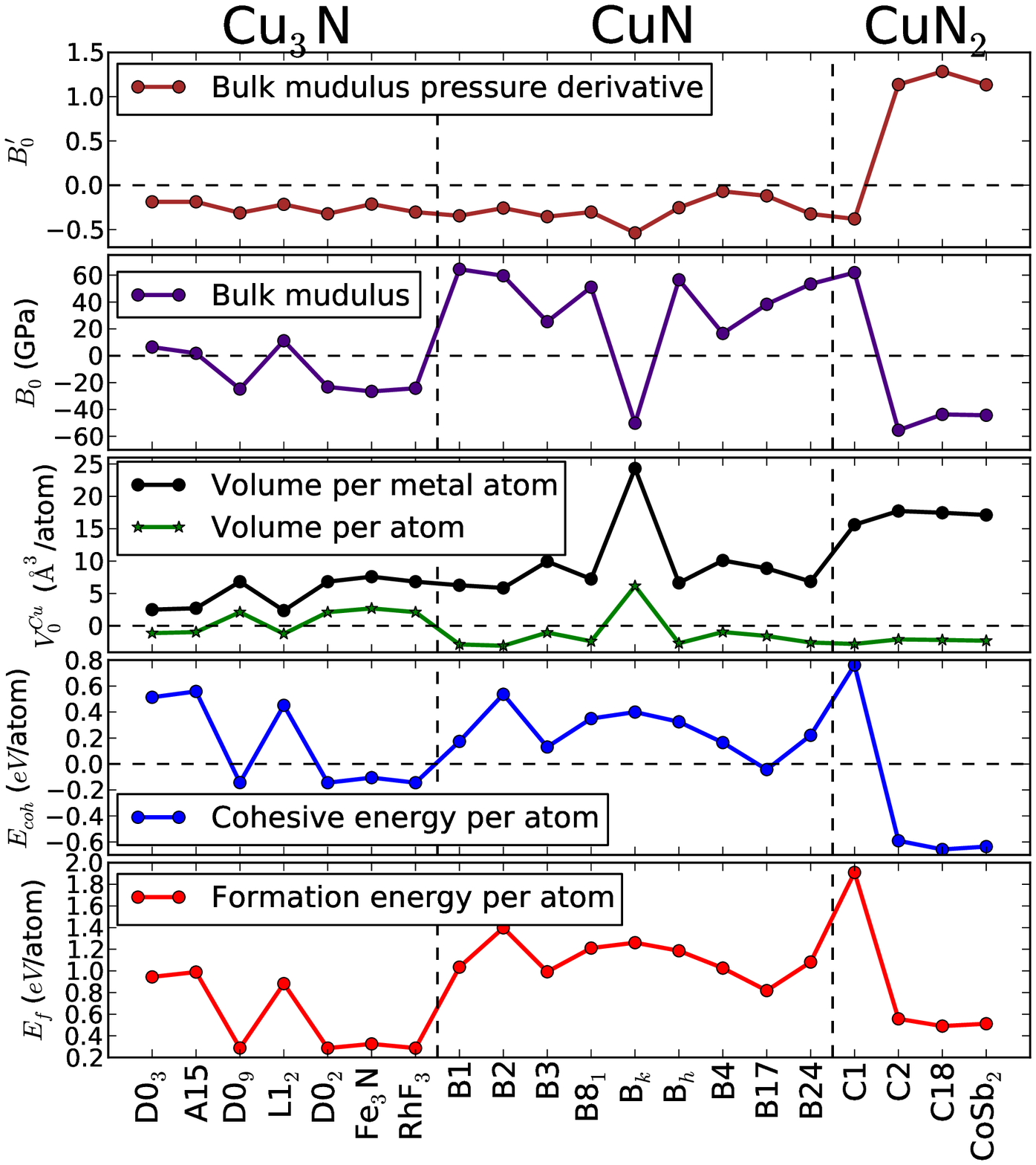}
\caption{\label{copper_nitrides_equilibrium_properties}(Color online.) Calculated equilibrium properties of the twenty studied phases of copper nitrides. All quantities are given relative to the corresponding ones of the \textit{fcc} crystalline elemental copper given in the first row of Table \ref{copper_nitrides_equilibrium_structural_properties}.}
\end{figure*}

To deeper analyze and to compare the obtained equilibrium properties of the three stoichiometries series with respect to one another, these quantities are depicted/visualized again in Fig. \ref{copper_nitrides_equilibrium_properties}. All quantities in this figure are given relative to the corresponding ones of the \textit{fcc} crystalline elemental copper given in Table \ref{copper_nitrides_equilibrium_structural_properties}. This will allow us to study the effect of nitridation on pure crystalline Cu \footnote{In Table \ref{copper_nitrides_equilibrium_structural_properties}, our computed properties of the elemental Cu are compared with experiment and with previous calculations as well. This may benchmark the accuracy of the rest of our calculations.}.
%
\subsection{Relative Stability: Cohesive Energy} 
Considering $E_{\text{coh}}$ in the Cu$_3$N series, one can use Fig. \ref{copper_nitrides_equilibrium_properties} to group these phases into two groups: a lower energy (more stable than the elemental Cu) group, containing D0$_9$, D0$_2$, RhF$_3$ and Fe$_3$N structures; and a higher energy (significantly less stable than the elemental Cu) group containing L1$_2$, D0$_3$ and A15 structures. The difference in cohesive energy between the least stable phase in the lower group (Fe$_3$N) and the relatively most stable phase in the higher group (L1$_2$) is $0.557 \; eV/\text{atom}$, as one can see from Table \ref{copper_nitrides_equilibrium_properties}. It is interesting to point out here that, except Cu$_3$N(Fe$_3$N), all phases in the first group are insulators, while all the less stable phases are metallic\footnote{The electronic structure of D0$_9$, D0$_2$ and RhF$_3$ phases are discussed in Sec. \ref{Electronic Properties}, while the rest are not shown here.}. 

Although this simple cubic D0$_{9}$ phase was indeed found to be stable, yet one interesting result we obtained is that, in their equilibrium, the Cu$_{3}$N(RhF$_{3}$) and the Cu$_{3}$N(D0$_{2}$) phases would be $0.001$ $eV/\text{atom}$ and $0.002$ $eV/\text{atom}$ more stable than Cu$_{3}$N(D0$_{9}$), respectively. Moreover, Fig. \ref{Cu3N1_ev_EOS} shows clearly that the $E(V)$ relations of Cu$_3$N in D0$_9$, D0$_2$ and in RhF$_3$ structures are almost identical. This marginal\footnote{In fact, the accuracy of the approximate $XC$ functional (PBE and others) does not really allow us to make a distinction among these.} difference in energy (Table \ref{copper_nitrides_equilibrium_structural_properties}) and the almost identical $E(V)$ curves (Fig. \ref{Cu3N1_ev_EOS}) may indicate the possiblity of the co-existence of these phases during the copper nitride synthesis process. However, this behaviour in the EOS could be attributed to the structural relationships between these three structures that being discussed in Ref. [\onlinecite[p. 265]{StructuralInChem}] and in Ref. [\onlinecite{D0_9_and_D0_2_structures}].

The bcc skutterudite structure (D0$_{2}$) can be derived from the more symmetric sc D0$_{9}$ structure by simply displacing four of the N atoms located on parallel edges of the Cu cube to its center. This is done for two adjacent Cu cubes but in two vertical displacement directions, as nicely explained in Ref. [\onlinecite{D0_9_and_D0_2_structures}]. On the other hand, to see the relation between D0$_{9}$ and RhF$_{3}$, it is better to think of D0$_{9}$ as built of Cu$_{6}$N octahedra (cf. Fig. I in Ref. [\onlinecite{Cu3N_2005_comp}]). A simple rotation of $60^{\circ}$ of an octaheron about a shared vertex with another octahedron brings the system to a structure in which Cu atoms are in hcp positions. Interested readers are referred to Ref. [\onlinecite[p. 266]{StructuralInChem}] for more details. Thus, both D0$_{2}$ and RhF$_{3}$ can simply be derived from D0$_{9}$. Hence it is not surprising that these structural relations reflect in their EOS's and in other physical properties.

The odd behaviour of the EOS of Fe$_3$N with the existence of two minima (Fig. \ref{Cu3N1_ev_EOS}) shows that the first minima (to the left) is a metastable local minimum that cannot be maintained as the system is decompressed. Cu ions are in the $6g$ Wyckoff positions: $(x,0,0), (0,x,0), (-x,-x,0), (-x,0,\frac{1}{2}), (0,-x,\frac{1}{2}), (x,x,\frac{1}{2})$; with $x=\frac{1}{3}$ to the left of the potential barrier (represented by the sharp peak in Fig. \ref{Cu3N1_ev_EOS}), and $x=\frac{1}{2}$ to the right of the peak.  It may be relevant to mention here that Wang and Xue \cite{CuN_NiN_2004_comp} obtained an additional local minimum at high pressure (lower volume) in the $E(V)$ EOS of CuN(B1).

In the CuN series, all phases show less binding than the Cu(fcc), except that the simple tetragonal structure of cooperite (B17) is slightly more stable, with $0.043 \; eV/\text{atom}$ lower $E_{\text{coh}}$. This structure, B17, was theoretically predicted to be the ground-state structure of PtN \cite{PtN_2006_comp_B17_structure_important}.

In the CuN$_2$ nitrogen-richest phase series, we can see from Table \ref{copper_nitrides_equilibrium_structural_properties} and from Fig. \ref{copper_nitrides_equilibrium_properties} that the phases of this group are significantly more stable than all the studied phases, except C1, which, in contrast, is the least stable among the twenty studied phases.

Comparing the relative stability of Cu$_3$N, CuN and CuN$_2$, we find from Table \ref{copper_nitrides_equilibrium_structural_properties} and from Fig. \ref{copper_nitrides_equilibrium_properties} that CuN$_2$(C18) is the most energetically stable phase with $0.526$ $eV/\text{atom}$ lower than the experimentally reported Cu$_3$N(D0$_9$) phase.
\subsection{Volume per Atom and Lattice Parameters} 
The equilibrium volume per atom $V_{0}$ is an average quantity referring to the volume of the unit cell divided by the number of all atoms in the unit cell regardless of the type of the contained atoms. $V_{0}$ is a quantity that is being used frequently in the literature in the calculations from the EOS and to compare the packing of the different considered phases, since $V_{0}$ is the inverse of the so-called number density. Our obtained numerical values are given in Table \ref{copper_nitrides_equilibrium_structural_properties} and visualized in Fig. \ref{copper_nitrides_equilibrium_properties}. Relative to the Cu(fcc), all phases tend to slightly lower the $V_{0}$ values except CuN(B$_{\text{k}}$) and the semiconducting Cu$_{3}$N phases.

To study the structural effect of the nitrogen on the hosting Cu lattice, we, instead of using the commonly used average $V_{0}$, introduce the volume per metal atom $V_{0}^{Cu}$. In the case of CuN and CuN$_{2}$ it is numerically equivalent to the volume per formula unit, while for Cu$_3$N it equals to $\text{(volume per formula unit)}/3$. Hence, this quantity ($V_{0}^{Cu}$ ) may be considered as a direct measure of the Cu-Cu bond length and, thus, as an indicator of the effect of nitridation on the mechanical properties of the elemental Cu. That is, for a given cohesive energy, an increase in $V_{0}^{Cu}$ may/should lead to a decrease in $B_{0}$ and vise versa, as will be seen when we discuss the trends in $B_{0}$ values.

In the same sub-window as $V_{0}$, obtained $V_{0}^{Cu}$ values are depicted relative to the Cu(fcc) in Fig. \ref{copper_nitrides_equilibrium_properties}. Having a look at this figure, one can see a \textit{general} behaviour: $V_{0}^{Cu}$ \textit{tends} to increase with the increase in the nitrogen content \textit{and} with the decrease in the structural symmetry. There is only one phase which has a clear odd bevaviour, that is CuN(B$_{\text{k}}$). It is worth to mention here that this B$_{\text{k}}$ is not an \textit{hcp} structure, and we have not optimized its $c/a$ ratio. Thus, this is the most open phase among all the investigated set. Nevertheless, all phases show an increase in $V_{0}^{Cu}$ relative to the elemental Cu, and thus Cu-Cu bond is longer in all these nitrides than in the elemental Cu. This cannot be seen directly from the $V_{0}$ values given in Tabel \ref{copper_nitrides_equilibrium_structural_properties}.
%
\subsection{\label{Bulk Modulus and its Pressure Derivative}Bulk Modulus and its Pressure Derivative}
Beside $E_{0}$ and $V_{0}$, the equilibrium bulk modulus 
\begin{equation} \label{B_eq}
B_{0} = -V \frac{\partial P}{\partial V}\Bigg|_{V=V_{0}} = -V \frac{\partial^{2} E}{\partial V^{2}}\Bigg|_{V=V_{0}}
\end{equation}
and its pressure derivative
\begin{eqnarray}\label{B_prime_eq}
\begin{split}
B^{\prime}_{0} = \frac{\partial B}{\partial P} \Bigg|_{P=0} &=& \frac{\partial B}{\partial V} \frac{\partial V}{\partial P} \Bigg|_{V=V_{0}} = \frac{1}{B_{0}} \left(  -V \frac{\partial B}{\partial V} \right)  \Bigg|_{V=V_{0}} \\
&=& \frac{1}{B_{0}} \left(  V \frac{\partial}{\partial V} (V \frac{\partial^{2} E}{\partial V^{2}}) \right)  \Bigg|_{V=V_{0}}
\end{split}
\end{eqnarray}
appear in Eq. \ref{3rd_BM_eos} as fitting parameters. The last parts to the right in Eq. \ref{B_eq} and Eq. 
\ref{B_prime_eq} show that the only DFT calculated quantity is the total energy $E$ (or, equivalently $E_{coh}$), and that $B_{0}$ and $B^{\prime}_{0}$ are a second- and a third-order energy derivative, respectively. Thus, $B_{0}$ and $B^{\prime}_{0}$ are directly related to the curvature of the shown $E(V)$ curves.

Eq. \ref{B_eq} and Eq. \ref{B_prime_eq} also tell us that if all phases have the same $E_{coh}$, the $B_{0}$ curve in Fig. \ref{copper_nitrides_equilibrium_properties} would become a mirror reflection-like with respect to the $V_{0}^{Cu}$ curve, and vise versa. In fact, such a general trend can be seen in Fig. \ref{copper_nitrides_equilibrium_properties}.

Compared to the parent Cu(fcc), the CuN phases tend to increase $B_{0}$. Such a conclusion has also been arrived at by Shimizu, Shirai and Suzuki \cite{CuN_1997_comp} who calculated $B_{0}$ for a series of 1:1 TMNs, including CuN. On the other hand, the considered CuN$_{2}$ phases are all, except C1, more compressible than the Cu(fcc). Considering the 1:3 phases, one can easily see that the trend in $E_{coh}$ manifests itself again and divides this series into two groups: a group of more compressible semiconductors containing D0$_9$, D0$_2$, RhF$_3$ and Fe$_3$N; and a group with almost no change in the Cu(fcc) bulk mudulus containing L1$_2$, D0$_3$ and A15 metallic phases.

Having a look at Fig. \ref{copper_nitrides_equilibrium_properties}, one may argue that, relative to Cu(fcc), the lower $V_{0}$ and the lower $E_{coh}$ of the CuN$_2$(C2), CuN$_2$(C18) and CuN$_2$(CoSb$_2$) phases must have led to higher $B_{0}$ values. Since this is not the case, we turn to our introduced $V_{0}^{Cu}$: Fig. \ref{copper_nitrides_equilibrium_properties} tells us that all these three nitrogen-rich phases have higher $V_{0}^{Cu}$ relative to Cu(fcc). Hence, $V_{0}^{Cu}$  won the competition with their relatively lower $E_{coh}$, leading to lower $B_{0}$. This, again,makes sense and justifies our introduction of $V_{0}^{Cu}$ when dealing with such nitrides.

Take, for example CuN$_2$(C18). One can notice from Table \ref{copper_nitrides_equilibrium_structural_properties} that, relative to Cu$_3$N(D0$_9$), CuN$_2$(C18) has about $30\%$ less atomic volume $V_{0}$ and about $14\%$ less $E_{coh}$, but resulting in about $30\%$ \textit{less} bulk modulus than Cu$_3$N(D0$_9$). However, if we consider our introduced $V_{0}^{Cu}$, rather than the commonly used $V_{0}$, one can see that CuN$_2$(C18) has about $57\%$ more $V_{0}^{Cu}$ than Cu$_3$N(D0$_9$), which explains the lesser $B_{0}$ value. In fact, CuN$_2$(C2), CuN$_2$(C18) and CuN$_2$(CoSb$_2$) share almost the same features when compared to Cu$_3$N(D0$_9$), Cu$_3$N(D0$_2$) and Cu$_3$N(RhF$_3$) as CuN$_2$(C18) when compared to Cu$_3$N(D0$_9$).

Compared to the other CuN$_{2}$ phases, the relatively greater value of $E_{coh}$ of C1 is overtaken by the relatively less $V_{0}^{Cu}$ value, resulting in a greater $B_{0}$ than all the three other CuN$_{2}$ phases. It is also worth to notice from Table \ref{copper_nitrides_equilibrium_structural_properties} that all the CuN phases, except the open B$_{\bf{k}}$, have higher bulk moduli than all Cu$_{3}$N and CuN$_{2}$ phases, except the least compressible phase, C1.

Hence, the isotropic elastic properties depend on, and are more appropriately described by, $V_{0}^{Cu}$ rather than $V_{0}$. Fig. \ref{copper_nitrides_equilibrium_properties} tells that the more the nitrogen content, the longer the Cu-Cu bond length. Thus nitrogen tends to open the hosting Cu lattice and to reduce the bulk mudulus. Nevertheless, bulk mudulus is a result of the competition between the change in $V_{0}^{Cu}$ and the change in $E_{coh}$.

Physically, the pressure dependence of $B_{0}$ can be quantified via its pressure derivative $B_{0}^{\prime}$ given by Eq. \ref{B_prime_eq}. Except for the last three CuN$_{2}$ phases in Fig. \ref{copper_nitrides_equilibrium_properties}, all phases show almost equal sensitivity. However, the change in $B_{0}$ of the elemental Cu would be greater than all these phases. On the other hand CuN$_{2}$(C2), CuN$_{2}$(C18) and CuN$_{2}$(CoSb$_{2}$) show high elastic sensitivity to any isotropic pressure. It is worth to mention here that this quantity is a measurable quantity \cite{B_prime_1997_theory_comp_n_exp}, but we cannot find any experimental value for the synthesized Cu$_{3}$N(D0$_{9}$) phase.

Table \ref{copper_nitrides_equilibrium_structural_properties} shows that our obtained $B_{0}$ and $B_{0}^{\prime}$ values agree well with many other theoretical works. However, clear differences between the current and, and among, the theoretically obtained values exist. This can be traced back to two factors:
\begin{itemize}
\item From Eq. \ref{B_eq} and Eq. \ref{B_prime_eq}, one needs only to calculate $E_{coh}$ from DFT. Hence, $B^{\prime}_{0}$ values depends on the DFT algorithm/method and functional. For example, in Table \ref{copper_nitrides_equilibrium_structural_properties}, the GGA (e.g. Refs. \onlinecite{CuN_AgN_AuN_2007_comp} and \onlinecite{CuN_CuN2_2011_comp}) calculated $B_{0}$ values of B1, B2 and B3 are all lower than the obtained LDA (e.g. Refs. \onlinecite{CuN_AgN_AuN_2007_comp} and \onlinecite{CuN_1997_comp}) corresponding ones. This is, in fact, a well-known feature of GGA compared to LDA. That is, relative to the latter,the former understimates the cohesion and thus overstimates $V_{0}$ resulting in a lower $B_{0}$.
\item Yet, there are considerable differences among the GGA values and among the LDA values! Recalling that $B_{0}$ and $B^{\prime}_{0}$ are obtained from the EOS fitting, the accuracy in the obtained values depends not only on the accuracy of the DFT calculated $E_{coh}$, but also on the numerical fitting; e.g. number of points around $V_{0}$ and how far these points are from both $V_{0}$ and $E_{0}$. Moreover, $B_{0}^{\prime}$ is numerically more sensitive than $B_{0}$ \cite{B_prime_Aug_2012_comp}, that why, we believe, there is no even clear/general trend/systematic in the calculated $B_{0}^{\prime}$ values of Table \ref{copper_nitrides_equilibrium_structural_properties}.
\end{itemize}
\subsection{Relative Stability: Formation Energy} 
Since the driving force for the formation of a solid may lie in the energy released when the elements condense to form the solid \cite{Ladd_1998}, it has become a common practice in \textit{ab initio} studies to calculate the so-called energy of formation in order to test the formation possibility of materials under consideration \cite{stability_killing_2008} as well as a measure/indicator of the relative stabilities of the phases under consideration \cite{Cu3N_Cu4N_Cu8N_2007_comp,PhysRevB.67.064108}.

Within first-principles calculations, the formation energy $E_f$ can be calculated from 
the difference of the cohesive energies (or enthalpies) of products $E_{\text{coh}}(\text{products})$ and reactants $E_{\text{coh}}(\text{reactants})$ \cite{Atkin,enthalpies_of_formation_2003}

\begin{eqnarray} \label{general E_f}
E_f = \sum E_{\text{coh}}(\text{products}) - \sum E_{\text{coh}}(\text{reactants});
\end{eqnarray}

where $E_{\text{coh}}(\text{reactants})$ should be given in the reference states of the reactants; that is, at their most stable state at specified temperature and pressure \cite{Atkin}.

In our case, if we assume that our product Cu$_m$N$_n$ results from the interaction between the gaseous diatomic molecular N$_{2}$ and the solid Cu metal in its reference \textit{fcc} A1 structure through/via the chemical reaction

\begin{eqnarray} \label{E_f reaction}
m \text{Cu}^{\text{solid}} + \frac{n}{2} \text{N}_2^{\text{gas}} \longrightarrow   \text{Cu}_m\text{N}_n^{\text{solid}},
\end{eqnarray}
then Eq. \ref{general E_f} above can be written for formation energy per atom as (cf. Refs. \onlinecite{Cu3N_Cu4N_Cu8N_2007_comp,PhysRevB.63.165116,PhysRevB.67.064108}):
\begin{eqnarray} \label{formation energy equation}
\begin{split}
E_f(\text{Cu}_m\text{N}_n^{\text{solid}}) = E_\text{coh}(\text{Cu}_m\text{N}_n^{\text{solid}})
\\ - \frac{  m E_\text{coh}(\text{Cu}^{\text{solid}}) + \frac{n}{2} E_\text{coh}(\text{N}_2^{\text{gas}})}{m + n},
\end{split}
\end{eqnarray}
where $m,n=1,2,3$ are the stoichiometric weights and $E_\text{coh}(\text{Cu}_m\text{N}_n^{\text{solid}})$ is, again, the cohesive energy per atom calculated via Eq. \ref{E_coh equation}\footnote{If $E_\text{coh}$ is used with a positive sign convention, i.e. negative of Eqs. \ref{general_E_coh equation} and \ref{E_coh equation}, then signs in Eq. \ref{formation energy equation} must be reversed.}.

To determine the equilibrium cohesive energy of the elemental metallic copper $E_\text{coh}(\text{Cu}^{\text{solid}})$ in its well-known \textit{fcc} A1 structure (space group Fm$\bar{3}$m No. 225) \cite{Wyckoff,Structure_of_Materials,Handbook_of_Mineralogy}, we followed the same procedures described in Sec.\ref{Electronic Optimization Details} and Sec.\ref{Geometry Relaxation and Relative Stabilities}. The obtained structural and cohesive properties of the bulk Cu are placed in the first row of Table \ref{copper_nitrides_equilibrium_structural_properties}, where they show excellent agreement with experiment and good agreement with the theoretically predicted ones.

The total energy of the gaseous diatomic molecular nitrogen ($E_\text{coh}(\text{N}_2^{\text{gas}})$), was obtained by placing one N atom at a corner of a cubic cell with $a=14$ \AA, while the second atom is displaced along the diagonal direction and was allowed to move during the relaxation. $E_{\text{cut}}$ of $800 \; eV$, $\Gamma$ point and Gaussian smearing method with width of $0.05$ $eV$ were used. Molecular and cohesive energies were calculated in the same way as described in Sec. \ref{Geometry Relaxation and Relative Stabilities}. Obtained cohesive and structural results are given in Table \ref{nitrogen_dimer_properties} and compared with experiment and with previous calculations therein.
\begin{table}[h]	
\caption{\label{nitrogen_dimer_properties}Calculated and experimental cohesive energy $E_\text{coh}(eV)$ and bond length $d(\text{\AA})$ of the gaseous diatomic molecular nitrogen ($\text{N}_2^{\text{gas}}$). The presented data are of the current work (\textit{Pres.}), experimentally reported (\textit{Expt.}) and of previous calculations (\textit{Comp.}).}
\resizebox{0.49\textwidth}{!}{
\begin{tabular}{llll}
\hline
						&Pres.			&Expt. 		&Comp.					\\
\hline \hline 
$E_\text{coh}(eV)$   	&$-10.392$	&  $-(9.797658 \pm 0.0061149)$\footnotemark[1]	& $10.49$\footnotemark[2], $11.75$\footnotemark[3], $11.57$\footnotemark[4], $10.69$\footnotemark[5]\\
$d(\text{\AA})$   		& $1.113$	& $(1.0976 \pm 0.0002)$\footnotemark[6]	& $1.102$\footnotemark[2], $1.085$\footnotemark[3], $1.095$\footnotemark[4], $1.095$\footnotemark[5]\\             	       		       \hline \hline	
\end{tabular}
}	

\footnotetext[1]{This bond strength in nitrogen diatomic molecule is taken from Ref. [\onlinecite[p. 9:55]{CRC_Handbook_82ed}] where it is given there as $(945.33\pm 0.59 \; \text{KJ/mol})$ with positive sign convention and at $298 \; K$. In p. 9:76 of the same reference, the force constant for bond stretching in  nitrogen diatomic molecule is given to be $22.95$ N/cm.}
\footnotetext[2]{Ref. [\onlinecite{PhysRevB.65.245212}], PBE(GGA)-LAPW, with spherical ground-state density of the N free atom.}
\footnotetext[3]{Ref. [\onlinecite{PhysRevB.65.245212}], LDA-PP, with spherical ground-state density of the N free atom.}
\footnotetext[4]{Ref. [\onlinecite{PhysRevB.65.245212}], LDA-LAPW, with spherical ground-state density of the N free atom.}
\footnotetext[5]{Ref. [\onlinecite{PhysRevB.65.245212}], PBE(GGA)-PP, with spherical ground-state density of the N free atom.}
\footnotetext[6]{Ref. [\onlinecite[p. S7]{Table_of_interatomic_distances_1958}].}
\end{table}	

The obtained formation energies $E_{f}$ of the twenty relaxed phases are given in Table \ref{copper_nitrides_equilibrium_structural_properties} and shown graphically in Fig. \ref{copper_nitrides_equilibrium_properties}. All these values are positive; which means that all these twenty phases are, in principle, thermodynamically unstable (endothermic). However, these results have to be interpreted with some caution:

\begin{itemize}	
\item Many other theoretical calculations found positive formation energy for experimentally synthesized transition metal nitrides; e.g. OsN$_2$ \cite[using PP and PBE-GGA]{positive_Ef_n_solid_N2_2007_comp_Scandolo}, PtN$_2$ \cite[using PP and PBE-GGA]{positive_Ef_n_PtN2_2006_comp_Scandolo} and InN \cite[using PP with LDA and different GGAs]{PhysRevB.65.245212}\cite[using PW91-GGA]{PhysRevB.59.5521}.

\item The fact that we obtained a positive formation energy for the successfully synthesized Cu$_{3}$N(D0$_{9}$) phase means that it may be possible that other phases can still be synthesized, and it may indicate that there is a problem with our calculations method (i.e. with the approximations) and/or with the physical conditions assumed for the calculations (i.e. pressure and temperature); see below.

\item These positive values are the result of static DFT calculations ($T = 0 \; K$) at equilbrium volume ($P = 0 \; \text{GPa}$);  while the fact is that most, if not all, of the successfully synthesized TMNs were obtained by subjecting their parent elements to extreme conditions of pressure and temperature ($c. f.$ Ref. [\onlinecite{positive_Ef_n_solid_N2_2007_comp_Scandolo}]).

\item Referring to Table \ref{nitrogen_dimer_properties}, the difference between our calculated cohesive energy of N$_{2}$ and experiment (Ref. [\onlinecite{CRC_Handbook_82ed}]) is about $- 0.297 \; eV/\text{atom}$; while for the bulk Cu the difference is about  $0.016 \; eV/\text{atom}$. Now, using Eq. \ref{formation energy equation} with this significant overestimation of $E_\text{coh}(\text{N}_2^{\text{gas}})$ and the reasonable value of $E_\text{coh}(\text{Cu}^{\text{solid}})$ will result in underestimation of $E_f(\text{Cu}_m\text{N}_n^{\text{solid}})$. This contribution has to be considered, as an apparent shortcoming of the PBE-GGA, whenever one deals with a dimeric crystal \cite{PhysRevB.65.245212,B_prime_Aug_2012_comp}.

\item Nevertheless, since all formation energies are calculated as the difference between the \textit{ab initio} cohesive energies, which in turn are calculated at the same level of accuracy, one can still use these formation energies to measure the \textit{relative} thermodynamic stabilities of these structures. That is, the lower the formation energy, the lower the propensity to dissociate back into the constituent elements Cu and N$_2$ ($c. f.$ Ref. [\onlinecite{positive_Ef_n_solid_N2_2007_comp_Scandolo}]).

\item Moreover, because Cu$_{3}$N(D0$_{9}$) has been synthesized, we can take it as a \textit{reference measure} of stability. It is also worth to recall here that experiment found Cu$_{3}$N(D0$_{9}$) to be metastable at room temperature \cite{Cu3N_2006_exp}.
\end{itemize}	
Relative to each other, and within each series, $E_{f}$ of the twenty phases shows almost the same trend as $E_{coh}$. However, the CuN phases tend to be relatively less stable than the Cu$_{3}$N and CuN$_{2}$ phases, except the odd C1 phase. In fact, C2, C18 and CoSb$_2$ are the most stable and share almost the same features when compared to Cu$_3$N(D0$_9$), Cu$_3$N(D0$_2$) and Cu$_3$N(RhF$_3$). This may agree well with Armenta and Soto \cite{Cu3N_Cu4N_Cu8N_2007_comp} who proved, from the study of formation energy, that the metallic phases of copper nitrides would be more stable than the semiconducting phase.
\subsection{More Comparison with Experiment and with Theory}
Comparing our obtained results with experiment, one can see from Table \ref{copper_nitrides_equilibrium_structural_properties} that the lattice parameter $a$ of Cu$_{3}$N(D0$_{9}$) was reproduced very well. Excellent agreement with previous calculations is also clear, though, with respect to experiment (or: with respect to each other), the common overestimation of $a$ by GGA and the underestimation of $a$ by LDA (c.f. Refs. \onlinecite{Cu3N_Cu4N_Cu3N2_2008_comp,Cu3MN_2007_comp,Cu3N_2005_comp}; and Refs. \onlinecite{PhysRevB.65.064302,PhysRevB.76.024309}) is showing up.

Using the full-potential (linearized) augmented plane waves plus local orbitals (FP-LAPW+lo) method within LDA and within GGA, Kanoun and Said \cite{CuN_AgN_AuN_2007_comp} studied the $E(V)$ EOS for CuN in the B1, B2, B3 and B4 structures. While within GGA, they found equilibrium lattice parameters which are in excellent agreement with ours, their obtained LDA lattice parameter values show the common underestimation with respect to our and their GGA values (see Table \ref{copper_nitrides_equilibrium_structural_properties}). Also, the relative stabilities of these phases they arrived at agree well with ours, and they concluded that B3 is the ground-state phase of CuN and is metalic.

Shimizu, Shirai and Suzuki \cite{CuN_1997_comp} performed first-principles calculations using full-potential linearized augmented-plane-wave (FLAPW) method in the framework of LDA and found that CuN(B1) is less than $0.20 \; eV/\text{atom}$ more stable than CuN(B3), while we found that CuN(B3) is $0.043 \; eV/\text{atom}$ (GGA) more stable than CuN(B1). Some of their findings are shown in Table \ref{copper_nitrides_equilibrium_structural_properties}; and, again, their predicted LDA lattice constants are slightly less than our GGA values, while their obtained bulk moduli are overestimated when compared to ours.

Using full-potential linear muffin-tin orbital (FP-LMTO) method within GGA(PBE), Wang et al. \cite{CuN_NiN_2004_comp} studied the $E(V)$ EOS of CuN in the B1, B2, B3, B8$_{1}$ and B4 structures. Their obtained equilibrium lattice parameters and bulk muduli are included in Table \ref{copper_nitrides_equilibrium_structural_properties} which show good agreement with ours. However, Fig. \ref{Cu1N1_ev_EOS} shows that, within this parameter space, equilibrium cohesive energy decreases as B2, B8$_{1}$, B4, B1 and B3. This is consistent with Wang et al. but B8$_{1}$ and B4 are swaped. Nevertheless, B3 is the most stable in both works, contrary to the findings of Shimizu, Shirai and Suzuki \cite{CuN_1997_comp}.

Whatever the case, in our wider parameter space, Fig. \ref{Cu1N1_ev_EOS} and Table \ref{copper_nitrides_equilibrium_structural_properties} reveal that CuN(B17) is $0.17 \; eV/\text{atom}$ and $0.21 \; eV/\text{atom}$ (GGA) more stable than CuN(B3) and CuN(B1), respectively. It may be worth to mention again here that B17 was theoretically predicted to be the ground-state structure of PtN \cite{PtN_2006_comp_B17_structure_important}.

Using norm-conserving ultra-soft pseudopotential within GGA and the so-called BFGS scheme for geometry optimization, Bouayed et al. \cite{CuN_CuN2_2011_comp} studied CuN in the B1 and B3 structures, and CuN$_2$ in the C1 structure. Although their obtained lattice constants (given in Table \ref{copper_nitrides_equilibrium_structural_properties}) are in good agreement with our findings, the noticable difference in bulk moduli may be traced back to the numerical fitting (see Sec. \ref{Bulk Modulus and its Pressure Derivative}).
%
%
\section{\label{Electronic Properties}Electronic Properties}
\begin{figure*}
\includegraphics[width=1.0\textwidth]{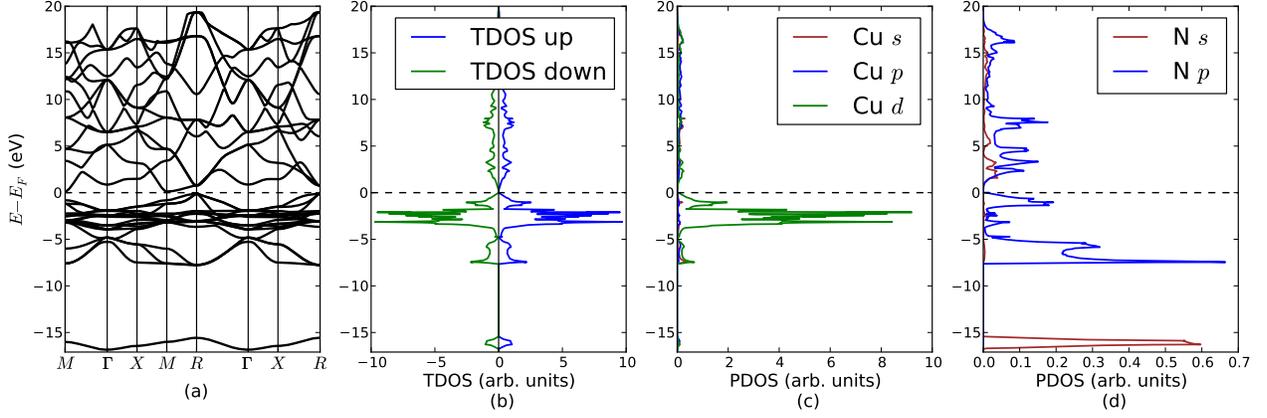}
\caption{\label{Cu3N1_D0_9_electronic_structure}(Color online.) DFT calculated electronic structure for Cu$_3$N in the D0$_9$ structure: \textbf{(a)} spin-projected total density of states (TDOS); \textbf{(b)} partial density of states (PDOS) of Cu($s, p, d$) orbitals in Cu$_3$N; \textbf{(c)} PDOS of N($s, p$) orbitals in Cu$_3$N, and \textbf{(d)} band structure along the high-symmetry $\mathbf{k}$-points which are labeled according to Ref. [\onlinecite{Bradley}]. Their coordinates w.r.t. the reciprocal lattice basis vectors are: $M(0.5, 0.5, 0.0)$, $\Gamma(0.0, 0.0, 0.0 )$, $X(0.0, 0.5, 0.0 )$, $R(0.5, 0.5, 0.5 )$.}
\end{figure*}

\begin{figure*}
\includegraphics[width=1.0\textwidth]{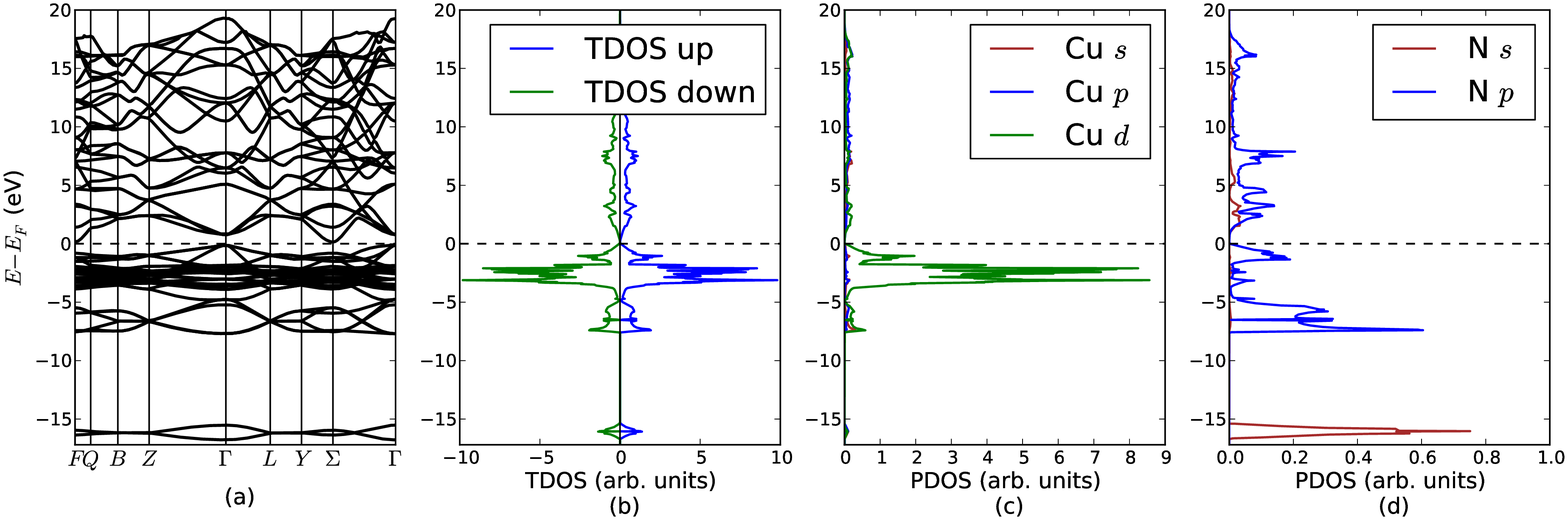}
\caption{\label{Cu3N1_RhF_3_electronic_structure}(Color online.) DFT calculated electronic structure for Cu$_3$N in the RhF$_3$ structure: \textbf{(a)} spin-projected total density of states (TDOS); \textbf{(b)} partial density of states (PDOS) of Cu($s, p, d$) orbitals in Cu$_3$N; \textbf{(c)} PDOS of N($s, p$) orbitals in Cu$_3$N, and \textbf{(d)} band structure along the high-symmetry $\mathbf{k}$-points which are labeled according to Ref. [\onlinecite{Bradley}]. Their coordinates w.r.t. the reciprocal lattice basis vectors are: $F(0.5, 0.5, 0.0)$, $Q(0.375, 0.625, 0.0)$, $B(0.5, 0.75, 0.25)$, $Z(0.5, 0.5, 0.5)$, $\Gamma(0.0,  0.0, 0.0)$, $L(0.0, 0.5, 0.0)$, $Y(0.25, 0.5, -.25)$, $\Sigma(0.0, 0.5, -.5)$.}
\end{figure*}

\begin{figure*}
\includegraphics[width=1.0\textwidth]{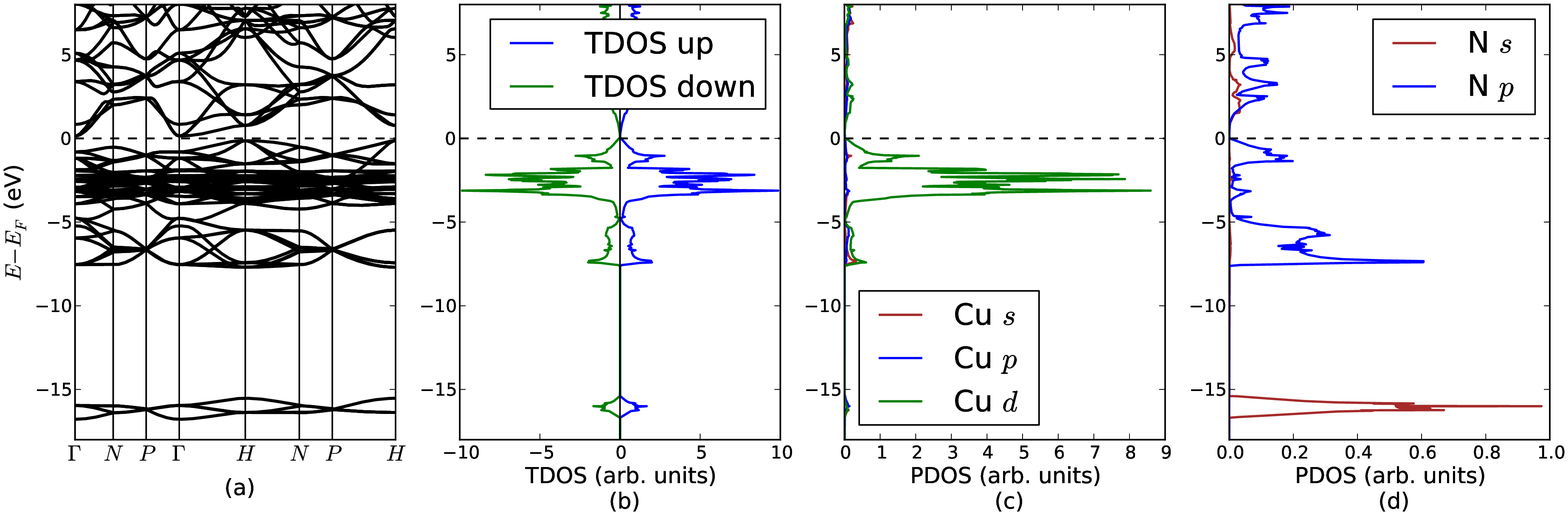}
\caption{\label{Cu3N1_D0_2_electronic_structure}(Color online.) DFT calculated electronic structure for Cu$_3$N in the D0$_2$ structure: \textbf{(a)} spin-projected total density of states (TDOS); \textbf{(b)} partial density of states (PDOS) of Cu($s, p, d$) orbitals in Cu$_3$N; \textbf{(c)} PDOS of N($s, p$) orbitals in Cu$_3$N, and \textbf{(d)} band structure along the high-symmetry $\mathbf{k}$-points which are labeled according to Ref. [\onlinecite{Bradley}]. Their coordinates w.r.t. the reciprocal lattice basis vectors are:  $\Gamma(0.0, 0.0, 0.0)$, $N(0.0, 0.0, 0.5)$, $P(0.25, 0.25, 0.25)$, $H(0.5, -.5, 0.5)$.}
\end{figure*}

\begin{figure*}
\includegraphics[width=1.0\textwidth]{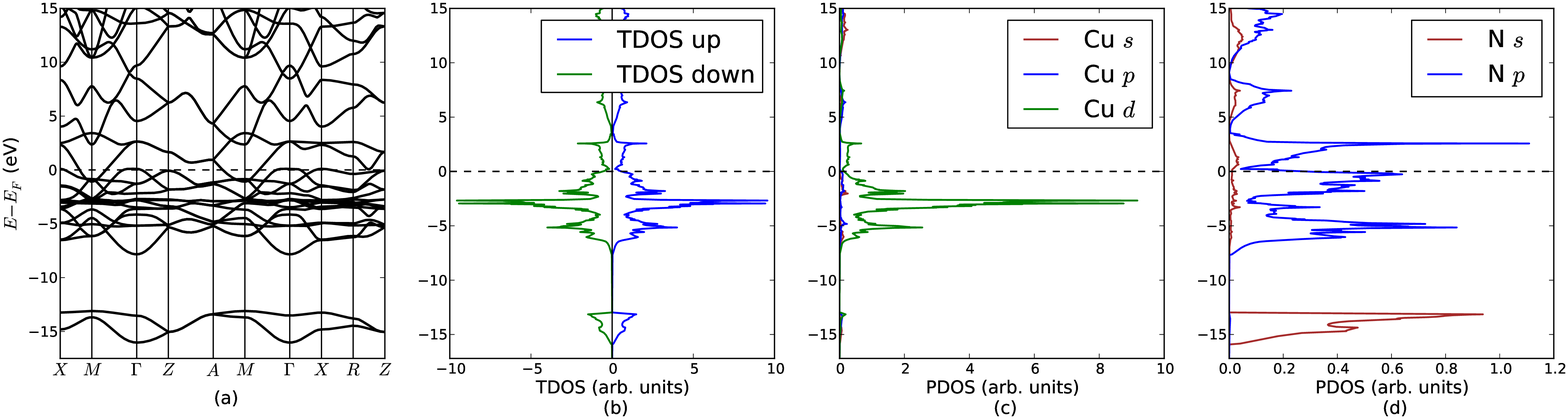}
\caption{\label{Cu1N1_B17_electronic_structure}(Color online.) DFT calculated electronic structure for CuN in the B17 structure: \textbf{(a)} spin-projected total density of states (TDOS); \textbf{(b)} partial density of states (PDOS) of Cu($s, p, d$) orbitals in CuN; \textbf{(c)} PDOS of N($s, p$) orbitals in CuN, and \textbf{(d)} band structure along the high-symmetry $\mathbf{k}$-points which are labeled according to Ref. [\onlinecite{Bradley}]. Their coordinates w.r.t. the reciprocal lattice basis vectors are: $X (0.0, 0.5, 0.0)$, $M (0.5, 0.5, 0.0)$, $\Gamma (0.0, 0.0, 0.0)$, $Z (0.0, 0.0, 0.5)$, $A (0.5, 0.5, 0.5)$, $R (0.0, 0.5, 0.5)$.}
\end{figure*}

\begin{figure*}
\includegraphics[width=1.0\textwidth]{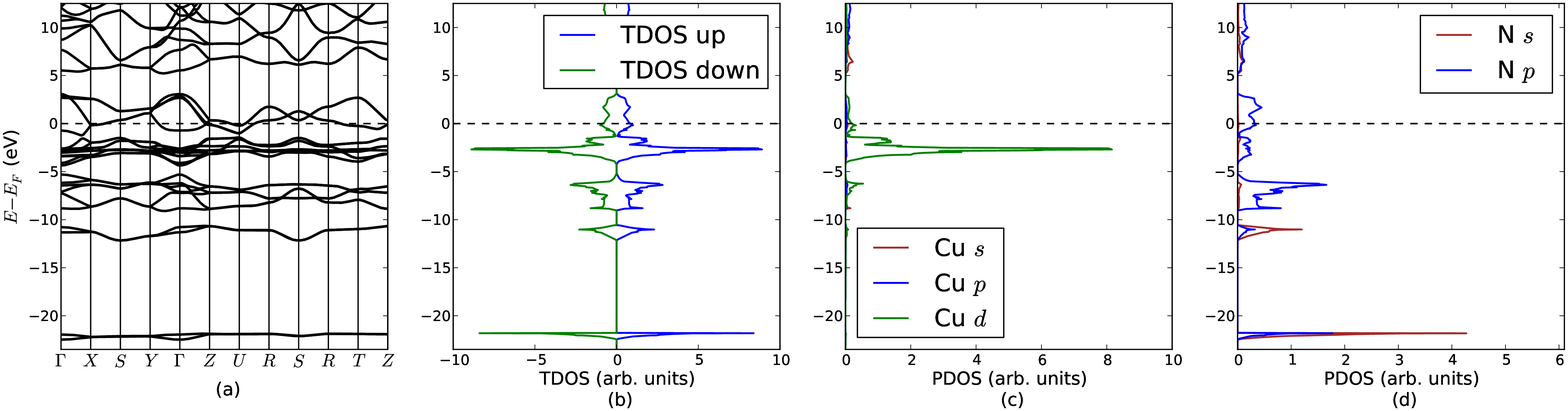}
\caption{\label{Cu1N2_C18_electronic_structure}(Color online.) DFT calculated electronic structure for CuN$_2$ in the C18 structure: \textbf{(a)} spin-projected total density of states (TDOS); \textbf{(b)} partial density of states (PDOS) of Cu($s, p, d$) orbitals in CuN$_2$; \textbf{(c)} PDOS of N($s, p$) orbitals in CuN$_2$, and \textbf{(d)} band structure along the high-symmetry $\mathbf{k}$-points which are labeled according to Ref. [\onlinecite{Bradley}]. Their coordinates w.r.t. the reciprocal lattice basis vectors are: $\Gamma( 0.0, 0.0, 0.0)$, $X( 0.0, 0.5, 0.0)$, $S( -.5, 0.5, 0.0)$, $Y( -.5, 0.0, 0.0)$, $Z( 0.0, 0.0, 0.5)$, $U( 0.0, 0.5, 0.5)$, $R( -.5, 0.5, 0.5)$, $T( -.5, 0.0, 0.5)$.}
\end{figure*}

Band structure (i.e. $\epsilon_{i}^{\sigma}(\mathbf{k})$ curves) and spin-projected total and partial (i.e. orbital resolved) density of states (DOS) of the energetically most stable phases are presented in Figs. \ref{Cu3N1_D0_9_electronic_structure}, \ref{Cu3N1_RhF_3_electronic_structure}, \ref{Cu3N1_D0_2_electronic_structure}, \ref{Cu1N1_B17_electronic_structure} and \ref{Cu1N2_C18_electronic_structure}. Spin-projected total density of states (TDOS) are shown in subfigure (b) in each case. In all cases, TDOS's are completely symmetrical in majority
and minority spins. That is, electrons occupy the majority and minority spin bands equally and
result in a zero total spin moment and a zero spin-polarization ratio: $\text{SPR}_{\text{DOS}}(E) = | \left( D_{\uparrow}(E) - D_{\downarrow}(E)\right) / \left( D_{\uparrow}(E) + D_{\downarrow}(E)\right)| $. That is why it was sufficient only to display spin-up partial density of states (PDOS) and spin-up band structures. To properly show details of the electronic structure of these phases, we plotted the energy bands along densely sampled high-symmetry string of neighbouring points in the $\mathbf{k}$-space; while displaying the Cu($s, p, d$) and N($s, p$) partial DOS allows us to extract information about the orbital character of these bands.

Energy bands $\epsilon_{i}^{\sigma}(\mathbf{k})$, total density of states (TDOS) and  partial (orbital-resolved) density of states (PDOS) of Cu$_3$N(D0$_9$) are  shown in Figs. \ref{Cu3N1_D0_9_electronic_structure}. It is clear that Cu$_3$N(D0$_9$) presents insulating character in its spin band. It has its valence band maximum (VBM) at $R$ which lies $0.13$ $eV$ below the Fermi energy $E_F$, and its conduction band minimum (CBM) at $M$ which also lies $0.13$ $eV$ above $E_F$, resulting in a narrow indirect gap of $0.26$ $eV$. By looking at the PDOS plots, energy bands can be divided into three parts: a deep band around $\thicksim -16$ $eV$ below $E_F$ consists mainly of N$(2s)$, a broad group of $12$ valence  bands with $\thicksim 8$ $eV$ of width that comes mostly from the $3d$ electrons of Cu plus smaller contribution from N$(2p)$, and the conduction bands.

Our obtained band structure of Cu$_3$N(D0$_9$) agrees \textit{qualitatively} well with many previous theoretical works \cite{Cu3N_1996_comp,Cu3MN_2007_comp,Cu3N_Cu4N_Cu3N2_2008_comp}; however, depending on the calculation method, the value of the indirect band gap of Cu$_3$N(D0$_9$) was predicted to be $0.23$ $eV$ (LAPW+LDA)\cite{Cu3N_1996_comp}, $0.25$ $eV$ (FP-LAPW+GGA(PBE))\cite{Cu3MN_2007_comp,Cu3N_2004_comp}, $0.355$ $eV$ (UPP-GGA(Perdew-Wang))\cite{Cu3N_Cu4N_Cu3N2_2008_comp} and $0.9$ $eV$ (LCAO+LDA)\cite{Cu3N_1996_comp}. Although our predicted band structure and gap value agree well with many theoretical calculations, the experimentally reported values of the energy gap $E_{g}$ of Cu$_3$N(D0$_9$) are larger, as will be discussed in Sec. \ref{GWA Calculations and Optical Properties}. This is a well known drawback of Kohn-Sham DFT-based calculations to understimate the band gap.

Fig. \ref{Cu3N1_RhF_3_electronic_structure} shows that the top of the valence band and the bottom of the conduction band of Cu$_3$N(RhF$_3$) are about to touch the Fermi level at $(\Gamma, -0.13 \; eV)$ and  $(\Sigma, 0.15 \; eV)$, respectively. Hence, we have an indirect band gap of $0.28 \; eV$ with $E_F$ slightly shifted towards the VBM.

The third most stable candidate in this 3:1 series is Cu$_3$N(D0$_2$). Its band structure (shown in Fig. \ref{Cu3N1_D0_2_electronic_structure}) has the VBM at $(H, -0.14 \; eV)$, and the CBM at $(\Gamma, 0.14 \; eV)$, resulting in an indirect energy gap of $0.28 \; eV$.

The orbital resolved density of states (PDOS) of both Cu$_3$N(D0$_2$) and Cu$_3$N(RhF$_3$) share the same qualitative features with those of Cu$_3$N(D0$_9$). As expected, the structural relation between these three phases are reflected into their electronic properties.

Energy bands $\epsilon_{i}^{\sigma}(\mathbf{k})$, total density of states (TDOS) and partial (orbital-resolved) density of states (PDOS) of CuN(B17) are  shown in Figs. \ref{Cu1N1_B17_electronic_structure}. It is clear that CuN(B17) would be a true metal at its equilibrium. The major contribution to the very low TDOS around Fermi energy $E_F$ comes from the $2p$ states of the N atoms. Beneath $E_F$ lies a band with $\thicksim 8$ $eV$ of width, in which one can notice that the main contribution is due to the mixture of Cu($3d$) states with N($2p$) states. The N($2s$) states dominate the deep lowest region, while the unoccupied states stem mainly from the N($2p$) states. Fermi surface intersects two bands: a lower one in the $M$-$\Gamma$-$X$-$R$ directions, and a higher band in the $\Gamma$-$M$-$A$ and $M$-$X$ directions. Hence, $E_F$ is not a continous surface contained entirely within the first BZ.

So, in CuN(B17), the nitridation effects on the pure Cu can be summarized from previous sections as: significant increase in the volume per atom $V_{0}$, increase in the bulk modulus resulting in a less compressible material than the pure metal, while the metallic character is preserved . Similar results for CuN(B3) were also arrived at by other researchers \cite{CuN_CuN2_2011_comp}.

It may be worth to mention here that B1 \cite{CuN_AgN_AuN_2007_comp,CuN_CuN2_2011_comp}, B2 \cite{CuN_AgN_AuN_2007_comp,CuN_CuN2_2011_comp}, B3 \cite{CuN_AgN_AuN_2007_comp} and B4 \cite{CuN_AgN_AuN_2007_comp} phases of CuN were also theoretically predicted to be metallic.

With $E_F$ crossing the finite TDOS, Fig. \ref{Cu1N2_C18_electronic_structure} shows that CuN$_2$(C18) is metallic at $0 \; K$. The PDOS reveals that the major contribution to the TDOS at $E_F$ comes from the N($2p$) states with minor contribution from the Cu($3d$) states. Compared to CuN(B17), a new feature of this 1:2 nitride is the contribution of N($2p$) states to N($2s$) states at the deep lowest region. However, variation in N($2s$) energy with respect to \textbf{k} is smaller than the variation of N($2p$) states, resulting in a narrower and higher PDOS. It may be instructive to mention here that CuN$_2$(C1) phase was also found to be metallic \cite{CuN_CuN2_2011_comp}.

A common feature between all the studied cases is the higly structured, intense and narrow series of peaks in the TDOS valance band corresponding to superposition of N($2p$)-states and Cu($3d$)-states. In their $\mathbf{k}$-space, Cu($3d$) energies show little variation with respect to $\mathbf{k}$; hence the Hove singularities-like sharp features.

To summarize this section, we have found that the most stable phases of CuN and CuN$_{2}$ are metallic, while Cu$_{3}$N is a semiconductor. This finding agrees well with literature, specially with Armenta and Soto \cite{Cu3N_Cu4N_Cu8N_2007_comp} who predicted theoretically that the semiconducting state is sensitive to the nitrogen concentration and changes to metallic when the composition is out of the ideal nitrogen to copper ratio, $x = 1/3$. Armenta and Soto, who studied the effect of introducing N atoms in one by one basis to the bulk cubic Cu matrix, also pointed out that as $x$ increases, the TDOS at $E_{F}$, due to both N and Cu atoms, increases as well. Concerning this point, our findings are in excellent agreement with theirs, since we found for the most stable phases (i.e. Cu$_{3}$N, CuN and CuN$_{2}$, respectively) that $\text{TDOS}(x= 1/3) = 0$, $\text{TDOS}(x= 1) \sim 0.70$ and $\text{TDOS}(x= 2) \sim 0.85$, in relative arbitrary units. 
%
%
%
%
\section{\label{GWA Calculations and Optical Properties}GWA Calculations and Optical Properties}
Although a qualitative agreement between DFT-calculated optical properties and experiment is possible, accurate quantitative description requires treatments beyond DFT level \cite{PAW_optics}. Another approach provided by many-body perturbation theory (MBPT) leads to a system of quasi-particle (QP) equations, which can be written for a periodic crystal as \cite{GWA_and_QP_review_1999,Kohanoff,JudithThesis2008}
\begin{eqnarray}	\label{QP equations}
\begin{split}
  \Bigg \{ - \frac{\hbar^{2}} {2m}  \nabla^{2} + \int d\mathbf{r}^{\prime} \frac{n(\mathbf{r}^{\prime})}{|\mathbf{r}-\mathbf{r}^{\prime}|} + V_{ext}(\mathbf{r}) \Bigg \} \psi_{i,\mathbf{k}}^{QP}(\mathbf{r}) \\  + \int d\mathbf{r}^{\prime} \Sigma(\mathbf{r},\mathbf{r}^{\prime};\epsilon_{i,\mathbf{k}}^{QP})  \psi_{i,\mathbf{k}}^{QP}(\mathbf{r}^{\prime}) = \epsilon_{i,\mathbf{k}}^{QP} \psi_{i,\mathbf{k}}^{QP}(\mathbf{r}).
\end{split}
\end{eqnarray}
Practically, the wave functions are taken from the DFT calculations. However, in consideration of computational cost, we used a less dense mesh of $\mathbf{k}$-points ($12 \times 12 \times 12$). The term $\Sigma(\mathbf{r},\mathbf{r}^{\prime};\epsilon_{i,\mathbf{k}}^{QP})$ is the self-energy which contains all the exchange and correlation effects, static and dynamic, including those neglected in our DFT reference system. In the so-called $GW$ approximation, $\Sigma$ is given in terms of Green’s function $G$ as
\begin{eqnarray}	\label{GW self-energy}
\begin{split}
\Sigma_{GW} = j \int d\epsilon^{\prime} G(\mathbf{r},\mathbf{r}^{\prime};\epsilon,\epsilon^{\prime}) W(\mathbf{r},\mathbf{r}^{\prime};\epsilon),
\end{split}
\end{eqnarray}
where the screened interaction $W$ is related to the bare Coulomb interaction $v$ through 
\begin{eqnarray}
\begin{split}
W(\mathbf{r},\mathbf{r}^{\prime};\epsilon) = j \int d\mathbf{r}_{1} \varepsilon^{-1}(\mathbf{r},\mathbf{r}_{1};\epsilon)v(\mathbf{r}_{1},\mathbf{r}^{\prime}),
\end{split}
\end{eqnarray}
with $\varepsilon$ the dielectric function. We followed the $GW_{0}$ self-consistent routine on $G$, in which the QP eigenvalues are updated in the calculations of $G$, while $W$ is kept at the DFT level. After the final iteration of $G$, $\varepsilon$ is calculated, within the so-called random phase approximation (RPA)\footnote{The physical meaning of the RPA is that electrons are considered to respond to the total (external plus induced) field independently \cite{The_GW_method_1998}.}, using the updated QP eigenvalues \cite{Kohanoff,JudithThesis2008,VASPguide}. From the real $\varepsilon_{re}(\omega)$ and the imaginary $\varepsilon_{im}(\omega)$ parts of this frequency-dependent microscopic dielectric tensor one can derive all the other frequency-dependent dielectric response functions, such as absorption coefficient $\alpha\left(\omega\right)$, reflectivity $R\left(\omega\right)$ and refractive index $n\left(\omega\right)$: 
\begin{eqnarray}	\label{R(omega)}
\begin{split}
R\left(\omega\right) = \left| \frac{\left[  \varepsilon_{re}\left(\omega\right) + j \varepsilon_{im}\left(\omega\right)  \right]^{\frac{1}{2}} - 1}{\left[  \varepsilon_{re}\left(\omega\right) + j \varepsilon_{im}\left(\omega\right)  \right]^{\frac{1}{2}} + 1} \right| ^{2}
\end{split}
\end{eqnarray}
\begin{eqnarray}	\label{alpha(omega)}
\begin{split}
\alpha\left(\omega\right) = \sqrt{2} \omega  \left(  \left[  \varepsilon_{re}^{2}\left(\omega\right) + j \varepsilon_{im}^{2}\left(\omega\right) \right]^{\frac{1}{2}}  - \varepsilon_{re}\left(\omega\right) \right)^{\frac{1}{2}} 
\end{split}
\end{eqnarray}
\begin{eqnarray}	\label{n(omega)}
\begin{split}
n\left(\omega\right) = \frac{1}{\sqrt{2}} \left(  \left[  \varepsilon_{re}^{2}\left(\omega\right) + \varepsilon_{im}^{2}\left(\omega\right) \right]^{\frac{1}{2}} + \varepsilon_{re}\left(\omega\right) \right)^{\frac{1}{2}} 
\end{split}
\end{eqnarray}

\begin{figure}	
\includegraphics[width=0.49\textwidth]{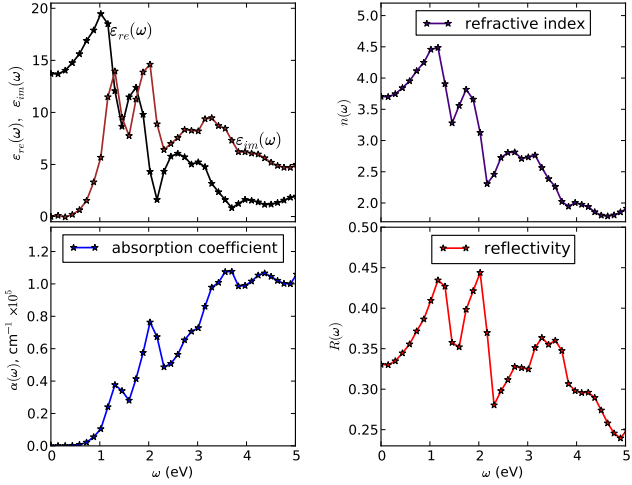}
\caption{\label{Cu3N1_D0_9_optical_constants}(Color online.) Normal-incidence frequency-dependent optical constants of Cu$_3$N(D0$_9$) obtained using Eqs. \ref{R(omega)}--\ref{n(omega)} and $GW_{0}$ eigenvalues.}
\end{figure}

Fig. \ref{Cu3N1_D0_9_optical_constants} displays the real and the imaginary parts of the frequency-dependent dielectric function $\varepsilon_{\text{RPA}}(\omega)$ of Cu$_3$N(D0$_9$) and the corresponding derived optical constants within the optical region\footnote{Recall that the optical region (visible spectrum) is about $(390 \sim 750)$ nm which corresponds to $(3.183 \sim 1.655) \; eV$.}.
The real part $\varepsilon_{re}(\omega)$ shows an upward trend before $0.3 \; eV$, reaches a maximum value at $1 \; eV$ and generally decreases after that. The imaginary part $\varepsilon_{im}(\omega)$ has two main peaks located at $\sim 1.3 \; eV$ and $\sim 2.0 \; eV$. Niu et al. \cite{Cu3N_Cu4N_2011_comp} carried out DFT(UPP-GGA) calculations and derived an $\varepsilon(\omega)$ spectrum with a real part that shows an upward trend before $1 \; eV$,  reaches the maximum value at $1 \; eV$ (as ours) and decreases after that. The imaginary part they obtained has two main peaks, in the range $0 \; eV$ to $5 \; eV$, located at $2.07 \; eV$ and $3.51 \; eV$. By analyzing the DOS, Niu et al. claimed that these two peaks are mainly due to the electron transition from the Cu($3d$) band to the conduction band.

Refractive index $n\left(\omega\right)$ spectrum (Fig. \ref{Cu3N1_D0_9_optical_constants}) shows almost the same frequency dependance as  $\varepsilon_{re}(\omega)$. Comparing our obtained $n\left(\omega\right)$ with the experimental results of Gordillo et al. \cite{Cu3N_2009_exp_optical_properties}, one can see a kind of general qualitative agreement between the experimental curve and the theoretical one, represented by the increase in $n$ with increase in the photon energy till reaching a maximum, then followed by a generally decreasing behaviour. However, the experimental peak of $n$ is at $\sim (1.4 \;eV , 3.65)$, while our obtained peak is at $\sim (1.2 \;eV , 4.5)$. Surprisingly, the DFT(UPP-GGA)-refractive index spectrum derived by Niu et al. \cite{Cu3N_Cu4N_2011_comp} shows far better agreement with the experimental results of Gordillo et al. \cite{Cu3N_2009_exp_optical_properties} and they got the peak at $\sim (1.6 \;eV , 3.9)$.

From the absorption coefficient $\alpha\left(\omega\right)$ spectrum (Fig. \ref{Cu3N1_D0_9_optical_constants}), it can be seen that Cu$_{3}$N(D0$_{9}$) starts absorbing photons with $\sim 0.75 \; eV$ energy. Hence, it is clear that $GW_{0}$ calculations give a band gap of $\sim 0.75 \; eV$, which is a significant correction to the obtained DFT value. Our presented $\alpha\left(\omega\right)$ spectrum agrees qualitatively well with the experimental work of Gordillo et al. \cite{Cu3N_2009_exp_optical_properties}, who, in the $\sim (0.6 \; - \; 1.4) \; eV$ region, obtained a smooth exponential-like curve. However, their obtained $\alpha\left(\omega\right)$ reaches $1 \times 10^{5} \; \text{cm}^{-1}$ in the visible range before $1.4 \; eV$, while ours (Fig. \ref{Cu3N1_D0_9_optical_constants}) never reaches such a value before $3.3 \; eV$. Niu et al. \cite{Cu3N_Cu4N_2011_comp} got a curve that reaches this value at $\sim 3.6 \; eV$. However, their $\alpha\left(\omega\right)$ spectrum starts to be non-zero from $\sim 0.71 \; eV$!

Gordillo et al. \cite{Cu3N_2009_exp_optical_properties} prepared nearly stoichiometric copper nitride  polycrystalline films having nitrogen contents of $(27 \pm 2)\%$ with lattice parameter $a = 3.8621$ \AA. They referred to it in their article as stoichiometric Cu$_{3}$N, and, at room temperature and with orientation along the $(1 \; 0 \; 0)$ crystallographic axis, they carried out some optical measurements and fitted the obtained data. From the fits, they managed to derive the refractive index and the absorption coefficient, while reflectance was measured directly. So, although our calculated optical properties show partial agreement with this experimental work, discrepancies may be attributed due to the lack of knowledge of the exact stoichiometry of the prepared samples. Wang, Nakamine, and Hayashi \cite{metallic_Cu3N_1998_exp} also prepared nearly stoichiometric Cu$_{3}$N films at $67 \; \text{Pa}$. Their experimentally obtained $\alpha\left(\omega\right)$ spectra reach $1 \times 10^{5} \; \text{cm}^{-1}$ at about $2.4 \; eV$. However, $\alpha$ leaves the zero level only at about $(1.8 \sim 1.9) \; eV$.

On the other hand, the used $E_{cut} = 290 \; eV$ and $8 \times 8 \times 8$ $\mathbf{k}$-mesh in the DFT(UPP-GGA) calculations by Niu et al. \cite{Cu3N_Cu4N_2011_comp} may not be sufficient to reproduce \textit{qualitatively} similar spectra as those we obtained from GWA calculations. No quantitative correspondence is to be expected.

Experimentally reported values of the Cu$_{3}$N(D0$_{9}$) optical gap spread over a wide range \cite{Cu3N_2006_exp,Cu3N_2009_exp_optical_properties}. Some of these values are: $(0.25 \sim 0.83) \; eV$ \cite{copper_nitrides_optical_properties_2001_exp}, $1.30 \; eV$ \cite{copper_nitrides_2001_exp,Cu3N_1995_exp}, $(1.7 \sim 1.84) \; eV$ \cite{optical_properties_Cu3N_2011_exp}, $1.85 \; eV$ \cite{nano_Cu3N_2005_exp} and $(1.8 \sim 1.9) \; eV$ \cite{metallic_Cu3N_1998_exp}. Hence, although GWA calculations give a band gap of $\sim 0.75 \; eV$, which is a significant correction to the obtained DFT value of $0.26 \; eV$, the difference between theory and experiment is still considerable. This is a well known problem with GWA calculations on top of DFT eigenvalues and eigenstates that correspond to a very small bandgap compared to experiment.

Given that standard DFT functionals severely underestimate the band gaps while the Hartree-Fock (HF) approximation overestimates them\cite{DFT_vs_GW_2002}, a potential solution is to combine local or semilocal DFT exchange with a portion of nonlocal exact exchange thereby constructing the so-called hybrid functional \cite{exact_X_Becke_1993}. Much improved band gaps can be obtained by screening the nonlocal HF-type portion of exchange potential with a suitable screening parameter \cite{Hybrid_DFT_2002,Hybrid_2008}. A more appropriate approach is to apply the partially self-consistent GW method on eigenvalues obtained using hybrid functionals or DFT+U schemes \cite{PhysRevLett.96.116405} which readily provide better band gaps for insulators \cite{arXiv_hybrid_2012}.
%
\section{Summary and Conclusions}
DFT-based first-principles calculations on bulk crystalline Cu$_3$N, CuN and CuN$_2$ over a series of twenty structural phases have been successfully carried out. The studied structural properties include energy-volume equation of state (EOS), equilibrium lattice structural parameters, cohesive and formation energies, relative phase stabilities, bulk modulus and its presssure derivative. Electronic characterization of the energetically most stable phases was done via the analysis of their band structure and their total and partial density of states (DOS). Further, we carried out GW$_{0}$ calculations within the random-phase approximation (RPA) to the dielectric tensor $\varepsilon_{\text{RPA}}(\omega)$. The frequency-dependent optical constants (absorption coefficient, reflectivity and refractive index spectra) of the experimentally reported phase Cu$_{3}$N(D0$_{9}$) were derived from the updated $\varepsilon_{\text{RPA}}(\omega)$. Obtained results were compared with experiment and/or with previous calculations whenever possible. The main conclusions which we can derive from all these calculations are the followings:
\begin{itemize}
\item The calculated lattice constants are in good agreement with experiment and with theory.
\item From the obtained cohesive energies, the energetically most stable phases are D0$_{9}$, B17, and C18 in the Cu$_{3}$N, CuN and CuN$_{2}$ stoichiometric series, respectively. However, other Cu$_{3}$N phases show similar stability to Cu$_{3}$N(D0$_{9}$) and may present during the nitridation process.
\item Including the successfully synthesized Cu$_{3}$N(D0$_{9}$) phase, all obtained formation energies are positive, yet they can be used to measure the relative thermodynamic stabilities of these phases.
\item Although CuN$_{2}$(C18) is the most bound phase, its tendency to decompose back into its elemental constituents is more than the less bound Cu$_{3}$N phases.
\item The volume dependence of the bulk modulus is more precisely described by the change in volume of the Cu sublattice rather than the common average atomic volume of the nitride.
\item The most stable Cu$_3$N phases are predicted to be indirect-gap semiconducing materials with lower bulk modulus than the pure metal, while CuN(B17) preserves the metallicity and improves the bulk modulus. However, the CuN$_{2}$(C18) phase substantially increases the compressibility while preserving the metallicity.
\item Our GWA calculated optical properties show partial agreement with experiment and with the available theoretical work (Ref. [\onlinecite{Cu3N_Cu4N_2011_comp}]). Discrepancies are probably due to the lack of knowledge of the exact stoichiometry of the prepared samples, and due to the big difference in the used plane waves cut-off energy and in the density of the $\mathbf{k}$-mesh. Convergence criterion with respect to these two quantities have not been referred to in Ref. [\onlinecite{Cu3N_Cu4N_2011_comp}].
\item Our GWA calculated energy gap of Cu$_{3}$N(D0$_{9}$) shows significant improvement over the calculated DFT value.
\end{itemize}
We hope that the present work would serve as a reference source for meaningful comparisons which may be made among the largely different calculations.
%
\section*{Acknowledgments}
All GW calculations and some DFT calculations were carried out using the infrastructure of the Centre for High Performance Computing (CHPC) in Cape Town. Suleiman would like to acknowledge the support he recieved from Wits, DAAD, AIMS and SUST. Many thanks to the ASESMA family, and special thanks to Dr Kris Delaney and Sinead Griffin for their invaluable help and useful discussions.
%
\bibliography{arXiv_copper_nitrides_article.bib}

\providecommand{\noopsort}[1]{}\providecommand{\singleletter}[1]{#1}%
\begin{thebibliography}{100}%
\makeatletter
\providecommand \@ifxundefined [1]{%
 \ifx #1\undefined \expandafter \@firstoftwo
 \else \expandafter \@secondoftwo
\fi
}%
\providecommand \@ifnum [1]{%
 \ifnum #1\expandafter \@firstoftwo
 \else \expandafter \@secondoftwo
\fi
}%
\providecommand \enquote [1]{``#1''}%
\providecommand \bibnamefont  [1]{#1}%
\providecommand \bibfnamefont [1]{#1}%
\providecommand \citenamefont [1]{#1}%
\providecommand\href[0]{\@sanitize\@href}%
\providecommand\@href[1]{\endgroup\@@startlink{#1}\endgroup\@@href}%
\providecommand\@@href[1]{#1\@@endlink}%
\providecommand \@sanitize [0]{\begingroup\catcode`\&12\catcode`\#12\relax}%
\@ifxundefined \pdfoutput {\@firstoftwo}{%
 \@ifnum{\z@=\pdfoutput}{\@firstoftwo}{\@secondoftwo}%
}{%
 \providecommand\@@startlink[1]{\leavevmode\special{html:<a href="#1">}}%
 \providecommand\@@endlink[0]{\special{html:</a>}}%
}{%
 \providecommand\@@startlink[1]{%
  \leavevmode
  \pdfstartlink
   attr{/Border[0 0 1 ]/H/I/C[0 1 1]}%
   user{/Subtype/Link/A<</Type/Action/S/URI/URI(#1)>>}%
  \relax
 }%
 \providecommand\@@endlink[0]{\pdfendlink}%
}%
\providecommand \url  [0]{\begingroup\@sanitize \@url }%
\providecommand \@url [1]{\endgroup\@href {#1}{\urlprefix}}%
\providecommand \urlprefix [0]{URL }%
\providecommand \Eprint[0]{\href }%
\@ifxundefined \urlstyle {%
  \providecommand \doi [1]{doi:\discretionary{}{}{}#1}%
}{%
  \providecommand \doi [0]{doi:\discretionary{}{}{}\begingroup
  \urlstyle{rm}\Url }%
}%
\providecommand \doibase [0]{http://dx.doi.org/}%
\providecommand \Doi[1]{\href{\doibase#1}}%
\providecommand \bibAnnote [3]{%
  \BibitemShut{#1}%
  \begin{quotation}\noindent
    \textsc{Key:}\ #2\\\textsc{Annotation:}\ #3%
  \end{quotation}%
}%
\providecommand \bibAnnoteFile [2]{%
  \IfFileExists{#2}{\bibAnnote {#1} {#2} {\input{#2}}}{}%
}%
\providecommand \typeout [0]{\immediate \write \m@ne }%
\providecommand \selectlanguage [0]{\@gobble}%
\providecommand \bibinfo [0]{\@secondoftwo}%
\providecommand \bibfield [0]{\@secondoftwo}%
\providecommand \translation [1]{[#1]}%
\providecommand \BibitemOpen[0]{}%
\providecommand \bibitemStop [0]{}%
\providecommand \bibitemNoStop [0]{.\EOS\space}%
\providecommand \EOS [0]{\spacefactor3000\relax}%
\providecommand \BibitemShut [1]{\csname bibitem#1\endcsname}%
\bibitem{the_first_Cu3N_1939_exp}%
  \BibitemOpen
  \bibfield{author}{%
  \bibinfo {author} {\bibfnamefont{R.}~\bibnamefont{Juza}}\ and\ \bibinfo
  {author} {\bibfnamefont{H.}~\bibnamefont{Hahn}},\ }%
  \bibfield{journal}{%
  \Doi{10.1002/zaac.19392410204}{\bibinfo {journal} {Zeitschrift für
  anorganische und allgemeine Chemie}}\ }%
  \textbf{\bibinfo {volume} {241}},\ \bibinfo {pages} {172} (\bibinfo {year}
  {1939}),\ ISSN \bibinfo {issn} {1521-3749},\
  \url{http://dx.doi.org/10.1002/zaac.19392410204}%
  \bibAnnoteFile{NoStop}{the_first_Cu3N_1939_exp}%
\bibitem{Cu3N_1996_comp}%
  \BibitemOpen
  \bibfield{author}{%
  \bibinfo {author} {\bibfnamefont{U.}~\bibnamefont{Hahn}}\ and\ \bibinfo
  {author} {\bibfnamefont{W.}~\bibnamefont{Weber}},\ }%
  \bibfield{journal}{%
  \Doi{10.1103/PhysRevB.53.12684}{\bibinfo {journal} {Phys. Rev. B}}\ }%
  \textbf{\bibinfo {volume} {53}},\ \bibinfo {pages} {12684} (\bibinfo {month}
  {May}\ \bibinfo {year} {1996}),\
  \url{http://link.aps.org/doi/10.1103/PhysRevB.53.12684}%
  \bibAnnoteFile{NoStop}{Cu3N_1996_comp}%
\bibitem{Cu3N_2006_exp}%
  \BibitemOpen
  \bibfield{author}{%
  \bibinfo {author} {\bibfnamefont{A.}~\bibnamefont{Ji}}, \bibinfo {author}
  {\bibfnamefont{R.}~\bibnamefont{Huang}}, \bibinfo {author}
  {\bibfnamefont{Y.}~\bibnamefont{Du}}, \bibinfo {author}
  {\bibfnamefont{C.}~\bibnamefont{Li}}, \bibinfo {author}
  {\bibfnamefont{Y.}~\bibnamefont{Wang}},\ and\ \bibinfo {author}
  {\bibfnamefont{Z.}~\bibnamefont{Cao}},\ }%
  \bibfield{journal}{%
  \Doi{10.1016/j.jcrysgro.2006.07.007}{\bibinfo {journal} {Journal of Crystal
  Growth}}\ }%
  \textbf{\bibinfo {volume} {295}},\ \bibinfo {pages} {79 } (\bibinfo {year}
  {2006}),\ ISSN \bibinfo {issn} {0022-0248},\
  \url{http://www.sciencedirect.com/science/article/pii/S0022024806006944}%
  \bibAnnoteFile{NoStop}{Cu3N_2006_exp}%
\bibitem{nano_Cu3N_2005_exp}%
  \BibitemOpen
  \bibfield{author}{%
  \bibinfo {author} {\bibfnamefont{Y.}~\bibnamefont{Du}}, \bibinfo {author}
  {\bibfnamefont{A.}~\bibnamefont{Ji}}, \bibinfo {author}
  {\bibfnamefont{L.}~\bibnamefont{Ma}}, \bibinfo {author}
  {\bibfnamefont{Y.}~\bibnamefont{Wang}},\ and\ \bibinfo {author}
  {\bibfnamefont{Z.}~\bibnamefont{Cao}},\ }%
  \bibfield{journal}{%
  \Doi{10.1016/j.jcrysgro.2005.03.077}{\bibinfo {journal} {Journal of Crystal
  Growth}}\ }%
  \textbf{\bibinfo {volume} {280}},\ \bibinfo {pages} {490 } (\bibinfo {year}
  {2005}),\ ISSN \bibinfo {issn} {0022-0248},\
  \url{http://www.sciencedirect.com/science/article/pii/S0022024805004264}%
  \bibAnnoteFile{NoStop}{nano_Cu3N_2005_exp}%
\bibitem{Cu3N_Cu4N_Cu8N_2007_comp}%
  \BibitemOpen
  \bibfield{author}{%
  \bibinfo {author} {\bibfnamefont{M.~G.}\ \bibnamefont{Moreno-Armenta}}\ and\
  \bibinfo {author} {\bibfnamefont{G.}~\bibnamefont{Soto}},\ }%
  \bibfield{journal}{%
  \Doi{10.1016/j.solidstatesciences.2007.10.003}{\bibinfo {journal} {Solid
  State Sciences}}\ }%
  \textbf{\bibinfo {volume} {10}},\ \bibinfo {pages} {573} (\bibinfo {year}
  {2008}),\ ISSN \bibinfo {issn} {1293-2558},\
  \url{http://www.sciencedirect.com/science/article/pii/S1293255807002920}%
  \bibAnnoteFile{NoStop}{Cu3N_Cu4N_Cu8N_2007_comp}%
\bibitem{copper_nitrides_2001_exp}%
  \BibitemOpen
  \bibfield{author}{%
  \bibinfo {author} {\bibfnamefont{S.}~\bibnamefont{Ghosh}}, \bibinfo {author}
  {\bibfnamefont{F.}~\bibnamefont{Singh}}, \bibinfo {author}
  {\bibfnamefont{D.}~\bibnamefont{Choudhary}}, \bibinfo {author}
  {\bibfnamefont{D.}~\bibnamefont{Avasthi}}, \bibinfo {author}
  {\bibfnamefont{V.}~\bibnamefont{Ganesan}}, \bibinfo {author}
  {\bibfnamefont{P.}~\bibnamefont{Shah}},\ and\ \bibinfo {author}
  {\bibfnamefont{A.}~\bibnamefont{Gupta}},\ }%
  \bibfield{journal}{%
  \Doi{10.1016/S0257-8972(01)01091-X}{\bibinfo {journal} {Surface and Coatings
  Technology}}\ }%
  \textbf{\bibinfo {volume} {142–144}},\ \bibinfo {pages} {1034 } (\bibinfo
  {year} {2001}),\ ISSN \bibinfo {issn} {0257-8972},\ \bibinfo {note}
  {<ce:title>Proceedings of the 7th International Conference on Plasma Surface
  Engineering</ce:title>},\
  \url{http://www.sciencedirect.com/science/article/pii/S025789720101091X}%
  \bibAnnoteFile{NoStop}{copper_nitrides_2001_exp}%
\bibitem{Cu3N_2009_exp_optical_properties}%
  \BibitemOpen
  \bibfield{author}{%
  \bibinfo {author} {\bibfnamefont{N.}~\bibnamefont{Gordillo}}, \bibinfo
  {author} {\bibfnamefont{R.}~\bibnamefont{Gonzalez-Arrabal}}, \bibinfo
  {author} {\bibfnamefont{A.}~\bibnamefont{Álvarez Herrero}},\ and\ \bibinfo
  {author} {\bibfnamefont{F.}~\bibnamefont{Agulló-López}},\ }%
  \bibfield{journal}{%
  \bibinfo {journal} {Journal of Physics D: Applied Physics}\ }%
  \textbf{\bibinfo {volume} {42}},\ \bibinfo {pages} {165101} (\bibinfo {year}
  {2009}),\ \url{stacks.iop.org/JPhysD/42/165101}%
  \bibAnnoteFile{NoStop}{Cu3N_2009_exp_optical_properties}%
\bibitem{Cu3N_2000_applications}%
  \BibitemOpen
  \bibfield{author}{%
  \bibinfo {author} {\bibfnamefont{R.}~\bibnamefont{Cremer}}, \bibinfo {author}
  {\bibfnamefont{M.}~\bibnamefont{Witthaut}}, \bibinfo {author}
  {\bibfnamefont{D.}~\bibnamefont{Neuschütz}}, \bibinfo {author}
  {\bibfnamefont{C.}~\bibnamefont{Trappe}}, \bibinfo {author}
  {\bibfnamefont{M.}~\bibnamefont{Laurenzis}}, \bibinfo {author}
  {\bibfnamefont{O.}~\bibnamefont{Winkler}},\ and\ \bibinfo {author}
  {\bibfnamefont{H.}~\bibnamefont{Kurz}},\ }%
  \bibfield{journal}{%
  \bibinfo {journal} {Microchimica Acta}\ }%
  \textbf{\bibinfo {volume} {133}},\ \bibinfo {pages} {299} (\bibinfo {year}
  {2000}),\ ISSN \bibinfo {issn} {0026-3672},\ \bibinfo {note}
  {10.1007/s006040070109},\ \url{http://dx.doi.org/10.1007/s006040070109}%
  \bibAnnoteFile{NoStop}{Cu3N_2000_applications}%
\bibitem{Cu3N_1996_exp}%
  \BibitemOpen
  \bibfield{author}{%
  \bibinfo {author} {\bibfnamefont{T.}~\bibnamefont{Maruyama}}\ and\ \bibinfo
  {author} {\bibfnamefont{T.}~\bibnamefont{Morishita}},\ }%
  \bibfield{journal}{%
  \Doi{10.1063/1.117978}{\bibinfo {journal} {Applied Physics Letters}}\ }%
  \textbf{\bibinfo {volume} {69}},\ \bibinfo {pages} {890} (\bibinfo {year}
  {1996}),\ \url{http://dx.doi.org/10.1063/1.117978}%
  \bibAnnoteFile{NoStop}{Cu3N_1996_exp}%
\bibitem{Cu3N_1990_exp}%
  \BibitemOpen
  \bibfield{author}{%
  \bibinfo {author} {\bibfnamefont{M.}~\bibnamefont{Asano}}, \bibinfo {author}
  {\bibfnamefont{K.}~\bibnamefont{Umeda}},\ and\ \bibinfo {author}
  {\bibfnamefont{A.}~\bibnamefont{Tasaki}},\ }%
  \bibfield{journal}{%
  \Doi{10.1143/JJAP.29.1985}{\bibinfo {journal} {Japanese Journal of Applied
  Physics}}\ }%
  \textbf{\bibinfo {volume} {29}},\ \bibinfo {pages} {1985} (\bibinfo {year}
  {1990}),\ \url{http://jjap.jsap.jp/link?JJAP/29/1985/}%
  \bibAnnoteFile{NoStop}{Cu3N_1990_exp}%
\bibitem{Cu3N_Ni3N_1993_exp_conference}%
  \BibitemOpen
  \bibfield{author}{%
  \bibinfo {author} {\bibfnamefont{L.}~\bibnamefont{Maya}}\ }%
  (\bibinfo {year} {1993})\ pp.\ \bibinfo {pages} {203--208},\
  \url{http://www.scopus.com/inward/record.url?eid=2-s2.0-0027211159&partnerID%
=40&md5=02970645b01832b3506a2c4b0b6f7126}%
  \bibAnnoteFile{NoStop}{Cu3N_Ni3N_1993_exp_conference}%
\bibitem{metallic_Cu3N_1998_exp}%
  \BibitemOpen
  \bibfield{author}{%
  \bibinfo {author} {\bibfnamefont{D.}~\bibnamefont{yuan Wang}}, \bibinfo
  {author} {\bibfnamefont{N.}~\bibnamefont{Nakamine}},\ and\ \bibinfo {author}
  {\bibfnamefont{Y.}~\bibnamefont{Hayashi}},\ }%
  \bibfield{journal}{%
  \Doi{10.1116/1.581314}{\bibinfo {journal} {Journal of Vacuum Science \&
  Technology A: Vacuum, Surfaces, and Films}}\ }%
  \textbf{\bibinfo {volume} {16}},\ \bibinfo {pages} {2084 } (\bibinfo {year}
  {1998}),\ \url{http://link.aip.org/link/?JVA/16/2084/1}%
  \bibAnnoteFile{NoStop}{metallic_Cu3N_1998_exp}%
\bibitem{Cu3N_Cu4N_2006_exp_just_read_it}%
  \BibitemOpen
  \bibfield{author}{%
  \bibinfo {author} {\bibfnamefont{C.}~\bibnamefont{Gallardo-Vega}}\ and\
  \bibinfo {author} {\bibfnamefont{W.}~\bibnamefont{de~la Cruz}},\ }%
  \bibfield{journal}{%
  \Doi{10.1016/j.apsusc.2005.10.007}{\bibinfo {journal} {Applied Surface
  Science}}\ }%
  \textbf{\bibinfo {volume} {252}},\ \bibinfo {pages} {8001 } (\bibinfo {year}
  {2006}),\ ISSN \bibinfo {issn} {0169-4332},\
  \url{http://www.sciencedirect.com/science/article/pii/S0169433205014844}%
  \bibAnnoteFile{NoStop}{Cu3N_Cu4N_2006_exp_just_read_it}%
\bibitem{copper_nitrides_optical_properties_2001_exp}%
  \BibitemOpen
  \bibfield{author}{%
  \bibinfo {author} {\bibfnamefont{J.}~\bibnamefont{Pierson}},\ }%
  \bibfield{journal}{%
  \Doi{10.1016/S0042-207X(01)00425-0}{\bibinfo {journal} {Vacuum}}\ }%
  \textbf{\bibinfo {volume} {66}},\ \bibinfo {pages} {59 } (\bibinfo {year}
  {2002}),\ ISSN \bibinfo {issn} {0042-207X},\
  \url{http://www.sciencedirect.com/science/article/pii/S0042207X01004250}%
  \bibAnnoteFile{NoStop}{copper_nitrides_optical_properties_2001_exp}%
\bibitem{Cu3N_Cu4N_Cu3N2_2008_comp}%
  \BibitemOpen
  \bibfield{author}{%
  \bibinfo {author} {\bibfnamefont{Z.}~\bibnamefont{Hou}},\ }%
  \bibfield{journal}{%
  \Doi{10.1016/j.solidstatesciences.2008.02.013}{\bibinfo {journal} {Solid
  State Sciences}}\ }%
  \textbf{\bibinfo {volume} {10}},\ \bibinfo {pages} {1651} (\bibinfo {year}
  {2008}),\ ISSN \bibinfo {issn} {1293-2558},\
  \url{http://www.sciencedirect.com/science/article/pii/S1293255808000691}%
  \bibAnnoteFile{NoStop}{Cu3N_Cu4N_Cu3N2_2008_comp}%
\bibitem{Cu3N_1989_exp}%
  \BibitemOpen
  \bibfield{author}{%
  \bibinfo {author} {\bibfnamefont{S.}~\bibnamefont{Terada}}, \bibinfo {author}
  {\bibfnamefont{H.}~\bibnamefont{Tanaka}},\ and\ \bibinfo {author}
  {\bibfnamefont{K.}~\bibnamefont{Kubota}},\ }%
  \bibfield{journal}{%
  \Doi{10.1016/0022-0248(89)90038-9}{\bibinfo {journal} {Journal of Crystal
  Growth}}\ }%
  \textbf{\bibinfo {volume} {94}},\ \bibinfo {pages} {567 } (\bibinfo {year}
  {1989}),\ ISSN \bibinfo {issn} {0022-0248},\
  \url{http://www.sciencedirect.com/science/article/pii/0022024889900389}%
  \bibAnnoteFile{NoStop}{Cu3N_1989_exp}%
\bibitem{Cu3N_2004_comp}%
  \BibitemOpen
  \bibfield{author}{%
  \bibinfo {author} {\bibfnamefont{M.~G.}\ \bibnamefont{Moreno-Armenta}},
  \bibinfo {author} {\bibfnamefont{A.}~\bibnamefont{Martı́nez-Ruiz}},\ and\
  \bibinfo {author} {\bibfnamefont{N.}~\bibnamefont{Takeuchi}},\ }%
  \bibfield{journal}{%
  \Doi{10.1016/j.solidstatesciences.2003.10.014}{\bibinfo {journal} {Solid
  State Sciences}}\ }%
  \textbf{\bibinfo {volume} {6}},\ \bibinfo {pages} {9} (\bibinfo {year}
  {2004}),\ ISSN \bibinfo {issn} {1293-2558},\
  \url{http://www.sciencedirect.com/science/article/pii/S1293255803002504}%
  \bibAnnoteFile{NoStop}{Cu3N_2004_comp}%
\bibitem{CuxNy_2009_ooooh}%
  \BibitemOpen
  \bibfield{author}{%
  \bibinfo {author} {\bibfnamefont{J.-N.}\ \bibnamefont{Ding}}, \bibinfo
  {author} {\bibfnamefont{N.-Y.}\ \bibnamefont{Yuan}}, \bibinfo {author}
  {\bibfnamefont{F.}~\bibnamefont{Li}}, \bibinfo {author}
  {\bibfnamefont{G.-Q.}\ \bibnamefont{Ding}}, \bibinfo {author}
  {\bibfnamefont{Z.-G.}\ \bibnamefont{Chen}}, \bibinfo {author}
  {\bibfnamefont{X.-S.}\ \bibnamefont{Chen}}, ,\ and\ \bibinfo {author}
  {\bibfnamefont{W.}~\bibnamefont{Lu}},\ }%
  \bibfield{journal}{%
  \Doi{10.1063/1.3257899}{\bibinfo {journal} {Journal of Chemical Physics}}\ }%
  \textbf{\bibinfo {volume} {131}},\ \bibinfo {pages} {174102} (\bibinfo {year}
  {2009}),\ ISSN \bibinfo {issn} {1089-7690},\
  \url{http://dx.doi.org/10.1063/1.3257899}%
  \bibAnnoteFile{NoStop}{CuxNy_2009_ooooh}%
\bibitem{Cu3N_thin_films_2005_exp}%
  \BibitemOpen
  \bibfield{author}{%
  \bibinfo {author} {\bibfnamefont{G.}~\bibnamefont{Yue}}, \bibinfo {author}
  {\bibfnamefont{P.}~\bibnamefont{Yan}},\ and\ \bibinfo {author}
  {\bibfnamefont{J.}~\bibnamefont{Wang}},\ }%
  \bibfield{journal}{%
  \Doi{10.1016/j.jcrysgro.2004.10.032}{\bibinfo {journal} {Journal of Crystal
  Growth}}\ }%
  \textbf{\bibinfo {volume} {274}},\ \bibinfo {pages} {464 } (\bibinfo {year}
  {2005}),\ ISSN \bibinfo {issn} {0022-0248},\
  \url{http://www.sciencedirect.com/science/article/pii/S0022024804012928}%
  \bibAnnoteFile{NoStop}{Cu3N_thin_films_2005_exp}%
\bibitem{copper_nitride_films_2003_exp}%
  \BibitemOpen
  \bibfield{author}{%
  \bibinfo {author} {\bibfnamefont{G.}~\bibnamefont{Soto}}, \bibinfo {author}
  {\bibfnamefont{J.}~\bibnamefont{Dı́az}},\ and\ \bibinfo {author}
  {\bibfnamefont{W.}~\bibnamefont{de~la Cruz}},\ }%
  \bibfield{journal}{%
  \Doi{10.1016/S0167-577X(03)00277-5}{\bibinfo {journal} {Materials Letters}}\
  }%
  \textbf{\bibinfo {volume} {57}},\ \bibinfo {pages} {4130 } (\bibinfo {year}
  {2003}),\ ISSN \bibinfo {issn} {0167-577X},\
  \url{http://www.sciencedirect.com/science/article/pii/S0167577X03002775}%
  \bibAnnoteFile{NoStop}{copper_nitride_films_2003_exp}%
\bibitem{1stPtN2004}%
  \BibitemOpen
  \bibfield{author}{%
  \bibinfo {author} {\bibfnamefont{E.}~\bibnamefont{Gregoryanz}}, \bibinfo
  {author} {\bibfnamefont{C.}~\bibnamefont{Sanloup}}, \bibinfo {author}
  {\bibfnamefont{M.}~\bibnamefont{Somayazulu}}, \bibinfo {author}
  {\bibfnamefont{J.}~\bibnamefont{Badro}}, \bibinfo {author}
  {\bibfnamefont{G.}~\bibnamefont{Fiquet}}, \bibinfo {author}
  {\bibfnamefont{H.}~\bibnamefont{kwang Mao}},\ and\ \bibinfo {author}
  {\bibfnamefont{R.~J.}\ \bibnamefont{Hemley}},\ }%
  \bibfield{journal}{%
  \Doi{10.1038/nmat1115}{\bibinfo {journal} {Nature Materials}}\ }%
  \textbf{\bibinfo {volume} {3}},\ \bibinfo {pages} {294 } (\bibinfo {year}
  {2004}),\ \url{http://www.nature.com/nmat/journal/v3/n5/full/nmat1115.html}%
  \bibAnnoteFile{NoStop}{1stPtN2004}%
\bibitem{Cu3N_Cu4N_Cu6N_2004_exp}%
  \BibitemOpen
  \bibfield{author}{%
  \bibinfo {author} {\bibfnamefont{F.}~\bibnamefont{Gulo}}, \bibinfo {author}
  {\bibfnamefont{A.}~\bibnamefont{Simon}}, \bibinfo {author}
  {\bibfnamefont{J.}~\bibnamefont{Köhler}},\ and\ \bibinfo {author}
  {\bibfnamefont{R.~K.}\ \bibnamefont{Kremer}},\ }%
  \bibfield{journal}{%
  \Doi{10.1002/anie.200353424}{\bibinfo {journal} {Angewandte Chemie
  International Edition}}\ }%
  \textbf{\bibinfo {volume} {43}},\ \bibinfo {pages} {2032} (\bibinfo {year}
  {2004}),\ ISSN \bibinfo {issn} {1521-3773},\
  \url{http://dx.doi.org/10.1002/anie.200353424}%
  \bibAnnoteFile{NoStop}{Cu3N_Cu4N_Cu6N_2004_exp}%
\bibitem{Cu4N_1989_exp}%
  \BibitemOpen
  \bibfield{author}{%
  \bibinfo {author} {\bibfnamefont{J.}~\bibnamefont{Blucher}}, \bibinfo
  {author} {\bibfnamefont{K.}~\bibnamefont{Bang}},\ and\ \bibinfo {author}
  {\bibfnamefont{B.}~\bibnamefont{Giessen}},\ }%
  \bibfield{journal}{%
  \Doi{10.1016/0921-5093(89)90110-X}{\bibinfo {journal} {Materials Science and
  Engineering: A}}\ }%
  \textbf{\bibinfo {volume} {117}},\ \bibinfo {pages} {L1} (\bibinfo {year}
  {1989}),\ ISSN \bibinfo {issn} {0921-5093},\
  \url{http://www.sciencedirect.com/science/article/pii/092150938990110X}%
  \bibAnnoteFile{NoStop}{Cu4N_1989_exp}%
\bibitem{StructuralInChem}%
  \BibitemOpen
  \bibfield{author}{%
  \bibinfo {author} {\bibfnamefont{A.~F.}\ \bibnamefont{Wells}},\ }%
  \emph{\bibinfo {title} {Structural Inorganic Chemistry}},\ \bibinfo {edition}
  {5th}\ ed.\ (\bibinfo {publisher} {Oxford University Press},\ \bibinfo {year}
  {1984})%
  \bibAnnoteFile{NoStop}{StructuralInChem}%
\bibitem{CuN_AgN_AuN_2007_comp}%
  \BibitemOpen
  \bibfield{author}{%
  \bibinfo {author} {\bibfnamefont{M.}~\bibnamefont{Kanoun}}\ and\ \bibinfo
  {author} {\bibfnamefont{S.}~\bibnamefont{Goumri-Said}},\ }%
  \bibfield{journal}{%
  \Doi{10.1016/j.physleta.2006.09.100}{\bibinfo {journal} {Physics Letters A}}\
  }%
  \textbf{\bibinfo {volume} {362}},\ \bibinfo {pages} {73 } (\bibinfo {year}
  {2007}),\ ISSN \bibinfo {issn} {0375-9601},\
  \url{http://www.sciencedirect.com/science/article/pii/S0375960106015337}%
  \bibAnnoteFile{NoStop}{CuN_AgN_AuN_2007_comp}%
\bibitem{Note1}%
  \BibitemOpen
  \bibinfo {note} {$Z$ here, and in what follows, refers to the number of
  chemical formula units per unit cell.}%
  \bibAnnoteFile{Stop}{Note1}%
\bibitem{Pearson}%
  \BibitemOpen
  \bibfield{author}{%
  \bibinfo {author} {\bibfnamefont{W.~B.}\ \bibnamefont{Pearson}},\ }%
  \emph{\bibinfo {title} {Handbook of Lattice Spacings \& Structures of Metals
  \& Alloys (International Series of Monographs on Metals Physics \& Physical
  Metallurgy)}}\ (\bibinfo {publisher} {Franklin Book Co},\ \bibinfo {year}
  {1964})\ ISBN \bibinfo {isbn} {0080090788},\
  \url{http://www.amazon.com/Handbook-Structures-International-Monographs-Meta%
llurgy/dp/0080090788}%
  \bibAnnoteFile{NoStop}{Pearson}%
\bibitem{AgN2_AuN2_PtN2_2005_comp}%
  \BibitemOpen
  \bibfield{author}{%
  \bibinfo {author} {\bibfnamefont{R.}~\bibnamefont{Yu}}\ and\ \bibinfo
  {author} {\bibfnamefont{X.~F.}\ \bibnamefont{Zhang}},\ }%
  \bibfield{journal}{%
  \Doi{10.1103/PhysRevB.72.054103}{\bibinfo {journal} {Phys. Rev. B}}\ }%
  \textbf{\bibinfo {volume} {72}},\ \bibinfo {pages} {054103} (\bibinfo {month}
  {Aug}\ \bibinfo {year} {2005}),\
  \url{http://link.aps.org/doi/10.1103/PhysRevB.72.054103}%
  \bibAnnoteFile{NoStop}{AgN2_AuN2_PtN2_2005_comp}%
\bibitem{CuN_1997_comp}%
  \BibitemOpen
  \bibfield{author}{%
  \bibinfo {author} {\bibfnamefont{H.}~\bibnamefont{Shimizu}}, \bibinfo
  {author} {\bibfnamefont{M.}~\bibnamefont{Shirai}},\ and\ \bibinfo {author}
  {\bibfnamefont{N.}~\bibnamefont{Suzuki}},\ }%
  \bibfield{journal}{%
  \Doi{10.1143/JPSJ.66.3147}{\bibinfo {journal} {Journal of the Physical
  Society of Japan}}\ }%
  \textbf{\bibinfo {volume} {66}},\ \bibinfo {pages} {3147} (\bibinfo {year}
  {1997}),\ \url{http://jpsj.ipap.jp/link?JPSJ/66/3147/}%
  \bibAnnoteFile{NoStop}{CuN_1997_comp}%
\bibitem{PtN_2006_comp_B17_structure_important}%
  \BibitemOpen
  \bibfield{author}{%
  \bibinfo {author} {\bibfnamefont{J.}~\bibnamefont{von Appen}}, \bibinfo
  {author} {\bibfnamefont{M.-W.}\ \bibnamefont{Lumey}},\ and\ \bibinfo {author}
  {\bibfnamefont{R.}~\bibnamefont{Dronskowski}},\ }%
  \bibfield{journal}{%
  \Doi{10.1002/anie.200600431}{\bibinfo {journal} {Angewandte Chemie
  International Edition}}\ }%
  \textbf{\bibinfo {volume} {45}},\ \bibinfo {pages} {4365} (\bibinfo {year}
  {2006}),\ ISSN \bibinfo {issn} {1521-3773},\
  \url{http://dx.doi.org/10.1002/anie.200600431}%
  \bibAnnoteFile{NoStop}{PtN_2006_comp_B17_structure_important}%
\bibitem{PtN2_2005_comp}%
  \BibitemOpen
  \bibfield{author}{%
  \bibinfo {author} {\bibfnamefont{R.}~\bibnamefont{Yu}}, \bibinfo {author}
  {\bibfnamefont{Q.}~\bibnamefont{Zhan}},\ and\ \bibinfo {author}
  {\bibfnamefont{X.~F.}\ \bibnamefont{Zhang}},\ }%
  \bibfield{journal}{%
  \Doi{10.1063/1.2168683}{\bibinfo {journal} {Applied Physics Letters}}\ }%
  \textbf{\bibinfo {volume} {88}},\ \bibinfo {eid} {051913} (\bibinfo {year}
  {2006}),\ \url{http://link.aip.org/link/?APL/88/051913/1}%
  \bibAnnoteFile{NoStop}{PtN2_2005_comp}%
\bibitem{SDFT_1972}%
  \BibitemOpen
  \bibfield{author}{%
  \bibinfo {author} {\bibfnamefont{U.}~\bibnamefont{von Barth}}\ and\ \bibinfo
  {author} {\bibfnamefont{L.}~\bibnamefont{Hedin}},\ }%
  \bibfield{journal}{%
  \Doi{doi:10.1088/0022-3719/5/13/012}{\bibinfo {journal} {Journal of Physics
  C: Solid State Physics}}\ }%
  \textbf{\bibinfo {volume} {5}},\ \bibinfo {pages} {1629} (\bibinfo {month}
  {Feb}\ \bibinfo {year} {1972}),\
  \url{http://iopscience.iop.org/0022-3719/5/13/012/}%
  \bibAnnoteFile{NoStop}{SDFT_1972}%
\bibitem{SDFT_Pant_1972}%
  \BibitemOpen
  \bibfield{author}{%
  \bibinfo {author} {\bibfnamefont{M.}~\bibnamefont{Pant}}\ and\ \bibinfo
  {author} {\bibfnamefont{A.}~\bibnamefont{Rajagopal}},\ }%
  \bibfield{journal}{%
  \Doi{10.1016/0038-1098(72)90934-9}{\bibinfo {journal} {Solid State
  Communications}}\ }%
  \textbf{\bibinfo {volume} {10}},\ \bibinfo {pages} {1157 } (\bibinfo {year}
  {1972}),\ ISSN \bibinfo {issn} {0038-1098},\
  \url{http://www.sciencedirect.com/science/article/pii/0038109872909349}%
  \bibAnnoteFile{NoStop}{SDFT_Pant_1972}%
\bibitem{Vasp_ref_PhysRevB.47.558_1993}%
  \BibitemOpen
  \bibfield{author}{%
  \bibinfo {author} {\bibfnamefont{G.}~\bibnamefont{Kresse}}\ and\ \bibinfo
  {author} {\bibfnamefont{J.}~\bibnamefont{Hafner}},\ }%
  \bibfield{journal}{%
  \Doi{10.1103/PhysRevB.47.558}{\bibinfo {journal} {Physical Review B}}\ }%
  \textbf{\bibinfo {volume} {47}},\ \bibinfo {pages} {558} (\bibinfo {month}
  {Jan}\ \bibinfo {year} {1993}),\
  \url{http://link.aps.org/doi/10.1103/PhysRevB.47.558}%
  \bibAnnoteFile{NoStop}{Vasp_ref_PhysRevB.47.558_1993}%
\bibitem{Vasp_ref_PhysRevB.49.14251_1994}%
  \BibitemOpen
  \bibfield{author}{%
  \bibinfo {author} {\bibfnamefont{G.}~\bibnamefont{Kresse}}\ and\ \bibinfo
  {author} {\bibfnamefont{J.}~\bibnamefont{Hafner}},\ }%
  \bibfield{journal}{%
  \Doi{10.1103/PhysRevB.49.14251}{\bibinfo {journal} {Physical Review B}}\ }%
  \textbf{\bibinfo {volume} {49}},\ \bibinfo {pages} {14251} (\bibinfo {month}
  {May}\ \bibinfo {year} {1994}),\
  \url{http://link.aps.org/doi/10.1103/PhysRevB.49.14251}%
  \bibAnnoteFile{NoStop}{Vasp_ref_PhysRevB.49.14251_1994}%
\bibitem{Vasp_cite_Kressw_1996}%
  \BibitemOpen
  \bibfield{author}{%
  \bibinfo {author} {\bibfnamefont{G.}~\bibnamefont{Kresse}}\ and\ \bibinfo
  {author} {\bibfnamefont{J.}~\bibnamefont{Furthmüller}},\ }%
  \bibfield{journal}{%
  \Doi{10.1016/0927-0256(96)00008-0}{\bibinfo {journal} {Computational
  Materials Science}}\ }%
  \textbf{\bibinfo {volume} {6}},\ \bibinfo {pages} {15 } (\bibinfo {year}
  {1996}),\ ISSN \bibinfo {issn} {0927-0256},\
  \url{http://www.sciencedirect.com/science/article/pii/0927025696000080}%
  \bibAnnoteFile{NoStop}{Vasp_cite_Kressw_1996}%
\bibitem{Vasp_PWs_Kresse_1996}%
  \BibitemOpen
  \bibfield{author}{%
  \bibinfo {author} {\bibfnamefont{G.}~\bibnamefont{Kresse}}\ and\ \bibinfo
  {author} {\bibfnamefont{J.}~\bibnamefont{Furthm\"uller}},\ }%
  \bibfield{journal}{%
  \Doi{10.1103/PhysRevB.54.11169}{\bibinfo {journal} {Physical Review B}}\ }%
  \textbf{\bibinfo {volume} {54}},\ \bibinfo {pages} {11169} (\bibinfo {month}
  {Oct}\ \bibinfo {year} {1996}),\
  \url{http://link.aps.org/doi/10.1103/PhysRevB.54.11169}%
  \bibAnnoteFile{NoStop}{Vasp_PWs_Kresse_1996}%
\bibitem{DFT_VASP_Hafner_2008}%
  \BibitemOpen
  \bibfield{author}{%
  \bibinfo {author} {\bibfnamefont{J.}~\bibnamefont{Hafner}},\ }%
  \bibfield{journal}{%
  \Doi{10.1002/jcc.21057}{\bibinfo {journal} {Journal of Computational
  Chemistry}}\ }%
  \textbf{\bibinfo {volume} {29}},\ \bibinfo {pages} {2044} (\bibinfo {year}
  {2008}),\ ISSN \bibinfo {issn} {1096-987X},\
  \url{http://dx.doi.org/10.1002/jcc.21057}%
  \bibAnnoteFile{NoStop}{DFT_VASP_Hafner_2008}%
\bibitem{PAW_Kresse_n_Joubert}%
  \BibitemOpen
  \bibfield{author}{%
  \bibinfo {author} {\bibfnamefont{G.}~\bibnamefont{Kresse}}\ and\ \bibinfo
  {author} {\bibfnamefont{D.~P.}\ \bibnamefont{Joubert}},\ }%
  \bibfield{journal}{%
  \Doi{10.1103/PhysRevB.59.1758}{\bibinfo {journal} {Physical Review B}}\ }%
  \textbf{\bibinfo {volume} {59}},\ \bibinfo {pages} {1758} (\bibinfo {month}
  {Jan}\ \bibinfo {year} {1999}),\
  \url{http://link.aps.org/doi/10.1103/PhysRevB.59.1758}%
  \bibAnnoteFile{NoStop}{PAW_Kresse_n_Joubert}%
\bibitem{A_birds_eye}%
  \BibitemOpen
  \bibfield{author}{%
  \bibinfo {author} {\bibfnamefont{K.}~\bibnamefont{Capelle}},\ }%
  \bibfield{journal}{%
  \bibinfo {journal} {{Brazilian Journal of Physics}}\ }%
  \textbf{\bibinfo {volume} {36}},\ \bibinfo {pages} {1318 } (\bibinfo {month}
  {12}\ \bibinfo {year} {2006}),\ ISSN \bibinfo {issn} {0103-9733},\
  \url{http://www.scielo.br/scielo.php?script=sci_arttext&pid=S0103-9733200600%
0700035&nrm=iso}%
  \bibAnnoteFile{NoStop}{A_birds_eye}%
\bibitem{HK_1964}%
  \BibitemOpen
  \bibfield{author}{%
  \bibinfo {author} {\bibfnamefont{P.}~\bibnamefont{Hohenberg}}\ and\ \bibinfo
  {author} {\bibfnamefont{W.}~\bibnamefont{Kohn}},\ }%
  \bibfield{journal}{%
  \Doi{10.1103/PhysRev.136.B864}{\bibinfo {journal} {Physical Review}}\ }%
  \textbf{\bibinfo {volume} {136}},\ \bibinfo {pages} {B864} (\bibinfo {month}
  {Nov}\ \bibinfo {year} {1964}),\
  \url{http://link.aps.org/doi/10.1103/PhysRev.136.B864}%
  \bibAnnoteFile{NoStop}{HK_1964}%
\bibitem{KS_1965}%
  \BibitemOpen
  \bibfield{author}{%
  \bibinfo {author} {\bibfnamefont{W.}~\bibnamefont{Kohn}}\ and\ \bibinfo
  {author} {\bibfnamefont{L.~J.}\ \bibnamefont{Sham}},\ }%
  \bibfield{journal}{%
  \Doi{10.1103/PhysRev.140.A1133}{\bibinfo {journal} {Physical Review}}\ }%
  \textbf{\bibinfo {volume} {140}},\ \bibinfo {pages} {A1133} (\bibinfo {month}
  {Nov}\ \bibinfo {year} {1965}),\
  \url{http://link.aps.org/doi/10.1103/PhysRev.140.A1133}%
  \bibAnnoteFile{NoStop}{KS_1965}%
\bibitem{XC_SDFT_1976}%
  \BibitemOpen
  \bibfield{author}{%
  \bibinfo {author} {\bibfnamefont{O.}~\bibnamefont{Gunnarsson}}\ and\ \bibinfo
  {author} {\bibfnamefont{B.~I.}\ \bibnamefont{Lundqvist}},\ }%
  \bibfield{journal}{%
  \Doi{10.1103/PhysRevB.13.4274}{\bibinfo {journal} {Physical Review B}}\ }%
  \textbf{\bibinfo {volume} {13}},\ \bibinfo {pages} {4274} (\bibinfo {month}
  {May}\ \bibinfo {year} {1976}),\
  \url{http://link.aps.org/doi/10.1103/PhysRevB.13.4274}%
  \bibAnnoteFile{NoStop}{XC_SDFT_1976}%
\bibitem{MP_k_mesh_1976}%
  \BibitemOpen
  \bibfield{author}{%
  \bibinfo {author} {\bibfnamefont{H.~J.}\ \bibnamefont{Monkhorst}}\ and\
  \bibinfo {author} {\bibfnamefont{J.~D.}\ \bibnamefont{Pack}},\ }%
  \bibfield{journal}{%
  \Doi{10.1103/PhysRevB.13.5188}{\bibinfo {journal} {Physical Review B}}\ }%
  \textbf{\bibinfo {volume} {13}},\ \bibinfo {pages} {5188} (\bibinfo {month}
  {Jun}\ \bibinfo {year} {1976}),\
  \url{http://link.aps.org/doi/10.1103/PhysRevB.13.5188}%
  \bibAnnoteFile{NoStop}{MP_k_mesh_1976}%
\bibitem{tetrahedron_method_theory_1971}%
  \BibitemOpen
  \bibfield{author}{%
  \bibinfo {author} {\bibfnamefont{O.}~\bibnamefont{Jepson}}\ and\ \bibinfo
  {author} {\bibfnamefont{O.}~\bibnamefont{Anderson}},\ }%
  \bibfield{journal}{%
  \Doi{10.1016/0038-1098(71)90313-9}{\bibinfo {journal} {Solid State
  Communications}}\ }%
  \textbf{\bibinfo {volume} {9}},\ \bibinfo {pages} {1763 } (\bibinfo {year}
  {1971}),\ ISSN \bibinfo {issn} {0038-1098},\
  \url{http://www.sciencedirect.com/science/article/pii/0038109871903139}%
  \bibAnnoteFile{NoStop}{tetrahedron_method_theory_1971}%
\bibitem{tetrahedron_method_theory_1972}%
  \BibitemOpen
  \bibfield{author}{%
  \bibinfo {author} {\bibfnamefont{G.}~\bibnamefont{Lehmann}}\ and\ \bibinfo
  {author} {\bibfnamefont{M.}~\bibnamefont{Taut}},\ }%
  \bibfield{journal}{%
  \Doi{10.1002/pssb.2220540211}{\bibinfo {journal} {physica status solidi
  (b)}}\ }%
  \textbf{\bibinfo {volume} {54}},\ \bibinfo {pages} {469} (\bibinfo {year}
  {1972}),\ ISSN \bibinfo {issn} {1521-3951},\
  \url{http://dx.doi.org/10.1002/pssb.2220540211}%
  \bibAnnoteFile{NoStop}{tetrahedron_method_theory_1972}%
\bibitem{ISMEAR5_1994}%
  \BibitemOpen
  \bibfield{author}{%
  \bibinfo {author} {\bibfnamefont{P.~E.}\ \bibnamefont{Bl\"ochl}}, \bibinfo
  {author} {\bibfnamefont{O.}~\bibnamefont{Jepsen}},\ and\ \bibinfo {author}
  {\bibfnamefont{O.~K.}\ \bibnamefont{Andersen}},\ }%
  \bibfield{journal}{%
  \Doi{10.1103/PhysRevB.49.16223}{\bibinfo {journal} {Physical Review B}}\ }%
  \textbf{\bibinfo {volume} {49}},\ \bibinfo {pages} {16223} (\bibinfo {month}
  {Jun}\ \bibinfo {year} {1994}),\
  \url{http://link.aps.org/doi/10.1103/PhysRevB.49.16223}%
  \bibAnnoteFile{NoStop}{ISMEAR5_1994}%
\bibitem{MP_smearing_1989}%
  \BibitemOpen
  \bibfield{author}{%
  \bibinfo {author} {\bibfnamefont{M.}~\bibnamefont{Methfessel}}\ and\ \bibinfo
  {author} {\bibfnamefont{A.~T.}\ \bibnamefont{Paxton}},\ }%
  \bibfield{journal}{%
  \Doi{10.1103/PhysRevB.40.3616}{\bibinfo {journal} {Physical Review B}}\ }%
  \textbf{\bibinfo {volume} {40}},\ \bibinfo {pages} {3616} (\bibinfo {month}
  {Aug}\ \bibinfo {year} {1989}),\
  \url{http://link.aps.org/doi/10.1103/PhysRevB.40.3616}%
  \bibAnnoteFile{NoStop}{MP_smearing_1989}%
\bibitem{XC_GGA_1988}%
  \BibitemOpen
  \bibfield{author}{%
  \bibinfo {author} {\bibfnamefont{A.~D.}\ \bibnamefont{Becke}},\ }%
  \bibfield{journal}{%
  \Doi{10.1103/PhysRevA.38.3098}{\bibinfo {journal} {Physical Review A}}\ }%
  \textbf{\bibinfo {volume} {38}},\ \bibinfo {pages} {3098} (\bibinfo {month}
  {Sep}\ \bibinfo {year} {1988}),\
  \url{http://link.aps.org/doi/10.1103/PhysRevA.38.3098}%
  \bibAnnoteFile{NoStop}{XC_GGA_1988}%
\bibitem{XC_GGA_applications_1992}%
  \BibitemOpen
  \bibfield{author}{%
  \bibinfo {author} {\bibfnamefont{J.~P.}\ \bibnamefont{Perdew}}, \bibinfo
  {author} {\bibfnamefont{J.~A.}\ \bibnamefont{Chevary}}, \bibinfo {author}
  {\bibfnamefont{S.~H.}\ \bibnamefont{Vosko}}, \bibinfo {author}
  {\bibfnamefont{K.~A.}\ \bibnamefont{Jackson}}, \bibinfo {author}
  {\bibfnamefont{M.~R.}\ \bibnamefont{Pederson}}, \bibinfo {author}
  {\bibfnamefont{D.~J.}\ \bibnamefont{Singh}},\ and\ \bibinfo {author}
  {\bibfnamefont{C.}~\bibnamefont{Fiolhais}},\ }%
  \bibfield{journal}{%
  \Doi{10.1103/PhysRevB.46.6671}{\bibinfo {journal} {Physical Review B}}\ }%
  \textbf{\bibinfo {volume} {46}},\ \bibinfo {pages} {6671} (\bibinfo {month}
  {Sep}\ \bibinfo {year} {1992}),\
  \url{http://link.aps.org/doi/10.1103/PhysRevB.46.6671}%
  \bibAnnoteFile{NoStop}{XC_GGA_applications_1992}%
\bibitem{XC_GGA_applications_1992_ERRATUM}%
  \BibitemOpen
  \bibfield{author}{%
  \bibinfo {author} {\bibfnamefont{J.~P.}\ \bibnamefont{Perdew}}, \bibinfo
  {author} {\bibfnamefont{J.~A.}\ \bibnamefont{Chevary}}, \bibinfo {author}
  {\bibfnamefont{S.~H.}\ \bibnamefont{Vosko}}, \bibinfo {author}
  {\bibfnamefont{K.~A.}\ \bibnamefont{Jackson}}, \bibinfo {author}
  {\bibfnamefont{M.~R.}\ \bibnamefont{Pederson}}, \bibinfo {author}
  {\bibfnamefont{D.~J.}\ \bibnamefont{Singh}},\ and\ \bibinfo {author}
  {\bibfnamefont{C.}~\bibnamefont{Fiolhais}},\ }%
  \bibfield{journal}{%
  \Doi{10.1103/PhysRevB.48.4978.2}{\bibinfo {journal} {Physical Review B}}\ }%
  \textbf{\bibinfo {volume} {48}},\ \bibinfo {pages} {4978} (\bibinfo {month}
  {Aug}\ \bibinfo {year} {1993}),\
  \url{http://link.aps.org/doi/10.1103/PhysRevB.48.4978.2}%
  \bibAnnoteFile{NoStop}{XC_GGA_applications_1992_ERRATUM}%
\bibitem{PBE_GGA_1996}%
  \BibitemOpen
  \bibfield{author}{%
  \bibinfo {author} {\bibfnamefont{J.~P.}\ \bibnamefont{Perdew}}, \bibinfo
  {author} {\bibfnamefont{K.}~\bibnamefont{Burke}},\ and\ \bibinfo {author}
  {\bibfnamefont{M.}~\bibnamefont{Ernzerhof}},\ }%
  \bibfield{journal}{%
  \Doi{10.1103/PhysRevLett.77.3865}{\bibinfo {journal} {Physical Review
  Letters}}\ }%
  \textbf{\bibinfo {volume} {77}},\ \bibinfo {pages} {3865} (\bibinfo {month}
  {Oct}\ \bibinfo {year} {1996}),\
  \url{http://link.aps.org/doi/10.1103/PhysRevLett.77.3865}%
  \bibAnnoteFile{NoStop}{PBE_GGA_1996}%
\bibitem{PBE_GGA_Erratum_1997}%
  \BibitemOpen
  \bibfield{author}{%
  \bibinfo {author} {\bibfnamefont{J.~P.}\ \bibnamefont{Perdew}}, \bibinfo
  {author} {\bibfnamefont{K.}~\bibnamefont{Burke}},\ and\ \bibinfo {author}
  {\bibfnamefont{M.}~\bibnamefont{Ernzerhof}},\ }%
  \bibfield{journal}{%
  \Doi{10.1103/PhysRevLett.78.1396}{\bibinfo {journal} {Physical Review
  Letters}}\ }%
  \textbf{\bibinfo {volume} {78}},\ \bibinfo {pages} {1396} (\bibinfo {month}
  {Feb}\ \bibinfo {year} {1997}),\
  \url{http://link.aps.org/doi/10.1103/PhysRevLett.78.1396}%
  \bibAnnoteFile{NoStop}{PBE_GGA_Erratum_1997}%
\bibitem{XC_PBE_1999}%
  \BibitemOpen
  \bibfield{author}{%
  \bibinfo {author} {\bibfnamefont{M.}~\bibnamefont{Ernzerhof}}\ and\ \bibinfo
  {author} {\bibfnamefont{G.~E.}\ \bibnamefont{Scuseria}},\ }%
  \bibfield{journal}{%
  \Doi{10.1063/1.478401}{\bibinfo {journal} {The Journal of Chemical Physics}}\
  }%
  \textbf{\bibinfo {volume} {110}},\ \bibinfo {pages} {5029 } (\bibinfo {year}
  {1999}),\ \url{http://link.aip.org/link/?JCP/110/5029/1}%
  \bibAnnoteFile{NoStop}{XC_PBE_1999}%
\bibitem{PAW_Blochl}%
  \BibitemOpen
  \bibfield{author}{%
  \bibinfo {author} {\bibfnamefont{P.~E.}\ \bibnamefont{Bl\"ochl}},\ }%
  \bibfield{journal}{%
  \Doi{10.1103/PhysRevB.50.17953}{\bibinfo {journal} {Physical Review B}}\ }%
  \textbf{\bibinfo {volume} {50}},\ \bibinfo {pages} {17953} (\bibinfo {month}
  {Dec}\ \bibinfo {year} {1994}),\
  \url{http://link.aps.org/doi/10.1103/PhysRevB.50.17953}%
  \bibAnnoteFile{NoStop}{PAW_Blochl}%
\bibitem{Davidson_original_article_1975}%
  \BibitemOpen
  \bibfield{author}{%
  \bibinfo {author} {\bibfnamefont{E.~R.}\ \bibnamefont{Davidson}},\ }%
  \bibfield{journal}{%
  \Doi{10.1016/0021-9991(75)90065-0}{\bibinfo {journal} {Journal of
  Computational Physics}}\ }%
  \textbf{\bibinfo {volume} {17}},\ \bibinfo {pages} {87 } (\bibinfo {year}
  {1975}),\ ISSN \bibinfo {issn} {0021-9991},\
  \url{http://www.sciencedirect.com/science/article/pii/0021999175900650}%
  \bibAnnoteFile{NoStop}{Davidson_original_article_1975}%
\bibitem{Born_Oppenheimer_1927}%
  \BibitemOpen
  \bibfield{author}{%
  \bibinfo {author} {\bibfnamefont{M.}~\bibnamefont{Born}}\ and\ \bibinfo
  {author} {\bibfnamefont{J.~R.}\ \bibnamefont{Oppenheimer}},\ }%
  \bibfield{journal}{%
  \bibinfo {journal} {Annalen der Physik}\ }%
  \textbf{\bibinfo {volume} {84}},\ \bibinfo {pages} {457} (\bibinfo {year}
  {1927})%
  \bibAnnoteFile{NoStop}{Born_Oppenheimer_1927}%
\bibitem{Richard_Martin}%
  \BibitemOpen
  \bibfield{author}{%
  \bibinfo {author} {\bibfnamefont{R.~M.}\ \bibnamefont{Martin}},\ }%
  \emph{\bibinfo {title} {Electronic Structure, Basic Theory and Practical
  Methods}}\ (\bibinfo {publisher} {Cambridge University Press},\ \bibinfo
  {year} {2004})\ \bibinfo {note} {c.f. Appendix L: Numerical Methods}%
  \bibAnnoteFile{NoStop}{Richard_Martin}%
\bibitem{Hellmann–Feynman_theorem}%
  \BibitemOpen
  \bibfield{author}{%
  \bibinfo {author} {\bibfnamefont{R.~P.}\ \bibnamefont{Feynman}},\ }%
  \bibfield{journal}{%
  \Doi{10.1103/PhysRev.56.340}{\bibinfo {journal} {Physical Review}}\ }%
  \textbf{\bibinfo {volume} {56}},\ \bibinfo {pages} {340} (\bibinfo {month}
  {Aug}\ \bibinfo {year} {1939}),\
  \url{http://link.aps.org/doi/10.1103/PhysRev.56.340}%
  \bibAnnoteFile{NoStop}{Hellmann–Feynman_theorem}%
\bibitem{Grimvall}%
  \BibitemOpen
  \bibfield{author}{%
  \bibinfo {author} {\bibfnamefont{G.}~\bibnamefont{Grimvall}},\ }%
  \emph{\bibinfo {title} {Thermophysical Properties of Materials}}\ (\bibinfo
  {publisher} {North Holland},\ \bibinfo {year} {1986})%
  \bibAnnoteFile{NoStop}{Grimvall}%
\bibitem{Note2}%
  \BibitemOpen
  \bibinfo {note} {Eq. \ref {E_coh equation} results in a negative $E_{coh}$.
  However, another convention with positive $E_{coh}$ is also common, where
  energy signs in Eq. \ref {E_coh equation} change.}%
  \bibAnnoteFile{Stop}{Note2}%
\bibitem{interatomic_potentials}%
  \BibitemOpen
  \bibfield{author}{%
  \bibinfo {author} {\bibfnamefont{Y.}~\bibnamefont{Mishin}}, \bibinfo {author}
  {\bibfnamefont{D.}~\bibnamefont{Farkas}}, \bibinfo {author}
  {\bibfnamefont{M.~J.}\ \bibnamefont{Mehl}},\ and\ \bibinfo {author}
  {\bibfnamefont{D.~A.}\ \bibnamefont{Papaconstantopoulos}},\ }%
  \bibfield{journal}{%
  \Doi{10.1103/PhysRevB.59.3393}{\bibinfo {journal} {Physical Review B}}\ }%
  \textbf{\bibinfo {volume} {59}},\ \bibinfo {pages} {3393} (\bibinfo {month}
  {Feb}\ \bibinfo {year} {1999}),\
  \url{http://link.aps.org/doi/10.1103/PhysRevB.59.3393}%
  \bibAnnoteFile{NoStop}{interatomic_potentials}%
\bibitem{VASPguide}%
  \BibitemOpen
  \bibfield{author}{%
  \bibinfo {author} {\bibfnamefont{G.}~\bibnamefont{Kresse}}, \bibinfo {author}
  {\bibfnamefont{M.}~\bibnamefont{Marsman}},\ and\ \bibinfo {author}
  {\bibfnamefont{J.}~\bibnamefont{Furthmuller}},\ }%
  \enquote{\bibinfo {title} {Vasp the guide},}\  (\bibinfo {year} {2011}),\
  \bibinfo {note} {available on-line at
  \url{http://cms.mpi.univie.ac.at/vasp/vasp/}. Last accessed October 2012.}%
  \bibAnnoteFile{Stop}{VASPguide}%
\bibitem{B_prime_Aug_2012_comp}%
  \BibitemOpen
  \bibfield{author}{%
  \bibinfo {author} {\bibfnamefont{K.}~\bibnamefont{Lejaeghere}}, \bibinfo
  {author} {\bibfnamefont{V.~V.}\ \bibnamefont{Speybroeck}}, \bibinfo {author}
  {\bibfnamefont{G.~V.}\ \bibnamefont{Oost}},\ and\ \bibinfo {author}
  {\bibfnamefont{S.}~\bibnamefont{Cottenier}},\ }%
  \enquote{\bibinfo {title} {Error bars for solid-state density-functional
  theory predictions: an overview by means of the ground-state elemental
  crystals},}\  (\bibinfo {month} {August}\ \bibinfo {year} {2012}),\
  \Eprint{http://arxiv.org/abs/1204.2733v2}{arXiv:1204.2733v2
  [cond-mat.mtrl-sci]}%
  \bibAnnoteFile{NoStop}{B_prime_Aug_2012_comp}%
\bibitem{Note3}%
  \BibitemOpen
  \bibinfo {note} {It is also well known that GGA may slightly lower the
  ground-state energy when a nonspherical ground-state density is allowed for
  (cf. Ref. [\protect \rev@citealpnum {PhysRevB.65.245212}] and Ref. 46
  therein).}%
  \bibAnnoteFile{Stop}{Note3}%
\bibitem{Ashcroft1976}%
  \BibitemOpen
  \bibfield{author}{%
  \bibinfo {author} {\bibfnamefont{N.~W.}\ \bibnamefont{Ashcroft}}\ and\
  \bibinfo {author} {\bibfnamefont{N.~D.}\ \bibnamefont{Mermin}},\ }%
  \emph{\bibinfo {title} {Solid State Physics}}\ (\bibinfo {publisher}
  {Philadelphia: Saunders College},\ \bibinfo {year} {1976})%
  \bibAnnoteFile{NoStop}{Ashcroft1976}%
\bibitem{eos_f90_code}%
  \BibitemOpen
  \bibfield{author}{%
  \bibinfo {author} {\bibfnamefont{J.~K.}\ \bibnamefont{Dewhurst}},\ }%
  \enquote{\bibinfo {title} {{EOS} version 1.2},}\  (\bibinfo {month} {August}\
  \bibinfo {year} {2005}),\ \bibinfo {note} {{E}quation of {S}tate (EOS)
  program for fitting energy-volume data {(http://elk.sourceforge.net,
  http://exciting.sourceforge.net)}.}%
  \bibAnnoteFile{Stop}{eos_f90_code}%
\bibitem{BM_3rd_eos}%
  \BibitemOpen
  \bibfield{author}{%
  \bibinfo {author} {\bibfnamefont{F.}~\bibnamefont{Birch}},\ }%
  \bibfield{journal}{%
  \Doi{10.1103/PhysRev.71.809}{\bibinfo {journal} {Physical Review}}\ }%
  \textbf{\bibinfo {volume} {71}},\ \bibinfo {pages} {809} (\bibinfo {month}
  {Jun}\ \bibinfo {year} {1947}),\
  \url{http://link.aps.org/doi/10.1103/PhysRev.71.809}%
  \bibAnnoteFile{NoStop}{BM_3rd_eos}%
\bibitem{David_Young_Phase_Diagrams_1991}%
  \BibitemOpen
  \bibfield{author}{%
  \bibinfo {author} {\bibfnamefont{D.~A.}\ \bibnamefont{Young}},\ }%
  \emph{\bibinfo {title} {Phase Diagrams of the Elements}}\ (\bibinfo
  {publisher} {University of California Press},\ \bibinfo {year} {1991})\ ISBN
  \bibinfo {isbn} {0520074831},\
  \url{http://www.amazon.com/Phase-Diagrams-Elements-David-Young/dp/0520074831%
}%
  \bibAnnoteFile{NoStop}{David_Young_Phase_Diagrams_1991}%
\bibitem{CRC_Handbook_82ed}%
  \BibitemOpen
  \emph{\bibinfo {title} {CRC Handbook of Chemistry and Physics, 82nd
  Edition}}\ (\bibinfo {publisher} {CRC Press},\ \bibinfo {year} {2001})\ ISBN
  \bibinfo {isbn} {0849304822}%
  \bibAnnoteFile{NoStop}{CRC_Handbook_82ed}%
\bibitem{Jerry_1974}%
  \BibitemOpen
  \bibfield{author}{%
  \bibinfo {author} {\bibfnamefont{J.}~\bibnamefont{Donohue}},\ }%
  \emph{\bibinfo {title} {The Structures of the Elements}}\ (\bibinfo
  {publisher} {John Wiley \& Sons Inc},\ \bibinfo {year} {1974})\ ISBN \bibinfo
  {isbn} {0471217883},\
  \url{http://www.amazon.com/Structures-Elements-Jerry-Donohue/dp/0471217883}%
  \bibAnnoteFile{NoStop}{Jerry_1974}%
\bibitem{elemental_metals_1996_comp}%
  \BibitemOpen
  \bibfield{author}{%
  \bibinfo {author} {\bibfnamefont{M.~J.}\ \bibnamefont{Mehl}}\ and\ \bibinfo
  {author} {\bibfnamefont{D.~A.}\ \bibnamefont{Papaconstantopoulos}},\ }%
  \bibfield{journal}{%
  \Doi{10.1103/PhysRevB.54.4519}{\bibinfo {journal} {Phys. Rev. B}}\ }%
  \textbf{\bibinfo {volume} {54}},\ \bibinfo {pages} {4519} (\bibinfo {month}
  {Aug}\ \bibinfo {year} {1996}),\
  \url{http://link.aps.org/doi/10.1103/PhysRevB.54.4519}%
  \bibAnnoteFile{NoStop}{elemental_metals_1996_comp}%
\bibitem{Wyckoff}%
  \BibitemOpen
  \bibfield{author}{%
  \bibinfo {author} {\bibfnamefont{R.~W.~G.}\ \bibnamefont{Wyckoff}},\ }%
  \emph{\bibinfo {title} {The Structure of Crystals}}\ (\bibinfo {publisher}
  {The Chemical Catalog Co., New York},\ \bibinfo {year} {1935})%
  \bibAnnoteFile{NoStop}{Wyckoff}%
\bibitem{PhysRevB.45.5777}%
  \BibitemOpen
  \bibfield{author}{%
  \bibinfo {author} {\bibfnamefont{M.}~\bibnamefont{Sigalas}}, \bibinfo
  {author} {\bibfnamefont{D.~A.}\ \bibnamefont{Papaconstantopoulos}},\ and\
  \bibinfo {author} {\bibfnamefont{N.~C.}\ \bibnamefont{Bacalis}},\ }%
  \bibfield{journal}{%
  \Doi{10.1103/PhysRevB.45.5777}{\bibinfo {journal} {Phys. Rev. B}}\ }%
  \textbf{\bibinfo {volume} {45}},\ \bibinfo {pages} {5777} (\bibinfo {month}
  {Mar}\ \bibinfo {year} {1992}),\
  \url{http://link.aps.org/doi/10.1103/PhysRevB.45.5777}%
  \bibAnnoteFile{NoStop}{PhysRevB.45.5777}%
\bibitem{Kittel1996}%
  \BibitemOpen
  \bibfield{author}{%
  \bibinfo {author} {\bibfnamefont{C.}~\bibnamefont{Kittel}},\ }%
  \emph{\bibinfo {title} {Introduction to Solid State Physics, 5th edition,}}\
  (\bibinfo {publisher} {John Wiley \& Sons, New York},\ \bibinfo {year}
  {1996})%
  \bibAnnoteFile{NoStop}{Kittel1996}%
\bibitem{cohesive_energy_DFT}%
  \BibitemOpen
  \bibfield{author}{%
  \bibinfo {author} {\bibfnamefont{P.~H.~T.}\ \bibnamefont{Philipsen}}\ and\
  \bibinfo {author} {\bibfnamefont{E.~J.}\ \bibnamefont{Baerends}},\ }%
  \bibfield{journal}{%
  \Doi{10.1103/PhysRevB.54.5326}{\bibinfo {journal} {Phys. Rev. B}}\ }%
  \textbf{\bibinfo {volume} {54}},\ \bibinfo {pages} {5326} (\bibinfo {month}
  {Aug}\ \bibinfo {year} {1996}),\
  \url{http://link.aps.org/doi/10.1103/PhysRevB.54.5326}%
  \bibAnnoteFile{NoStop}{cohesive_energy_DFT}%
\bibitem{elemental_metals_2008_comp}%
  \BibitemOpen
  \bibfield{author}{%
  \bibinfo {author} {\bibfnamefont{E.}~\bibnamefont{Zarechnaya}}, \bibinfo
  {author} {\bibfnamefont{N.}~\bibnamefont{Skorodumova}}, \bibinfo {author}
  {\bibfnamefont{S.}~\bibnamefont{Simak}}, \bibinfo {author}
  {\bibfnamefont{B.}~\bibnamefont{Johansson}},\ and\ \bibinfo {author}
  {\bibfnamefont{E.}~\bibnamefont{Isaev}},\ }%
  \bibfield{journal}{%
  \Doi{10.1016/j.commatsci.2007.12.018}{\bibinfo {journal} {Computational
  Materials Science}}\ }%
  \textbf{\bibinfo {volume} {43}},\ \bibinfo {pages} {522 } (\bibinfo {year}
  {2008}),\ ISSN \bibinfo {issn} {0927-0256},\
  \url{http://www.sciencedirect.com/science/article/pii/S0927025608000037}%
  \bibAnnoteFile{NoStop}{elemental_metals_2008_comp}%
\bibitem{B_prime_1997_theory_comp_n_exp}%
  \BibitemOpen
  \bibfield{author}{%
  \bibinfo {author} {\bibfnamefont{S.}~\bibnamefont{Raju}}, \bibinfo {author}
  {\bibfnamefont{E.}~\bibnamefont{Mohandas}},\ and\ \bibinfo {author}
  {\bibfnamefont{V.}~\bibnamefont{Raghunathan}},\ }%
  \bibfield{journal}{%
  \bibinfo {journal} {J. Phys. Chem Solids}\ }%
  \textbf{\bibinfo {volume} {58}},\ \bibinfo {pages} {1367} (\bibinfo {year}
  {1997})%
  \bibAnnoteFile{NoStop}{B_prime_1997_theory_comp_n_exp}%
\bibitem{Cu3MN_2007_comp}%
  \BibitemOpen
  \bibfield{author}{%
  \bibinfo {author} {\bibfnamefont{M.~G.}\ \bibnamefont{Moreno-Armenta}},
  \bibinfo {author} {\bibfnamefont{W.~L.}\ \bibnamefont{Pérez}},\ and\
  \bibinfo {author} {\bibfnamefont{N.}~\bibnamefont{Takeuchi}},\ }%
  \bibfield{journal}{%
  \Doi{10.1016/j.solidstatesciences.2006.12.002}{\bibinfo {journal} {Solid
  State Sciences}}\ }%
  \textbf{\bibinfo {volume} {9}},\ \bibinfo {pages} {166} (\bibinfo {year}
  {2007}),\ ISSN \bibinfo {issn} {1293-2558},\
  \url{http://www.sciencedirect.com/science/article/pii/S1293255806002858}%
  \bibAnnoteFile{NoStop}{Cu3MN_2007_comp}%
\bibitem{optical_properties_Cu3N_2011_exp}%
  \BibitemOpen
  \bibfield{author}{%
  \bibinfo {author} {\bibfnamefont{J.}~\bibnamefont{Xiao}}, \bibinfo {author}
  {\bibfnamefont{Y.}~\bibnamefont{Li}},\ and\ \bibinfo {author}
  {\bibfnamefont{A.}~\bibnamefont{Jiang}},\ }%
  \bibfield{journal}{%
  \Doi{10.1016/S1005-0302(11)60082-0}{\bibinfo {journal} {Journal of Materials
  Science \& Technology}}\ }%
  \textbf{\bibinfo {volume} {27}},\ \bibinfo {pages} {403 } (\bibinfo {year}
  {2011}),\ ISSN \bibinfo {issn} {1005-0302},\
  \url{http://www.sciencedirect.com/science/article/pii/S1005030211600820}%
  \bibAnnoteFile{NoStop}{optical_properties_Cu3N_2011_exp}%
\bibitem{Cu3N_2005_comp}%
  \BibitemOpen
  \bibfield{author}{%
  \bibinfo {author} {\bibfnamefont{W.}~\bibnamefont{Yu}}, \bibinfo {author}
  {\bibfnamefont{L.}~\bibnamefont{Li}},\ and\ \bibinfo {author}
  {\bibfnamefont{C.}~\bibnamefont{Jin}},\ }%
  \bibfield{journal}{%
  \bibinfo {journal} {Journal of Materials Science}\ }%
  \textbf{\bibinfo {volume} {40}},\ \bibinfo {pages} {4661 } (\bibinfo {year}
  {2005}),\ ISSN \bibinfo {issn} {0022-2461},\ \bibinfo {note}
  {10.1007/s10853-005-0638-3},\
  \url{http://dx.doi.org/10.1007/s10853-005-0638-3}%
  \bibAnnoteFile{NoStop}{Cu3N_2005_comp}%
\bibitem{Cu3N_Cu4N_2011_comp}%
  \BibitemOpen
  \bibfield{author}{%
  \bibinfo {author} {\bibfnamefont{J.~G.}\ \bibnamefont{Niu}}, \bibinfo
  {author} {\bibfnamefont{W.}~\bibnamefont{Gao}}, \bibinfo {author}
  {\bibfnamefont{X.~P.}\ \bibnamefont{Dong}}, \bibinfo {author}
  {\bibfnamefont{L.}~\bibnamefont{Guan}},\ and\ \bibinfo {author}
  {\bibfnamefont{F.}~\bibnamefont{Xie}},\ }%
  \bibfield{journal}{%
  \Doi{10.4028/www.scientific.net/AMR.150-151.1290}{\bibinfo {journal}
  {Advanced Materials Research {(}Volumes 150--151{)}}}\ }%
  \textbf{\bibinfo {volume} {Advances in Composites}},\ \bibinfo {pages} {1290}
  (\bibinfo {month} {October}\ \bibinfo {year} {2010}),\
  \url{www.scientific.net}%
  \bibAnnoteFile{NoStop}{Cu3N_Cu4N_2011_comp}%
\bibitem{Cu3N_1995_exp}%
  \BibitemOpen
  \bibfield{author}{%
  \bibinfo {author} {\bibfnamefont{T.}~\bibnamefont{Maruyama}}\ and\ \bibinfo
  {author} {\bibfnamefont{T.}~\bibnamefont{Morishita}},\ }%
  \bibfield{journal}{%
  \Doi{10.1063/1.359868}{\bibinfo {journal} {Journal of Applied Physics}}\ }%
  \textbf{\bibinfo {volume} {78}},\ \bibinfo {pages} {4104} (\bibinfo {year}
  {1995}),\ \url{http://dx.doi.org/10.1063/1.359868}%
  \bibAnnoteFile{NoStop}{Cu3N_1995_exp}%
\bibitem{Cu3N_Ni3N_1993_exp}%
  \BibitemOpen
  \bibfield{author}{%
  \bibinfo {author} {\bibfnamefont{L.}~\bibnamefont{Maya}},\ }%
  \bibfield{journal}{%
  \Doi{http://dx.doi.org/10.1116/1.578778}{\bibinfo {journal} {Journal of
  Vacuum Science \& Technology A}}\ }%
  \textbf{\bibinfo {volume} {11}},\ \bibinfo {pages} {604} (\bibinfo {month}
  {May}\ \bibinfo {year} {1993}),\ \url{http://dx.doi.org/10.1116/1.578778}%
  \bibAnnoteFile{NoStop}{Cu3N_Ni3N_1993_exp}%
\bibitem{CuN_CuN2_2011_comp}%
  \BibitemOpen
  \bibfield{author}{%
  \bibinfo {author} {\bibfnamefont{N.}~\bibnamefont{Kanoun-Bouayed}}, \bibinfo
  {author} {\bibfnamefont{M.}~\bibnamefont{BenaliKanoun}},\ and\ \bibinfo
  {author} {\bibfnamefont{S.}~\bibnamefont{Goumri-Said}},\ }%
  \bibfield{journal}{%
  \bibinfo {journal} {Central European Journal of Physics}\ }%
  \textbf{\bibinfo {volume} {9}},\ \bibinfo {pages} {205} (\bibinfo {year}
  {2011}),\ ISSN \bibinfo {issn} {1895-1082},\ \bibinfo {note}
  {10.2478/s11534-010-0063-3},\
  \url{http://dx.doi.org/10.2478/s11534-010-0063-3}%
  \bibAnnoteFile{NoStop}{CuN_CuN2_2011_comp}%
\bibitem{CuN_NiN_2004_comp}%
  \BibitemOpen
  \bibfield{author}{%
  \bibinfo {author} {\bibfnamefont{W.}~\bibnamefont{Hong-Bo}}\ and\ \bibinfo
  {author} {\bibfnamefont{X.}~\bibnamefont{De-Sheng}},\ }%
  \bibfield{journal}{%
  \Doi{doi:10.1088/0256-307X/21/8/056}{\bibinfo {journal} {Chinese Physics
  Letters}}\ }%
  \textbf{\bibinfo {volume} {21}},\ \bibinfo {pages} {1612} (\bibinfo {year}
  {2004}),\ \url{http://iopscience.iop.org/0256-307X/21/8/056}%
  \bibAnnoteFile{NoStop}{CuN_NiN_2004_comp}%
\bibitem{Note4}%
  \BibitemOpen
  \bibinfo {note} {In Table \ref
  {copper_nitrides_equilibrium_structural_properties}, our computed properties
  of the elemental Cu are compared with experiment and with previous
  calculations as well. This may benchmark the accuracy of the rest of our
  calculations.}%
  \bibAnnoteFile{Stop}{Note4}%
\bibitem{Note5}%
  \BibitemOpen
  \bibinfo {note} {The electronic structure of D0$_9$, D0$_2$ and RhF$_3$
  phases are discussed in Sec. \ref {Electronic Properties}, while the rest are
  not shown here.}%
  \bibAnnoteFile{Stop}{Note5}%
\bibitem{Note6}%
  \BibitemOpen
  \bibinfo {note} {In fact, the accuracy of the approximate $XC$ functional
  (PBE and others) does not really allow us to make a distinction among
  these.}%
  \bibAnnoteFile{Stop}{Note6}%
\bibitem{D0_9_and_D0_2_structures}%
  \BibitemOpen
  \bibfield{author}{%
  \bibinfo {author} {\bibfnamefont{M.}~\bibnamefont{Llunell}}, \bibinfo
  {author} {\bibfnamefont{S.}~\bibnamefont{Alvarez}},\ and\ \bibinfo {author}
  {\bibfnamefont{P.}~\bibnamefont{Alemany}},\ }%
  \bibfield{journal}{%
  \Doi{10.1039/A708200B}{\bibinfo {journal} {Journal of the Chemical Society,
  Dalton Transactions}},\ \bibinfo {pages} {1195}}%
   (\bibinfo {year} {1998}),\ \url{http://dx.doi.org/10.1039/A708200B}%
  \bibAnnoteFile{NoStop}{D0_9_and_D0_2_structures}%
\bibitem{Ladd_1998}%
  \BibitemOpen
  \bibfield{author}{%
  \bibinfo {author} {\bibfnamefont{M.~F.~C.}\ \bibnamefont{Ladd}},\ }%
  \emph{\bibinfo {title} {Introduction to Physical Chemistry}}\ (\bibinfo
  {publisher} {Cambridge University Press},\ \bibinfo {year} {1998})\ ISBN
  \bibinfo {isbn} {0521578817},\
  \url{http://www.amazon.com/Introduction-Physical-Chemistry-Mark-Ladd/dp/0521%
578817}%
  \bibAnnoteFile{NoStop}{Ladd_1998}%
\bibitem{stability_killing_2008}%
  \BibitemOpen
  \bibfield{author}{%
  \bibinfo {author} {\bibfnamefont{N.}~\bibnamefont{Wang}}, \bibinfo {author}
  {\bibfnamefont{W.-Y.}\ \bibnamefont{Yu}}, \bibinfo {author}
  {\bibfnamefont{B.-Y.}\ \bibnamefont{Tang}}, \bibinfo {author}
  {\bibfnamefont{L.-M.}\ \bibnamefont{Peng}},\ and\ \bibinfo {author}
  {\bibfnamefont{W.-J.}\ \bibnamefont{Ding}},\ }%
  \bibfield{journal}{%
  \bibinfo {journal} {Journal of Physics D: Applied Physics}\ }%
  \textbf{\bibinfo {volume} {41}},\ \bibinfo {pages} {195408} (\bibinfo {year}
  {2008}),\ \url{http://stacks.iop.org/0022-3727/41/i=19/a=195408}%
  \bibAnnoteFile{NoStop}{stability_killing_2008}%
\bibitem{PhysRevB.67.064108}%
  \BibitemOpen
  \bibfield{author}{%
  \bibinfo {author} {\bibfnamefont{C.}~\bibnamefont{Stampfl}}\ and\ \bibinfo
  {author} {\bibfnamefont{A.~J.}\ \bibnamefont{Freeman}},\ }%
  \bibfield{journal}{%
  \Doi{10.1103/PhysRevB.67.064108}{\bibinfo {journal} {Physical Review B}}\ }%
  \textbf{\bibinfo {volume} {67}},\ \bibinfo {pages} {064108} (\bibinfo {month}
  {Feb}\ \bibinfo {year} {2003}),\
  \url{http://link.aps.org/doi/10.1103/PhysRevB.67.064108}%
  \bibAnnoteFile{NoStop}{PhysRevB.67.064108}%
\bibitem{Atkin}%
  \BibitemOpen
  \bibfield{author}{%
  \bibinfo {author} {\bibfnamefont{P.}~\bibnamefont{Atkins}}\ and\ \bibinfo
  {author} {\bibfnamefont{J.~D.}\ \bibnamefont{Paula}},\ }%
  \emph{\bibinfo {title} {Atkins' Physical Chemistry}}\ (\bibinfo {publisher}
  {Oxford University Press},\ \bibinfo {year} {2009})\ ISBN \bibinfo {isbn}
  {0199543372},\
  \url{http://www.amazon.com/Atkins-Physical-Chemistry-Julio-Peter/dp/01995433%
72}%
  \bibAnnoteFile{NoStop}{Atkin}%
\bibitem{enthalpies_of_formation_2003}%
  \BibitemOpen
  \bibfield{author}{%
  \bibinfo {author} {\bibfnamefont{C.}~\bibnamefont{Colinet}},\ }%
  \bibfield{journal}{%
  \Doi{10.1016/S0966-9795(03)00147-X}{\bibinfo {journal} {Intermetallics}}\ }%
  \textbf{\bibinfo {volume} {11}},\ \bibinfo {pages} {1095 } (\bibinfo {year}
  {2003}),\ ISSN \bibinfo {issn} {0966-9795},\ \bibinfo {note} {special Issue
  Dedicated to the International Meeting on Thermodynamics of Alloys -- TOFA
  2002},\
  \url{http://www.sciencedirect.com/science/article/pii/S096697950300147X}%
  \bibAnnoteFile{NoStop}{enthalpies_of_formation_2003}%
\bibitem{PhysRevB.63.165116}%
  \BibitemOpen
  \bibfield{author}{%
  \bibinfo {author} {\bibfnamefont{H.~W.}\ \bibnamefont{Hugosson}}, \bibinfo
  {author} {\bibfnamefont{P.}~\bibnamefont{Korzhavyi}}, \bibinfo {author}
  {\bibfnamefont{U.}~\bibnamefont{Jansson}}, \bibinfo {author}
  {\bibfnamefont{B.}~\bibnamefont{Johansson}},\ and\ \bibinfo {author}
  {\bibfnamefont{O.}~\bibnamefont{Eriksson}},\ }%
  \bibfield{journal}{%
  \Doi{10.1103/PhysRevB.63.165116}{\bibinfo {journal} {Physical Review B}}\ }%
  \textbf{\bibinfo {volume} {63}},\ \bibinfo {pages} {165116} (\bibinfo {month}
  {Apr}\ \bibinfo {year} {2001}),\
  \url{http://link.aps.org/doi/10.1103/PhysRevB.63.165116}%
  \bibAnnoteFile{NoStop}{PhysRevB.63.165116}%
\bibitem{Note7}%
  \BibitemOpen
  \bibinfo {note} {If $E_\protect \text {coh}$ is used with a positive sign
  convention, i.e. negative of Eqs. \ref {general_E_coh equation} and \ref
  {E_coh equation}, then signs in Eq. \ref {formation energy equation} must be
  reversed.}%
  \bibAnnoteFile{Stop}{Note7}%
\bibitem{Structure_of_Materials}%
  \BibitemOpen
  \bibfield{author}{%
  \bibinfo {author} {\bibfnamefont{M.~D.}\ \bibnamefont{Graef}}\ and\ \bibinfo
  {author} {\bibfnamefont{M.~E.}\ \bibnamefont{McHenry}},\ }%
  \emph{\bibinfo {title} {Structure of materials : An Introduction to
  Crystallography, Diffraction and Symmetry}}\ (\bibinfo {publisher} {Cambridge
  University Press},\ \bibinfo {year} {2007})\
  \url{http://www.cambridge.org/9780521651516}%
  \bibAnnoteFile{NoStop}{Structure_of_Materials}%
\bibitem{Handbook_of_Mineralogy}%
  \BibitemOpen
  \emph{\bibinfo {title} {Handbook of Mineralogy}},\ edited by\ \bibinfo
  {editor} {\bibfnamefont{J.~W.}\ \bibnamefont{Anthony}}, \bibinfo {editor}
  {\bibfnamefont{R.~A.}\ \bibnamefont{Bideaux}}, \bibinfo {editor}
  {\bibfnamefont{K.~W.}\ \bibnamefont{Bladh}},\ and\ \bibinfo {editor}
  {\bibfnamefont{M.~C.}\ \bibnamefont{Nichols}}\ (\bibinfo {publisher}
  {Mineralogical Society of America, Chantilly, VA 20151-1110, USA})\ \bibinfo
  {note} {available on-line at \url{http://www.handbookofmineralogy.org/}}%
  \bibAnnoteFile{NoStop}{Handbook_of_Mineralogy}%
\bibitem{PhysRevB.65.245212}%
  \BibitemOpen
  \bibfield{author}{%
  \bibinfo {author} {\bibfnamefont{M.}~\bibnamefont{Fuchs}}, \bibinfo {author}
  {\bibfnamefont{J.~L.~F.}\ \bibnamefont{Da~Silva}}, \bibinfo {author}
  {\bibfnamefont{C.}~\bibnamefont{Stampfl}}, \bibinfo {author}
  {\bibfnamefont{J.}~\bibnamefont{Neugebauer}},\ and\ \bibinfo {author}
  {\bibfnamefont{M.}~\bibnamefont{Scheffler}},\ }%
  \bibfield{journal}{%
  \Doi{10.1103/PhysRevB.65.245212}{\bibinfo {journal} {Phys. Rev. B}}\ }%
  \textbf{\bibinfo {volume} {65}},\ \bibinfo {pages} {245212} (\bibinfo {month}
  {Jun}\ \bibinfo {year} {2002}),\
  \url{http://link.aps.org/doi/10.1103/PhysRevB.65.245212}%
  \bibAnnoteFile{NoStop}{PhysRevB.65.245212}%
\bibitem{Table_of_interatomic_distances_1958}%
  \BibitemOpen
  \emph{\bibinfo {title} {Table of interatomic distances and configuration in
  molecules and ions}},\ edited by\ \bibinfo {editor} {\bibfnamefont{L.~E.}\
  \bibnamefont{Sutton}}\ (\bibinfo {publisher} {The Chemical Society, London,
  UK},\ \bibinfo {year} {1958})%
  \bibAnnoteFile{NoStop}{Table_of_interatomic_distances_1958}%
\bibitem{positive_Ef_n_solid_N2_2007_comp_Scandolo}%
  \BibitemOpen
  \bibfield{author}{%
  \bibinfo {author} {\bibfnamefont{J.~A.}\ \bibnamefont{Montoya}}, \bibinfo
  {author} {\bibfnamefont{A.~D.}\ \bibnamefont{Hernandez}}, \bibinfo {author}
  {\bibfnamefont{C.}~\bibnamefont{Sanloup}}, \bibinfo {author}
  {\bibfnamefont{E.}~\bibnamefont{Gregoryanz}},\ and\ \bibinfo {author}
  {\bibfnamefont{S.}~\bibnamefont{Scandolo}},\ }%
  \bibfield{journal}{%
  \Doi{10.1063/1.2430631}{\bibinfo {journal} {Applied Physics Letters}}\ }%
  \textbf{\bibinfo {volume} {90}},\ \bibinfo {eid} {011909} (\bibinfo {year}
  {2007}),\ \url{http://link.aip.org/link/?APL/90/011909/1}%
  \bibAnnoteFile{NoStop}{positive_Ef_n_solid_N2_2007_comp_Scandolo}%
\bibitem{positive_Ef_n_PtN2_2006_comp_Scandolo}%
  \BibitemOpen
  \bibfield{author}{%
  \bibinfo {author} {\bibfnamefont{A.~F.}\ \bibnamefont{Young}}, \bibinfo
  {author} {\bibfnamefont{J.~A.}\ \bibnamefont{Montoya}}, \bibinfo {author}
  {\bibfnamefont{C.}~\bibnamefont{Sanloup}}, \bibinfo {author}
  {\bibfnamefont{M.}~\bibnamefont{Lazzeri}}, \bibinfo {author}
  {\bibfnamefont{E.}~\bibnamefont{Gregoryanz}},\ and\ \bibinfo {author}
  {\bibfnamefont{S.}~\bibnamefont{Scandolo}},\ }%
  \bibfield{journal}{%
  \Doi{10.1103/PhysRevB.73.153102}{\bibinfo {journal} {Phys. Rev. B}}\ }%
  \textbf{\bibinfo {volume} {73}},\ \bibinfo {pages} {153102} (\bibinfo {month}
  {Apr}\ \bibinfo {year} {2006}),\
  \url{http://link.aps.org/doi/10.1103/PhysRevB.73.153102}%
  \bibAnnoteFile{NoStop}{positive_Ef_n_PtN2_2006_comp_Scandolo}%
\bibitem{PhysRevB.59.5521}%
  \BibitemOpen
  \bibfield{author}{%
  \bibinfo {author} {\bibfnamefont{C.}~\bibnamefont{Stampfl}}\ and\ \bibinfo
  {author} {\bibfnamefont{C.~G.}\ \bibnamefont{Van~de Walle}},\ }%
  \bibfield{journal}{%
  \Doi{10.1103/PhysRevB.59.5521}{\bibinfo {journal} {Phys. Rev. B}}\ }%
  \textbf{\bibinfo {volume} {59}},\ \bibinfo {pages} {5521} (\bibinfo {month}
  {Feb}\ \bibinfo {year} {1999}),\
  \url{http://link.aps.org/doi/10.1103/PhysRevB.59.5521}%
  \bibAnnoteFile{NoStop}{PhysRevB.59.5521}%
\bibitem{PhysRevB.65.064302}%
  \BibitemOpen
  \bibfield{author}{%
  \bibinfo {author} {\bibfnamefont{S.}~\bibnamefont{Narasimhan}}\ and\ \bibinfo
  {author} {\bibfnamefont{S.}~\bibnamefont{de~Gironcoli}},\ }%
  \bibfield{journal}{%
  \Doi{10.1103/PhysRevB.65.064302}{\bibinfo {journal} {Phys. Rev. B}}\ }%
  \textbf{\bibinfo {volume} {65}},\ \bibinfo {pages} {064302} (\bibinfo {month}
  {Jan}\ \bibinfo {year} {2002}),\
  \url{http://link.aps.org/doi/10.1103/PhysRevB.65.064302}%
  \bibAnnoteFile{NoStop}{PhysRevB.65.064302}%
\bibitem{PhysRevB.76.024309}%
  \BibitemOpen
  \bibfield{author}{%
  \bibinfo {author} {\bibfnamefont{B.}~\bibnamefont{Grabowski}}, \bibinfo
  {author} {\bibfnamefont{T.}~\bibnamefont{Hickel}},\ and\ \bibinfo {author}
  {\bibfnamefont{J.}~\bibnamefont{Neugebauer}},\ }%
  \bibfield{journal}{%
  \Doi{10.1103/PhysRevB.76.024309}{\bibinfo {journal} {Phys. Rev. B}}\ }%
  \textbf{\bibinfo {volume} {76}},\ \bibinfo {pages} {024309} (\bibinfo {month}
  {Jul}\ \bibinfo {year} {2007}),\
  \url{http://link.aps.org/doi/10.1103/PhysRevB.76.024309}%
  \bibAnnoteFile{NoStop}{PhysRevB.76.024309}%
\bibitem{Bradley}%
  \BibitemOpen
  \bibfield{author}{%
  \bibinfo {author} {\bibfnamefont{C.~J.}\ \bibnamefont{Bradley}}\ and\
  \bibinfo {author} {\bibfnamefont{A.~P.}\ \bibnamefont{Cracknell}},\ }%
  \emph{\bibinfo {title} {The Mathematical Theory of Symmetry in Solids:
  Representation Theory for Point Groups and Space Groups}}\ (\bibinfo
  {publisher} {Oxford: Clarendon Press},\ \bibinfo {year} {1972})%
  \bibAnnoteFile{NoStop}{Bradley}%
\bibitem{PAW_optics}%
  \BibitemOpen
  \bibfield{author}{%
  \bibinfo {author} {\bibnamefont{{M. Gajdo\ifmmode \check{s}\else \v{s}\fi{}
  and K. Hummer and G. Kresse and J. Furthm\"uller and F. Bechstedt}}},\ }%
  \bibfield{journal}{%
  \Doi{10.1103/PhysRevB.73.045112}{\bibinfo {journal} {Physical Review B}}\ }%
  \textbf{\bibinfo {volume} {73}},\ \bibinfo {pages} {045112} (\bibinfo {month}
  {Jan}\ \bibinfo {year} {2006}),\
  \url{http://link.aps.org/doi/10.1103/PhysRevB.73.045112}%
  \bibAnnoteFile{NoStop}{PAW_optics}%
\bibitem{GWA_and_QP_review_1999}%
  \BibitemOpen
  \bibfield{author}{%
  \bibinfo {author} {\bibfnamefont{W.~G.}\ \bibnamefont{Aulbur}}, \bibinfo
  {author} {\bibfnamefont{L.}~\bibnamefont{J\"onsson}},\ and\ \bibinfo {author}
  {\bibfnamefont{J.~W.}\ \bibnamefont{Wilkins}}\ }%
  (\bibinfo {publisher} {Academic Press},\ \bibinfo {year} {1999})\ pp.\
  \bibinfo {pages} {1 -- 218},\
  \url{http://www.sciencedirect.com/science/article/pii/S0081194708602489}%
  \bibAnnoteFile{NoStop}{GWA_and_QP_review_1999}%
\bibitem{Kohanoff}%
  \BibitemOpen
  \bibfield{author}{%
  \bibinfo {author} {\bibfnamefont{J.}~\bibnamefont{Kohanoff}},\ }%
  \emph{\bibinfo {title} {Electronic Structure Calculations for Solids and
  Molecules : Theory and Computational Methods}}\ (\bibinfo {publisher}
  {Cambridge University Press; Cambridge},\ \bibinfo {year} {2006})%
  \bibAnnoteFile{NoStop}{Kohanoff}%
\bibitem{JudithThesis2008}%
  \BibitemOpen
  \bibfield{author}{%
  \bibinfo {author} {\bibfnamefont{J.}~\bibnamefont{Harl}},\ }%
  \emph{\bibinfo {title} {The Linear Response Function in Density Functional
  Theory: Optical Spectra and Improved Description of the Electron
  Correlation}},\ Ph.D. thesis,\ \bibinfo {school} {University of Vienna}
  (\bibinfo {year} {2008}),\ \url{http://othes.univie.ac.at/2622/}%
  \bibAnnoteFile{NoStop}{JudithThesis2008}%
\bibitem{Note8}%
  \BibitemOpen
  \bibinfo {note} {The physical meaning of the RPA is that electrons are
  considered to respond to the total (external plus induced) field
  independently \cite {The_GW_method_1998}.}%
  \bibAnnoteFile{Stop}{Note8}%
\bibitem{Note9}%
  \BibitemOpen
  \bibinfo {note} {Recall that the optical region (visible spectrum) is about
  $(390 \sim 750)$ nm which corresponds to $(3.183 \sim 1.655) \protect
  \tmspace +\thickmuskip {.2777em} eV$.}%
  \bibAnnoteFile{Stop}{Note9}%
\bibitem{DFT_vs_GW_2002}%
  \BibitemOpen
  \bibfield{author}{%
  \bibinfo {author} {\bibfnamefont{G.}~\bibnamefont{Onida}}, \bibinfo {author}
  {\bibfnamefont{L.}~\bibnamefont{Reining}},\ and\ \bibinfo {author}
  {\bibfnamefont{A.}~\bibnamefont{Rubio}},\ }%
  \bibfield{journal}{%
  \Doi{10.1103/RevModPhys.74.601}{\bibinfo {journal} {Rev. Mod. Phys.}}\ }%
  \textbf{\bibinfo {volume} {74}},\ \bibinfo {pages} {601} (\bibinfo {month}
  {Jun}\ \bibinfo {year} {2002}),\
  \url{http://link.aps.org/doi/10.1103/RevModPhys.74.601}%
  \bibAnnoteFile{NoStop}{DFT_vs_GW_2002}%
\bibitem{exact_X_Becke_1993}%
  \BibitemOpen
  \bibfield{author}{%
  \bibinfo {author} {\bibfnamefont{A.~D.}\ \bibnamefont{Becke}},\ }%
  \bibfield{journal}{%
  \Doi{10.1063/1.464913}{\bibinfo {journal} {The Journal of Chemical Physics}}\
  }%
  \textbf{\bibinfo {volume} {98}},\ \bibinfo {pages} {5648} (\bibinfo {year}
  {1993}),\ \url{http://link.aip.org/link/?JCP/98/5648/1}%
  \bibAnnoteFile{NoStop}{exact_X_Becke_1993}%
\bibitem{Hybrid_DFT_2002}%
  \BibitemOpen
  \bibfield{author}{%
  \bibinfo {author} {\bibfnamefont{K.~N.}\ \bibnamefont{Kudin}}, \bibinfo
  {author} {\bibfnamefont{G.~E.}\ \bibnamefont{Scuseria}},\ and\ \bibinfo
  {author} {\bibfnamefont{R.~L.}\ \bibnamefont{Martin}},\ }%
  \bibfield{journal}{%
  \Doi{10.1103/PhysRevLett.89.266402}{\bibinfo {journal} {Phys. Rev. Lett.}}\
  }%
  \textbf{\bibinfo {volume} {89}},\ \bibinfo {pages} {266402} (\bibinfo {month}
  {Dec}\ \bibinfo {year} {2002}),\
  \url{http://link.aps.org/doi/10.1103/PhysRevLett.89.266402}%
  \bibAnnoteFile{NoStop}{Hybrid_DFT_2002}%
\bibitem{Hybrid_2008}%
  \BibitemOpen
  \bibfield{author}{%
  \bibinfo {author} {\bibfnamefont{E.~N.}\ \bibnamefont{Brothers}}, \bibinfo
  {author} {\bibfnamefont{A.~F.}\ \bibnamefont{Izmaylov}}, \bibinfo {author}
  {\bibfnamefont{J.~O.}\ \bibnamefont{Normand}}, \bibinfo {author}
  {\bibfnamefont{V.}~\bibnamefont{Barone}},\ and\ \bibinfo {author}
  {\bibfnamefont{G.~E.}\ \bibnamefont{Scuseria}},\ }%
  \bibfield{journal}{%
  \Doi{10.1063/1.2955460}{\bibinfo {journal} {The Journal of Chemical
  Physics}}\ }%
  \textbf{\bibinfo {volume} {129}},\ \bibinfo {eid} {011102} (\bibinfo {year}
  {2008}),\ \url{http://link.aip.org/link/?JCP/129/011102/1}%
  \bibAnnoteFile{NoStop}{Hybrid_2008}%
\bibitem{PhysRevLett.96.116405}%
  \BibitemOpen
  \bibfield{author}{%
  \bibinfo {author} {\bibfnamefont{W.-G.}\ \bibnamefont{Yin}}, \bibinfo
  {author} {\bibfnamefont{D.}~\bibnamefont{Volja}},\ and\ \bibinfo {author}
  {\bibfnamefont{W.}~\bibnamefont{Ku}},\ }%
  \bibfield{journal}{%
  \Doi{10.1103/PhysRevLett.96.116405}{\bibinfo {journal} {Phys. Rev. Lett.}}\
  }%
  \textbf{\bibinfo {volume} {96}},\ \bibinfo {pages} {116405} (\bibinfo {month}
  {Mar}\ \bibinfo {year} {2006}),\
  \url{http://link.aps.org/doi/10.1103/PhysRevLett.96.116405}%
  \bibAnnoteFile{NoStop}{PhysRevLett.96.116405}%
\bibitem{arXiv_hybrid_2012}%
  \BibitemOpen
  \bibfield{author}{%
  \bibinfo {author} {\bibfnamefont{C.}~\bibnamefont{Franchini}}, \bibinfo
  {author} {\bibfnamefont{R.}~\bibnamefont{Kovacik}}, \bibinfo {author}
  {\bibfnamefont{M.}~\bibnamefont{Marsman}}, \bibinfo {author}
  {\bibfnamefont{S.~S.}\ \bibnamefont{Murthy}}, \bibinfo {author}
  {\bibfnamefont{J.}~\bibnamefont{He}}, \bibinfo {author}
  {\bibfnamefont{C.}~\bibnamefont{Ederer}},\ and\ \bibinfo {author}
  {\bibfnamefont{G.}~\bibnamefont{Kresse}},\ }%
  \enquote{\bibinfo {title} {Maximally localized wannier functions in
  {LaMnO}$_{3}$ within {PBE+U}, hybrid functionals, and partially
  self-consistent {GW}: an efficient route to construct ab-initio tight-binding
  parameters for eg perovskites},}\  (\bibinfo {month} {May}\ \bibinfo {year}
  {2012}),\ \Eprint{http://arxiv.org/abs/1111.1528v2}{arXiv:1111.1528v2
  [cond-mat.str-el]}%
  \bibAnnoteFile{NoStop}{arXiv_hybrid_2012}%
\end{thebibliography}%

\end{document}